\documentclass{jfm}
\usepackage{graphicx}

\usepackage{amsfonts}
\usepackage{amsmath}

\usepackage{natbib}
\usepackage{hyperref}
\hypersetup{breaklinks,colorlinks=true,citecolor=blue,linkcolor=red,}

\def\beqn{\begin{equation}}
\def\eeqn{\end{equation}}
\def\b{\textcolor{blue}}

\title{A Hydrodynamical Thermal Irradiated Wind from the Outer Thin Accretion Disk in Low-luminosity Active Galactic Nuclei}
\date{}

\author{Nagendra Kumar \aff{1}\aff{2}
  \corresp{\email{nagendra.bhu@gmail.com}; * current address }
\corresp{ GEC Bhojpur, Arrah-802301, India}
}
\affiliation{\aff{1*}Aazad Path, New Bengali Tola, Mithapur, Patna-800001;  \aff{2} Department of Physics, Indian Institute of Science, Bangalore 560012, India}

\begin{document}
\maketitle

\begin{abstract}
  Evidently, low-luminosity active galactic nuclei  (LLAGNs) are comprised of an inner advective disk and an outer geometrically thin disk. Wind is inevitable in LLAGNs, mainly interpreted in an  indirect way, also the evidence is growing for the presence of wind in the outer thin disk. We present a hydrodynamics (HD) model for wind  from the outer thin disk, where the main driver is the inner disk irradiation (which is parameterized by a number $x$ in hydrostatic equilibrium equation) and the heating mechanism is  photoionization. The model works for  low-intensity irradiation or from a height $z_s$ in the optically thin medium. We solve the model equations in cylindrical coordinates  along the $z$-axis for a given radius $r$ with assuming a tiny vertical speed $v_z$ ($\ll c_s$ sound speed).  The sonic point conditions assure an isobaric regime above the sonic height ($z^{max}$); in addition to the height $z_f (\ll z^{max}$), the radial pressure gradient also supports the fluid rotation, and both jointly assure a wind ejection from the $z^{max}$ with fluid speed.
The $z^{max}$ increases with $x$, and beyond a large $z^{max}$ (say $z^{max}_t$ corresponding to maximum $x$), there is no physical solution.
We start the computation from the outer radius $r_o^{thin}$ to the inner $r_{in}^{thin}$ with a Bondi mass accretion rate $\dot{M}_{Bondi}$, to explore the $r$ dependency of the  mass inflow rate $\dot{M}$ and wind properties. We constrain the model by fixing $\dot{M}$ at $r_{in}^{thin}$ from the observations of NGC 1097 and check the feasibility of the model by comparing the energetics with the observed bolometric luminosity. The wind is an equatorial with a viewing angle $i>85$ degrees and capable to generate red/blueshifted lines, which would be a general characteristics for LLAGNs.
\end{abstract}




\section{Introduction}\label{sec:intro}
Active galactic nuclei (AGNs)  harbor a supermassive black hole (SMBH) 
that accretes the surrounding materials via an accretion disk.
The broadband spectral energy distribution (SED) of AGNs 
indicates that based on their luminosity,  AGNs could be classified into low-luminosity AGNs (LLAGNs) and luminous AGNs. In general, $L_{bol} < 0.001 L_{Edd}$ characterizes the LLAGNs \cite[see][]{Ho2008}, where $L_{bol}$ is the X-ray bolometric luminosity and $L_{Edd}$ is the Eddington luminosity.
The broadband SED of LLAGNs does not exhibit the optical-ultraviolet 'big blue bump'; rather, it shows a mid- or  near-IR 'red bump' with a steep optical-UV slope, commonly, without having a torus
(\citealp[e.g.,][]{Perlman-etal2007, Ho2008} (see for review); \citealp[][]{Elitzur-Ho2009, Gu-Cao2009, Younes-etal2012, She-etal2018, Younes-etal2019}). 
It also reflects that the underlying accretion process in LLAGNs is different from that in the luminous AGNs. 
The X-ray emission of LLAGNs cannot be explained by a standard geometrically thin disk, as the temperature of the inner region of the disk is comparatively too small to generate the X-ray emission, even for the mass accretion rate $\dot{M}_{bol}$
\big($= \frac{L_{bol}}{ \eta c^2}$, where c is the speed of light and $\eta$ is the efficiency) \cite[e.g.,][]{Lasota-etal1996, Narayan-etal1998, Quataert-etal1999}.
An advective-type accretion flow \cite[e.g.,][]{Narayan-Yi1994, Narayan-Yi1995, Chakrabarti-Titarchuk1995}, especially a radiatively inefficient accretion flow (RIAF; \citealp[e.g.,][see for review]{Yuan-etal2003, Yuan-Narayan2014}) is vital for the X-ray generation.
Many LLAGNs display broad double-peaked (red- $\&$ blueshifted) H$\alpha$ and H$\beta$ emission lines, which affirms the presence of a thin accretion disk in the outer part of the accretion disk (\citealt{Storchi-etal2003,Storchi-etal2017, Lewis-etal2010}, \citealt{Asmus-etal2011}). Recently, \cite{Murchikova-etal2019} have discovered a cooler ionized gas thin disk ($\sim 10^4$ K) around Sgr A* within 4 $\times 10^4 R_g$ by observing a broad double-peaked H30$\alpha$ recombination emission line, where $R_g$ is the gravitational radius. 
Therefore, the LLAGNs have both types of accretion flows, a hot flow (or advective flow in the inner region of the accretion disk) and a thin disk (cold flow in the outer region) with transition radii $R_{tr}$ $\sim$ 500 -- 2000 $R_g$ (\citealt{Nemmen-etal2006, Nemmen-etal2014, Ho2008, Schimoia-etal2015, Storchi-etal2017, Reb-etal2018}). 
However, the transition from the cold disk to  the hot disk is still not well understood.

The high spatial resolution and sensitivity of ihe  Chandra X-ray Observatory
facilitates the measurement of the nuclear X-ray emission and diffuse X-ray emission of its surrounding hot interstellar medium (ISM).
Particularly, the diffuse X-ray emission provides the Bondi accretion radius $r_{acc}$ (or sphere of influence; the gravitational pull of the SMBH dominates over the internal thermal energy of the gas) and Bondi mass accretion rate $\dot{M}_{Bondi}$ at $r_{acc}$ by estimating the temperature ($T_{ism}$) and electron number density ($n_{ism}$) of the gas from the spectrum of diffused (unresolved) X-ray emission and the surface brightness.
$r_{acc}$ and $\dot{M}_{Bondi}$ are written for an SMBH of mass $M_{c}$ as \cite[e.g.,][]{DiMatteo-etal2003}
\beqn \label{eq:rbondi}
r_{acc} \approx 0.05\left(\frac{0.8 keV}{kT_{ism}}\right)\left(\frac{M_{c}}{10^9 M_\odot}\right) kpc
\eeqn
\beqn \label{eq:mdbondi}
\dot{M}_{Bondi} \approx 7 \times 10^{23} \left(\frac{n_{ism}}{0.17 cm^{-3}}\right)\left(\frac{0.8 keV}{kT_{ism}}\right)^{3/2} \left(\frac{M_{c}}{10^9M_\odot}\right)^2 g s^{-1}
\eeqn
In LLAGNs, the total bolometric luminosity ($L^T_{bol}$) is many orders of magnitude less than the luminosity generated in the accretion disk with the Bondi mass accretion rate ($L_{Bondi} = \eta \dot{M}_{Bondi} c^2$, \citealp[e.g.,][]{Pellegrini2005, Soria-etal2006, Russell-etal2013}).
$L^T_{bol} \ll L_{Bondi}$ can be understood in RIAF  as (i) the accretion rate is much less than the $\dot{M}_{Bondi}$ or (ii) the accretion rate is of the order of $\dot{M}_{Bondi}$ with a very low radiative efficiency or an outflow solution.
Both arguments are degenerate over the broadband spectrum of LLAGNs, this apparent degeneracy usually can be lifted out by submillimeter polarization and Faraday rotation measurements, which generally predict a very small mass accretion rate in comparison to the Bondi mass accretion rate 
near to the SMBH  ($r < 100 R_g$, 
\citealp[][]{Quataert-Gruzinov2000, Marrone-etal2007, Sharma-etal2007, Feng-etal2016}).
To describe the SED of LLAGNs the more preferable model is the RIAF solution with outflow; here the outflow is parameterized in terms of mass accretion rate, which decreases with decreasing radius, or $\dot{M}(r) = \dot{M}(r_o) \left(\frac{r}{r_o}\right)^s$, where $r_o$ is the outer radius of RIAF (e.g., \citealt[and references therein]{Nemmen-etal2006, Wang-etal2013}, see also \citealt{Blandford-Begelman1999, Narayan-etal2000, Becker-etal2001}).
Numerous advection-disk-based hydrodynamical (HD) and magnetohydrodynamical (MHD) numerical simulations have been performed, and 
their estimated 
ranges of $s$ are $\sim(0.5-1)$ \cite[][and references therein]{Stone-etal1999, Yuan-etal2015}. 
In general, by SED modeling in ADAF/ RIAF for LLAGNs, the estimated ranges of $s$ and $r_o$ are $\sim(0.3 - 1)$ and $\sim(100 - 10^5 R_g)$, respectively; here the smaller $r_o$ ($< 10^3 R_g$) is estimated from the observed double-peaked H$\alpha$ line \cite[][]{Yuan-etal2009, Nemmen-etal2014}. 

The wind outflow is inevitable in LLAGNs. It cannot be limited only in the inner region of the disk (or RIAF/ ADAF), but it will also be  launched from the outer region  (or thin disk).
Usually, the winds are identified through a blueshifted UV and X-ray absorption/emission lines, which require a high-resolution  spectroscopy, and  possibly a luminous AGN (atleast a moderate one). \cite{Crenshaw-Kraemer2012} have studied a few moderately luminous AGNs (including one LLAGN) and obtained a wind speed of $\sim$500 km/s (i.e., for a disk wind, the wind would be launched from the outer region; see also, \citealt{Tombesi-etal2014} for a wind detection in radio-loud galaxies and note that many LLAGNs are also radio-loud \citealt{Terashima-Wilson2003}).
Recently, \cite{Goold-etal2023} claimed the detection of an outflow from two nearby LLAGNs (NGC 1052 and Sombrero) in the JWST survey by analyzing the blueshifted emission line in the mid-IR band.
A hot wind has been detected in two LLAGNs (M81 and NGC 7213) by \cite{Shi-etal2021, Shi-etal2022}  by identifying blueshifted (also redshifted) emission lines of Fe XXVI Ly$\alpha$ and Fe XXV K$\alpha$ using Chandra. The authors claim that these lines are generated in the hot accretion flow that is also present beyond  10$^6 R_g$, and disfavor the alternative explanations like AGN photoionization and stellar activities. 
In LLAGNs the evidence for wind outflow is mainly indirect, e.g., the absorption column density in X-rays increases with increasing luminosity \cite[][]{She-etal2018}, and these are bisymmetric emission features in the  H$\alpha$-EW resolved map (\citealt{Cheung-etal2016}, see also \citealt{Roy-etal2018}) and Faraday rotation in the jet (\citealt{Park-etal2019}, see also, for jet-driven wind, \citealt{May-etal2018}).
In this work, we aim to explore the wind launching mechanism from the outer region of the disk.

In thin disks (AGNs), apart from magnetic-driven wind \cite[e.g.,][]{Reynolds-2012, Chakravorty-etal2016} the wind can be a line-driven wind \cite[e.g.,][]{Murray-etal1995, Proga-etal2000} or a Compton heated thermal wind \cite[]{Begelman-etal1983}.
\cite{Giustini-Proga2019} have shown that the line-driven wind cannot be possible in LLAGNs even for $\dot{M} <0.01\dot{M}_{Edd}$ with $M_{BH} <10^8 M_\odot$.
The Compton-heated thermal wind is also generated at the outer disk due to the irradiation from the inner disk region. The heating mechanism of the outer disk (due to the irradiation) is a Compton scattering process, and when its  temperature increases by a sufficient large value that the thermal velocity exceeds the escape velocity (the corresponding limiting temperature is termed as Compton temperature, $T_{C}$ $\sim 10^7 - 10^8 K$), a wind is launched from this site. And, in general, for a temperature $< 10^{4.5} K$ the Compton heating is not possible; only photoionization heating is preferable \cite[]{Begelman-etal1983}. The temperature $T_c$ of the thin disk at radius $r = 2000R_g$ varies from $\sim$4300 to $\sim$8600 K by varying $\dot{M} = 0.001$ to $0.01\dot{M}_{Edd}$ for $M_{c}=10^8M_{\odot}$ and $\alpha$ = 0.1. Hence, in LLAGNs, the temperature of the outer region of disk is always $< 10^4 K$, therefore, at this site, the Compton scattering process cannot be triggered, or in other words, the Compton-heated thermal wind is not possible in LLAGNs. In addition, the high-energy photons can photoionize the disk, but the recombination rate is comparatively high at low temperatures \cite[see, e.g.,][]{Verner-Ferland1996}, so it cannot ignite the Compton-heated thermal wind. 
Recently, \citet[][hereafter Paper I]{Kumar-Mukhopadhyay2021} studied a thermal-irradiation-induced  wind outflow from the outer region of the thin disk ($r>1000R_g$), where the inner region of the disk irradiates the outer region and the dominant heating mechanism at the outer region is the photoionization. The wind is launched from a sonic height with fluid speed for a given radius. The sonic height increases with an increase in the magnitude of the irradiation.  The wind is mainly an equatorial wind. They explored the wind characteristics for a range of $r$ for the X-ray binaries with considering a constant mass accretion rate.
In LLAGNs, the outer disk must be irradiated by the inner region. Thus, the thermal-irradiation-induced wind mechanism is  plausible in the outer region of
LLAGNs.

In this paper, we extend the Paper I study. We develop the model for the LLAGNs, and due to the low-intensity irradiation from the inner region of the disk (or RIAF), in the present model, the irradiation effect is accounted for from the optically thin regime  of the outer region of the disk (a region above the scale height of the thin disk). We explore the mass accretion loss due to the wind  outflow as a function of radius; for this, we start the computation from three-forth of the Bondi accretion radius (or the outer radius of the thin disk) with the Bondi mass accretion rate. 
In general, in LLAGNs, the wind is ejected almost along the disk plane (with viewing angle $i >$ 85 degrees). 
In the next section, we briefly review the model and solution procedure. 
In section \S 3, we describe the general model results at a fixed radius and examine the assumptions and validity of the solutions.
In section \S 4, we study the mass inflow rate (and wind characteristics) as a function of radius starting from the outer radius of the thin disk (or the Bondi accretion radius).
Finally in section \S 5 we constrain the model results with observations, followed by a summary in section \S 6.

\section{Model}

To study the wind outflow in the outer region of the disk of LLAGNs, we consider a 2.5D  accretion disk formalism in cylindrical coordinates ($r,\phi,z$), and the accretion flow is  
a steady
\big($\frac{\partial}{\partial t}$ $\equiv$ 0\big) and an axisymmetric 
\big($\frac{\partial}{\partial \phi}$ $\equiv$ 0\big).
The hydrodynamics equations are (\citealp[e.g.,][]{Bisnovatyi-Lovelace2001}; Paper I)

\beqn\label{eq:conti}
\frac{1}{r}\frac{\partial (r\rho v_r)}{\partial r} + \frac{\partial (\rho v_z)}{\partial z} = 0,
\eeqn
\beqn \label{eq:radmom}
v_r \frac{\partial v_r}{\partial r} + v_z \frac{\partial v_r}{\partial z} - \frac{\lambda^2}{r^3} + \frac{1}{\rho}\frac{\partial p}{\partial r} + F_r  =\frac{1}{\rho}\frac{\partial W_{rz}}{\partial z},
\eeqn
\beqn \label{eq:azimom}
v_r \frac{\partial \lambda }{\partial r} + v_z \frac{\partial \lambda }{\partial z} = \frac{r}{\rho}\left[\frac{1}{r^2}\frac{\partial (r^2 W_{\phi r})}{\partial r} + \frac{\partial W_{\phi z} }{\partial z}\right],
\eeqn
\beqn \label{eq:varmom}
v_r \frac{\partial v_z }{\partial r} + v_z \frac{\partial v_z}{\partial z} + \frac{1}{\rho} \frac{\partial p}{\partial z} + F_z = \frac{1}{r\rho}\frac{\partial rW_{zr}}{\partial r},
\eeqn
\beqn \label{eq:energ}
\frac{v_r}{\Gamma_3 - 1} \left[\frac{\partial p}{\partial r}- \Gamma_1\frac{p}{\rho}\frac{\partial \rho}{\partial r}\right] + \frac{v_z}{\Gamma_3 -1}\left[\frac{\partial p}{\partial z} - \Gamma_1 \frac{p}{\rho} \frac{\partial \rho}{\partial z}\right] = 0. 
\eeqn
Here, $v_r$, $v_\phi$ and $v_z$ are the radial, azimuthal, and vertical veocity components, and $\lambda$ (=$r v_\phi$) is the specific angular momentum. $\rho$ is the mass density, and $p$ is the fluid pressure. $F_r$ and $F_z$ are the magnitudes of the radial and vertical
components of Newtonian gravitational force by the compact object,
respectively. $\Gamma_1$ and $\Gamma_3$ are adiabatic exponents. Equation (\ref{eq:conti}) is the equation of continuity, equations (\ref{eq:radmom}-\ref{eq:varmom}) are the
momentum balance equations, and equation (\ref{eq:energ}) is the energy equation.

\subsection{The Model's assumption}

 We adopt a gas-pressure-dominated regime,  $p \gg p_{rad}$, where $p_{rad}$ is the radiation pressure. The equation of state is $p$ = $k\rho T/\mu m_p$, where $k$ is the Boltzmann constant, $m_p$ is the mass of the proton, $\mu$ is the mean molecular weight, and $T$ is the temperature.
The sound speed of the medium is $c_s\sim \sqrt{p/\rho}$.

\noindent{\tt \textbf {Viscosity :}}
We adopt the $\alpha$-prescriptions of \cite{Shakura-Sunyaev1973} for tangential
shear stress $W_{\phi r}$, it is expressed as $W_{\phi r}$ $\left(=\eta r \frac{\partial \Omega}{\partial r}\right)$  = $\alpha p$, where $\eta =\alpha c_s h \ \rho$ is the dynamical viscosity, $\alpha$ is the Shakura-Sunyaev viscosity parameter,
$\Omega$ is the Keplerian angular velocity, and $h$ is the scale height of the Keplerian disk at radius $r$.
We also account for another tangential shear stress  $W_{\phi z}$ in the calculations and approximate it in terms of $W_{\phi r}$,  $\frac{W_{\phi z}}{W_{\phi r}} \approx$
 $\frac{\partial \Omega}{\partial z}\left / \frac{\partial \Omega}{\partial r} \right.$  $\approx$ $\frac{z}{r}$.
Other shearing stresses, $W_{r z}$ and $W_{z r}$ due to the motion of $v_r$ and $v_z$, respectively are comparatively negligible, as $v_z$, $v_r$ $\ll$ $v_\phi$, and we assume that $W_{rz}$ = $W_{zr}$ $\approx$ 0, or  
\beqn \label{eq:rzstress}
\frac{\partial v_r}{\partial z} + \frac{\partial v_z}{\partial r} = 0.
\eeqn
Here, in the notation of the viscous shearing stress W$_{ij}$, the first subscript is for the direction of the stress, and the second is for the outward normal to the surface on which it acts. In addition, we assume that the $\alpha$-prescriptions for $W_{\phi r}$ are also valid at any height above the scale height, $W_{\phi r} (z)$ = $\alpha p(z)$. 

\noindent{\tt \textbf{Opacity and heating and cooling :}} Like the Keplerian disk, we assume that the medium is optically thick and the total optical depth $\tau$ at midplane is very large, i.e., $\tau^{tot} \gg 1$, where $\tau$ at height $z$ is $\tau(z)= \int_z^\infty \kappa \rho dz$, and $\kappa$ is the opacity.
The annular rings are in the local thermodynamic equilibrium, so the annular rings radiate like a blackbody. 
As our interest in solutions is
focused on the outer region, we consider that the opacity is due to the free-free absorption $\sigma_{ff}$, which
is the Rosseland mean opacity. As discussed in Section \ref{sec:intro}, the outer thin disk is at a low temperature ($< 10^4K$). At low temperature,  the source of opacity is mainly the molecules and dust grains, and the corresponding magnitude of the Rosseland opacity is sufficient to make the medium optically thick \cite[see, e.g.,][]{Alexander-Ferguson1994}.
The heat is generated  predominantly by the viscous process,
and the disk immediately cools locally in the vertical direction by blackbody emission. Hence, on the right-hand side of the energy equation (\ref{eq:energ}), $q^+ - q^- =$ 0, where $q^+$ is the rate of heat generation per unit volume and $q^-$ is the rate of radiated energy density.


\noindent {\tt \textbf{Hydrostatic equilibrium and external disk irradiation:}}
In a Keplerian disk \cite[e.g.,][]{Shakura-Sunyaev1973,Frank-etal2002}, it is thought that there is no flow in the vertical $z$-direction; i.e., the disk is in
hydrostatic equilibrium vertically. In addition, the disk has a concave shape, the inner disk can shine on the outer region of the disk. In this work, we consider the irradiation of the outer disk by the inner disk. In principle, the irradiation  can
introduce a radiation pressure $p^{irr}_{rad}$, 
and can unbalance the vertical hydrostatic equilibrium.
The irradiated flux $\epsilon^{irr}$ at the outer radius $r$ by the inner disk of luminosity $L_{bol}$ can be  expressed by equation (\ref{eq:irr_en}). The radiation pressure is defined as  $p^{irr}_{rad}$ $\sim$ $\epsilon^{irr}/c$. We find that $p^{irr}_{rad}$ $\ll$ $p$ for $L_{bol}$ $<$ $10^{45}$erg/s and $\dot{M} \sim 0.0001\dot{M}_{Edd}$, $M_c$ $>$ $ 5 \times 10^6 M_{\odot}$, where  $\dot{M}$ is the mass accretion rate and $M_c$ is mass of the compact object. Therefore, in the outer region, the radiation pressure due to the irradiation is negligible in comparison to the gas pressure.

  The deviation from the vertical hydrostatic equilibrium of the fluid is expressed as
\beqn \label{eq:hydro}
\frac{1}{\rho}\frac{\partial p}{\partial z}= -(1-x)F_z
\eeqn
Here,  $x$ ($<$ 1) is a number, and for $x$ = 0, the disk is in vertical
hydrostatic equilibrium. After rearranging the terms, the equation (\ref{eq:hydro}) can be expressed for $x$ at a given height $z$ as $x = 1 + \left.\left(\frac{1}{\rho F_z}\frac{\Delta p}{\Delta z}\right)\right|_{z}$, where $\Delta p$ = [$p(z+\Delta z) - p(z)$] and $\Delta z$ is a small increment at height $z$.
We can note here that for a given height $z$, the pressure will increase with $x$, which indicates an external heating effects as mentioned in the above paragraph (see the detailed discussions in Paper I).


\subsection{Solution procedure and Discussions} \label{sub:sol}

Combining  the  equation of continuity (equation \ref{eq:conti}), the momentum balance equations (\ref{eq:radmom}-\ref{eq:varmom}), the
energy equations (\ref{eq:energ}) and equations (\ref{eq:rzstress})-(\ref{eq:hydro}) we obtain
\begin{align} 
\label{eq:master}
\nonumber
\frac{\partial v_z}{\partial z}\left[\frac{v_z^2-v_r^2}{v_r}\frac{(-\alpha r)\Gamma_1c_s^2}{v_r^2-\Gamma_1c_s^2}\right] &=\frac{3 W_{r\phi}}{\rho}+ \alpha z\frac{1}{\rho} \frac{\partial p}{\partial z} -v_r \frac{\partial \lambda}{\partial r}-v_z \frac{\partial \lambda}{\partial z} \nonumber \\ 
+& \alpha r \left[v_z\frac{xF_z}{v_r} +f^r_{bal} -\frac{v_r^2}{r}- 
\frac{1}{\rho} \frac{\partial p}{\partial z}\frac{v_zv_r}{\Gamma_1 c_s^2} \right]\frac{\Gamma_1 c_s^2}{ v_r^2-\Gamma_1 c_s^2 }
,   
\end{align}

\noindent where $f^r_{bal}$ = $-\frac{\lambda^2}{r^3}+F_r$.
Above, $\frac{\partial v_z}{\partial z} $ is expressed in terms of $\frac{\partial \lambda}{\partial r} $, and $\frac{\partial \lambda}{\partial z} $. For the
unique solution, we compute the derivatives
of $\lambda(z)$ as a function of height at
a given radius
\big(i.e.,$\frac{\partial \lambda}{\partial r}, \frac{\partial \lambda}{\partial z}$\big) using equation (\ref{eq:radmom}) 
by neglecting the higher order derivatives.
In addition, to obtain the unique solution, one has to know one more variable prior to the computation. For this we assume that $\frac{1}{\rho}\frac{\partial \rho}{\partial r} (r,z) = \frac{1}{\rho}\frac{\partial \rho}{\partial r}(r)$. This assumption can be justified as, since the fractional change of density between radius $r$ and $r+\Delta r$ $\left.\frac{\Delta \rho}{\rho}\right|_r = \frac{\rho(r+\Delta r,z)-\rho(r,z)}{\rho(r,z)}$ is a very small quantity, so one may assume that it does not vary with height, i.e., $\frac{\rho(r+\Delta r,z)-\rho(r,z)}{\rho(r,z)} = \frac{\rho(r+\Delta r,z=0)-\rho(r,z=0)}{\rho(r,z=0)}$.
We solve the partial differentials $\frac{\partial v_z}{\partial z}$, $\frac{\partial v_r}{\partial z}$ and $\frac{\partial c_s}{\partial z} $  simultaneously.

\noindent{\tt \textbf{(a) Initial conditions and Irradiation height:}} 
We solve the governing equations along the $z$-axis 
at a given launching radius $r$ from the midplane.
To start, we assume a very small initial vertical speed on the midplane ($z=0$) in comparison to the sound speed and define its magnitude in  ratio of radial velocity magnitude as 
\beqn v_z = f_v |v_r|, \eeqn
where $f_v$ is a number. We notice that the condition $v_z(z=0) \ll c_s$ 
assures these conditions,
$v_z\frac{\partial v_r}{\partial z}$, $v_r\frac{\partial v_r}{\partial r}$ $\ll$ $\frac{1}{\rho}\frac{\partial p}{\partial r}$ (while, $\frac{1}{\rho}\frac{\partial p}{\partial r}$  $\ll$ $F_r$ already) and $v_z\frac{\partial v_z}{\partial z}$, $v_r\frac{\partial v_z}{\partial r}$ $\ll$ $\frac{1}{\rho}\frac{\partial p}{\partial z}$.
Therefore, at least near to the midplane, the governing equations (\ref{eq:conti})-(\ref{eq:energ}) of the disk  become equivalent to the Keplerian disk \citet[][]{Shakura-Sunyaev1973}. 
With this equivalency of the present formalism to the Keplerian disk, we
initialize the flow variable to its respective Keplerian value, especially the
outer region solution, where the opacity comes mainly from the free-free absorption \cite[e.g.,][]{Shakura-Sunyaev1973, Novikov-Thorne1973, Frank-etal2002}.
The initial values of the flow variables would be a function of $\dot{M}$, $M_c$ and $\alpha$. 
In short, we start to solve the governing equations for wind outflow along the
$z$-axis from the midplane of the disk at radius $r$ and we initialize the flow variables to their respective  Keplerian value of the outer region. 
We  adopt 
a positive sign convention; i.e., the radially inward direction is negative,
and the vertical outward direction is positive. In this sign convention the radial inflow velocity $v_r$ is negative and vertical outflow velocity $v_z$ is positive. We consider a negative $\alpha$ to  ensure the angular momentum conservation
in this sign convention as prescribed by \citet{Bisnovatyi-Lovelace2001}. 


\noindent{\tt Irradiation height:} In Paper I for simplicity, we obtained the model solutions considering $x$ (or the irradiation effect) from the midplane.
In general, we found that for a given parameter set and given $x$
the fluid temperature increases in such a way that the pressure and density both decrease.
As the required energy to increase the medium temperature depends on the opacity (or mainly optically thick versus thin) of the medium, and it is measured by using the flux formula of the  blackbody ($\propto T^4$) for  the optically thick medium, while for the optically thin medium, it is measured in terms of enhancement in internal energy density ($\propto T)$.
However, the present formalism works only for $p^{irr}_{rad} \ll p$, in addition, for LLAGNs (even for $L_{bol}< 0.01L_{Edd}$), we find that $p^{irr}_{rad} \ll p_{rad} \big(=\frac{\sigma T_c^4}{c}\big)$ for $r<2\times 10^6R_g$, $\dot{M} >0.0005\dot{M}_{Edd}$ and $M_c =10^8 M_\odot$. Or, in other words, the irradiated flux from the inner region is significantly smaller than the radiated flux of the disk ($\sigma T_c^4$) at that $r$.
Hence, in the above circumstance, the model  will work only in the optically thin medium.
Therefore, we have to account for the irradiation effect from the appropriate height $z_s$ above the midplane rather than the midplane (which is an optically thick and used in Paper I). 
In the literature, the authors have a different choices for the height $z_s$,  where heat due to irradiation is deposited, e.g., disk scale height ($z_s=h$) or disk photosphere ($\tau(z=z_s) =2/3$), or disk surface ($\Sigma(z>z_s) = 0$, where $\Sigma$ is the surface density at height $z$, $\Sigma(z)=\int_z^\infty \rho dz$) \cite[e.g.,][and references therein]{Hubeny1990, King-Ritter1998, Dubus-etal1999}.
For convenience we choose $z_s =h$.
However, in general, an effective equilibrium $z_s$ must be established, where an irradiated energy is almost deposited into the medium, depending on the irradiated intensity and other factors of the diffusing nature of the irradiated photon. Below this height $z<z_s$, there is no any irradiation effect.
As the irradiated photon moves from the optically thin to thick medium, the diffusing mean free path of this photon \big($=\frac{1}{\rho \kappa}$\big) will decrease, and finally, after travelling appropriate distance from the top, the photon will deposit its owns energy to the medium.    
On the other hand, the photons emitted from the disk 
do not lose energy comparatively as it moves from an optically thick medium to a thin medium.
Due to the uncertainty over the $z_s$ height, we also  check the results with  $z_s =1.5h$.
In summary, due to the low irradiation intensity, the present formalism is only applicable in the optically thin medium 
with the base of the wind at height $z_s$ (from the midplane).
We solve the model equations for $z<z_s$ with $x=0$ and for $z>z_s$ with a given $x$.

\noindent{\tt \textbf{(b) Critical point :}}
For $v_z(z) \ = v_r(z)$, the equation (\ref{eq:master}) has a singular point. At that $z$, a smooth velocity field would be obtained when the right-hand side of equation (\ref{eq:master}) is 0, i.e.,
\begin{align}
\frac{3 W_{r\phi}}{\rho}+ \alpha z\frac{1}{\rho} \frac{\partial p}{\partial z}&-v_r \frac{\partial \lambda}{\partial r}-v_z \frac{\partial \lambda}{\partial z} = -\alpha r \left[v_z\frac{xF_z}{v_r} +f^r_{bal} \right. \nonumber \\
& \qquad \left.   -\frac{v_r^2}{r}-  \frac{1}{\rho} \frac{\partial p}{\partial z}\frac{v_zv_r}{\Gamma_1 c_s^2} \right]\frac{\Gamma_1 c_s^2}{ v_r^2-\Gamma_1 c_s^2 }
\end{align}
\noindent For $v_r$ $\ll$ $\Gamma_1 c_s^2$, $x \ll$ 1, and with equation (\ref{eq:azimom},)  the above condition is 
always satisfied. Hence, $\frac{\partial v_z}{\partial z}$ is smooth at that height, where $v_z(z) \ = v_r(z)$ and $v_r(z)^2 \ll \Gamma_1 c_s^2$. 

\noindent{\tt \textbf{(c) Solution behavior at height where, $v_r^2 \rightarrow \Gamma_1 c_s^2$ : }}
For $v_r^2 \rightarrow \Gamma_1 c_s^2$, the corresponding height is termed as a sonic point, and equation (\ref{eq:master}) reduces
as
 \begin{equation} \label{eq:max}
  \frac{\partial v_z}{\partial z}\left[\frac{v_z^2-v_r^2}{v_r}\right] =  -v_z\frac{xF_z}{v_r} -f^r_{bal}  +\frac{v_r^2}{r} + \frac{1}{\rho} \frac{\partial p}{\partial z}\frac{v_zv_r}{\Gamma_1 c_s^2},
\end{equation} 
 which has a singular point for $v_z = v_r$. Thus, for a smooth velocity field at singular point, we have

 \begin{equation}\label{eq:cond1}
f^r_{bal} \approx -v_z\frac{xF_z}{v_r}  +\frac{v_r^2}{r} + \frac{1}{\rho} \frac{\partial p}{\partial z} \quad \ \ \quad      \text{or} \quad  \quad \ \ \frac{1}{\rho} \frac{\partial p}{\partial r} + \frac{1}{\rho} \frac{\partial p}{\partial z} \approx -\frac{v_r^2}{r}.
\end{equation} 
Here, $f^r_{bal}$ = $-\frac{1}{\rho} \frac{\partial p}{\partial r}$. For $x$ $\ll$ 1 and using equation (\ref{eq:hydro}) one can have magnitude-wise
$\left|\frac{1}{\rho} \frac{\partial p}{\partial z}\right| = |F_z|$. Since at the outer region,
$\frac{v_r^2}{r}$ $<$ $\frac{1}{\rho} \frac{\partial p}{\partial z}$, we can write
\begin{equation}\label{eq:cond0}
 \left|\frac{1}{\rho} \frac{\partial p}{\partial r}\right| \approx \left|\frac{1}{\rho} \frac{\partial p}{\partial z}\right| = |F_z|. 
\end{equation}
$\frac{1}{\rho} \frac{\partial p}{\partial r}$ can be expressed  by using equations (\ref{eq:conti}), (\ref{eq:radmom}) and (\ref{eq:energ}) as
\begin{equation}\label{eq:cond2}
   \frac{1}{\rho} \frac{\partial p}{\partial r}\left(\frac{v_r^2}{\Gamma_1 c_s^2}-1\right)=v_z \frac{\partial v_r}{\partial z}+f^r_{bal}-\frac{v_r^2}{r}-v_r \frac{\partial v_z}{\partial z}-\frac{1}{\rho} \frac{\partial p}{\partial z} \frac{v_r v_z}{\Gamma_1 c_s^2}.
   \end{equation}
\noindent Using equations (\ref{eq:cond1}) and (\ref{eq:cond2}), we note $v_z \frac{\partial v_r}{\partial z} \approx v_r \frac{\partial v_z}{\partial z} $.
Finally, we find the  relations $\left|\frac{1}{\rho} \frac{\partial p}{\partial z}\right| \approx \left|v_z \frac{\partial v_z}{\partial z}\right|$ and  $\left|\frac{1}{\rho} \frac{\partial p}{\partial r}\right| \approx \left|v_r \frac{\partial v_r}{\partial r}\right|$ by using equations (\ref{eq:varmom}) and (\ref{eq:radmom}) magnitude-wise, respectively.
In summary, at a height where $v_r$ or $v_z$ is comparable to the sound speed (or sonic height/point), we mainly find two results (i)  $\left|\frac{1}{\rho} \frac{\partial p}{\partial z}\right| \approx \left|v_z \frac{\partial v_z}{\partial z}\right|$ and  $\left|\frac{1}{\rho} \frac{\partial p}{\partial r}\right| \approx \left|v_r \frac{\partial v_r}{\partial r}\right|$, and (ii) $\frac{1}{\rho} \frac{\partial p}{\partial r} + \frac{1}{\rho} \frac{\partial p}{\partial z} \approx -\frac{v_r^2}{r}$.

\noindent{\tt \textbf{(d) Sign flip of $\frac{\partial p}{\partial r}$ :}}
In the Keplerian disk, $\frac{1}{\rho}\frac{\partial p}{\partial r}$ \big($\ll$ $F_r (z=0)$\big)  acts in a radially outward direction, so in the present sign convention, its sign is positive. 
The quantity $\left(-F_r(z)+\left.\frac{\lambda^2}{r^3}\right|_{z=0}+\left.\frac{1}{\rho}\frac{\partial p}{\partial r}\right|_{z=0}\right)$  flips the sign  around $z = 0.92 h$, which is
equivalent to the quantity $\frac{1}{\rho}\frac{\partial p}{\partial r} (z)$ for a constant $\lambda$ within the scale height (here, it should be noted that in the Keplerian disk,   $\left. -F_r (z=0) +\left.\frac{\lambda^2}{r^3}\right|_{z=0} \rightarrow 0 \right)$.
To compare this, we evaluate $\frac{1}{\rho}\frac{\partial p}{\partial r}$  as a function of height at a fixed radius for $x$ = 0. We notice the sign flip of $\frac{\partial p}{\partial r}$ around 0.83$h$, 0.87$h$, and 0.90 $h$ for $f_v$ = 2.0, 1.02, and 0.1, respectively. Hence, it is consistent with the result of the Keplerian disk.
In general, the sign flip of $\frac{\partial p}{\partial r}$ (or $\frac{\partial p}{\partial r} = 0$) at height $z = z_f$ can be determined by using equation (\ref{eq:azimom}), as
 \beqn \label{eq:sign}
3\alpha c_s^2  = - \left(\alpha z_f \frac{1}{\rho} \frac{\partial p}{\partial z}-v_r \frac{\partial \lambda}{\partial r}-v_z \frac{\partial \lambda}{\partial z}\right). 
 \eeqn
 \noindent{\tt \textbf{(e) Force term in radial direction:}}
 In equation (\ref{eq:radmom}), for $v_r(z), v_z(z) \ll \Gamma c_s^2$, the terms $\left(v_r \frac{\partial v_r}{\partial r}-v_z \frac{\partial v_r}{\partial z}\right)$ are negligible in comparison to the $\frac{\lambda^2}{r^3}$ (or even, $\frac{1}{\rho}\frac{\partial p}{\partial r}$\big). Using equation (\ref{eq:rzstress}), these terms can be expressed as $\left( \frac{\partial (v_r^2 - v_z^2)}{\partial r}\right)$, thus, for $v_r^2 \rightarrow \Gamma_1 c_s^2$, where $v_r = v_z$, this will tend to 0. Hence,  the force term in the radial direction can be effectively expressed as
\beqn \label{eq:lamb}
 F_r + \frac{1}{\rho} \frac{\partial p}{\partial r} = \frac{\lambda^2}{r^3}.
\eeqn
Here, we highlight with discussion point (d) that for $z < z_f$, $\frac{\partial p}{\partial r}$ acts opposite to the direction of gravity by a compact object, while for $z >z_f$ in the direction of the gravity. 


\noindent{\tt \textbf{(f) Isobaric regime and wind launching :}}
As mentioned earlier, the external heating raises the temperature (or enhances the internal energy of the fluid), and in the interested region, the pressure is the gas-dominated.
The increment in fluid velocity happens due to the expense of the internal energy, and the acceleration is driven by the pressure gradient.
At the sonic point, see discussion point (c), we obtain a condition (i) that
asserts that the kinetic energy of the fluid is now comparable to the internal energy. Therefore, the fluid meets  the equipartition of the energy states, and there is no pressure gradient (no acceleration); i.e., above the sonic point, an isobaric regime exists.  
We term the sonic height as the maximum reachable  height due to an acceleration and denote it by $z^{max}$ (see Paper I for the detailed discussions).


Later, we will find that the height $z_f$ is far below to the height $z^{max}$, i.e., $z_f \ll z^{max}$. Near the sonic height, the $ \frac{1}{\rho} \frac{\partial p}{\partial r}$ acts in the direction of $F_r$ and both $F_r$ and $ \frac{1}{\rho} \frac{\partial p}{\partial r}$ are supporting the fluid rotations.
Above the sonic height, the term $ \frac{1}{\rho} \frac{\partial p}{\partial r}$  is absent, and only $F_z$ and $F_r$ act on the fluid particle. Hence, at $z^{max}$ if $\frac{1}{\rho} \frac{\partial p}{\partial r} \ll F_r$ then $F_r$ alone is able to
support the rotation, and the fluid is rotationally bound.
In another case, where the magnitude of $\frac{1}{\rho} \frac{\partial p}{\partial r}$ is significant in comparison to the $|F_r|$, then $F_r$ is not able to support the rotation alone, and the fluid materials get ejected from the disk at height $z^{max}$ with speed $\sqrt{v_r^2+v_\phi^2+v_z^2}$.
Hence, in the last case $\left(\frac{1}{\rho} \frac{\partial p}{\partial r} \nless\nless F_r \right)$ the wind outflow is launched at radius $r$ from the sonic height $z^{max}$ with speed $v_{wind} = \sqrt{v_r^2+v_\phi^2+v_z^2}$; otherwise, the fluid is rotationally bound.  
Above the height $z^{max}$, in the present work, there is a no point of interest,
 we perform calculations up to the height near to $z^{max}$. 

\noindent{\tt \textbf{(g) Mass loss by wind :}}
In Paper I, we explore the general
characteristics of wind outflow for a wide range of the launching radii, simply assuming a fixed accretion rate.
As expected, the mass accretion rate will decrease with decreasing $r$ due to a wind outflow.
The mass accretion rate at radius $r-\Delta r$, $\dot{M} (r-\Delta r)$, can be written as (with $\frac{\Delta r}{r}\ll 1$),
\beqn \label{eq:m-out}
\dot{M} (r-\Delta r) =  \dot{M} (r) - \dot{M}_{out} (r)
\eeqn
where $\dot{M}_{out}$ = $2(2\pi r \Delta r v_{wind} \rho_w)$ is the mass outflow rate at radius $r$, $\rho_w$ is the density at the wind ejection height ${z^{max}}$, and $\dot{M} (r)$ is the mass accretion rate at radius $r$.

\noindent{\tt \textbf{(h) Energetic for wind  :}} 
The present model works in the optically thin regime, and the driver of the wind is the pressure gradient. 
The fluid acceleration occurs at the expense of the internal energy where the internal energy rises due to the irradiation.
Hence, by comparing the enhancement in the internal energy for a given $x$ (i.e., the modeled value) with the irradiated energy, one can constrain the range of the free parameters, like $z^{max}$ and $f_v$, or in general, the required range of the mass accretion rate at the transition radius $R_{tr}$, to produce the X-ray emission in the RIAF for a known Bondi mass accretion rate and Bondi accretion radius.

The vertically averaged internal energy density  for a given $x$, $ u^{x}$, at launching radius $r$ can be determined as
\beqn \label{eq:int_en}
u^{x} = \frac{1}{z^{max}} \int \frac{3}{2}(c_s(z)^2)\rho(z) dz
\eeqn

%
%

The irradiated flux $\epsilon^{irr}$ at radius $r$ and height $z_s =h$ by the inner region of bolometric luminosity $L_{bol}$ can be expressed (see, e.g., \cite{Frank-etal2002}) as
\beqn \label{eq:irr_en}
\epsilon^{irr}   = \frac{L_{bol}}{4\pi r^2} (1-\beta) C_{sph}.
 \eeqn
Here, we approximate the distance between the inner and outer region $r-r_{in}$ to
 $r$ for $r_{in} \ll r$. $\beta$ is
 the albedo, and $C_{sph} (\approx \frac{h}{8r}$, see \cite{King-Ritter1998})
 measures the normally irradiated energy on the surface $2\pi r dr$ at height
 $h$. 
 The total irradiated energy in annular area $2\pi r \Delta r$ in time $t_w$  is $\epsilon^{irr}t_w (2\pi r \Delta r)$, where $t_w$ is the time the interval  in which the fluid rises from the height $z_s$ to the sonic point $z^{max}$.
 By dividing the volume  $(2\pi r \Delta r(z^{max}-z_s))$ in total energy, the total irradiated energy density is $\frac{\epsilon^{irr}}{\langle v_z \rangle}$, where $\langle v_z \rangle$ = $\frac{(z^{max}-z_s)}{t_w}$= $\frac{1}{z^{max}}\int v_z dz$ is the vertically averaged $v_z$.

 For a comparison of the irradiated one with the modeled value, in place of using the energy density, we consider a flux. 
 For that, we define the vertically averaged internal energy flux $\epsilon^x$ for a given $x$ as 
\beqn \label{eq:flu_en}\epsilon^x = \langle v_z \rangle u^{x}, \eeqn
where, $\langle v_z \rangle$ is the denominator term of the total irradiated energy density.
The vertically averaged  enhancement in the internal energy flux $ \epsilon_{exess}^{x}$ for a given $x$ at launching radius r can be determined as
\beqn \label{eq:int_enhance} 
 \epsilon_{exess}^{x} =   \left( \left. \frac{2\pi}{z^{max}} \langle v_z \rangle \int \frac{3}{2}(c_s(z)^2)\rho(z) dz \right|_{\text{arbitrary} \ x} - \left.\frac{2\pi}{z^{max}} \langle v_z \rangle u^{x} \right|_{x=0} \right). 
 \eeqn 
where, the $z^{max}$ and   $\langle v_z \rangle$ for a given $x$, are different from the respective values  for $x$ = 0. 
 

\section{General Results}
The solutions  are mainly characterized by model free parameters, the initial vertical speed (which is parameterized by $f_v$) and the index of external heating $x$, and by disk parameters $\dot{M}$, $M_c$.
We first examine the general behavior of solutions at fixed launching radius,  like Paper I, for an SMBH. Next, we extend the Paper I work by studying the decrement of the mass accretion rate due to the mass loss by a wind outflow as a function of radius.
As we noted earlier, the wind ejection/sonic height $z^{max}$ depends on $x$ and we see later in this section that the $z^{max}$ increases with $x$. Thus, to explore the wind properties, we consider $z^{max}$ as a parameter in place of $x$, and for a fixed viewing angle $i$ we parameterize the $z^{max}$ in the ratio of $r$ as
\beqn \label{eq:i}
z^{max} = f_z\ r,\eeqn
where $f_z$ is a number and $i = \tan^{-1}(f_z)$.
To explore it without a loss of generality, we consider an SMBH of mass $M_c = 10^8 M_\odot$ and a coefficient of viscosity $\alpha$ = 0.1, and for fix the launching radius, we take $r = 2000 R_g$.

\subsection{Vertical Disk Structure for $x= $ 0 }\label{sec:x0}
\begin{figure}
\centering
\begin{tabular}{lcr}\hspace{-0.9cm}
  \includegraphics[width=0.44\textwidth]{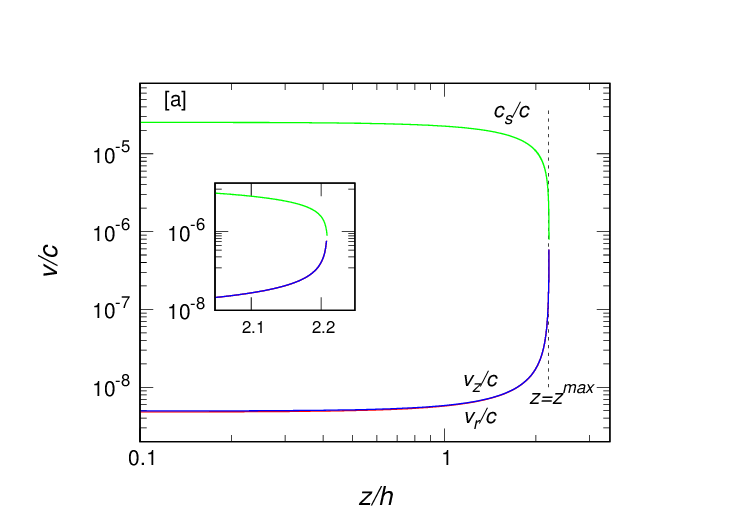}&\hspace{-1.60cm}
  \includegraphics[width=0.44\textwidth]{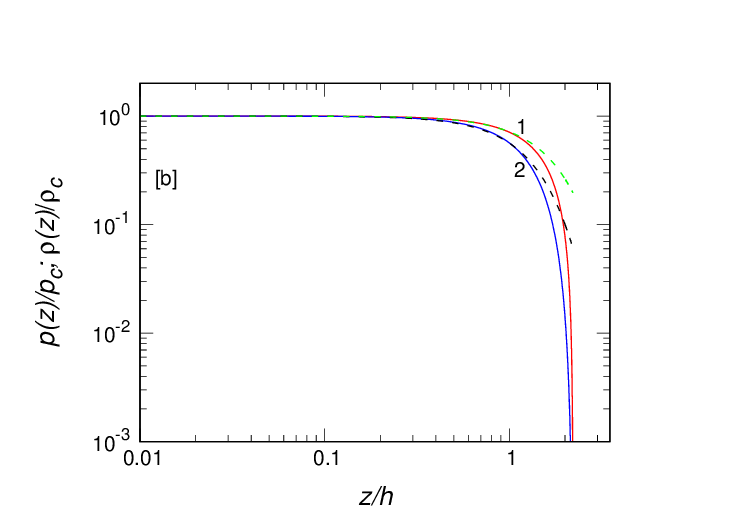}&\hspace{-1.590cm}
   \includegraphics[width=0.44\textwidth]{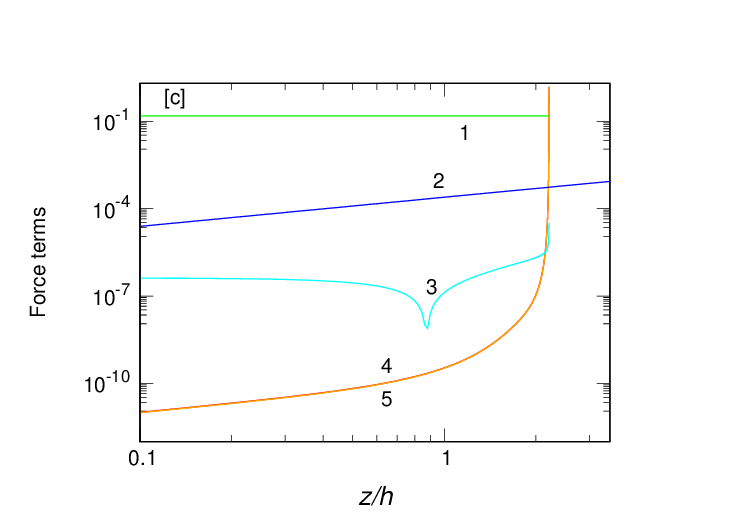}\\ 
\end{tabular}\vspace{-0.3cm}
\caption{The solutions of model equations for $x$ = 0, $r$ = 2000$R_g$, and $f_v$ =1.
The left panel is for three different velocities ($v_z$, $|v_r|$, $c_s$) as
functions of $z$ (measured in units of the Keplerian scale height $h$, here $r/h$ $\sim$118). The middle panel is for pressure $p/p_c$ and density $\rho/ \rho_c$, which  are shown by solid curves 2 and 1, respectively. The dashed curves 2 and 1 are for model curves  $\exp{\left(\frac{-z^2}{2(0.92h)^2}\right)}$ and $\exp{\left(\frac{-z^2}{2(1.2h)^2}\right)}$, respectively.  The right panel shows the comparison between $v_r \frac{\partial v_r}{\partial r}$, $v_z \frac{\partial v_z}{\partial z}$ and force terms  $\frac{1}{\rho} \frac{\partial p}{\partial r}$, $F_z$, and $F_r$, which  are shown by the curves 5, 4, 3, 2, and 1, respectively. In the left panel, we have marked the $z^{max}$ or sonic height  by a vertical line. 
}
\label{fig:x0}
\end{figure}

As stated, the present formalism is equivalent to the Keplerian disk, at least near the midplane in the limit of $v_z \ll c_s$. 
Here, we examine the validation of this equivalency in the vertical direction 
first for $x=$ 0.
In the Keplerian disk, for an isothermal disk, the pressure and density profiles are expressed as \cite[][]{Pringle1981},  $\rho(z,r) /\rho_c(r) = p(z,r)/p_c(r) = \exp \left(\frac{-z^2}{2h^2}\right)$,
where $p_c (r)$ and $\rho_c (r)$ are the pressure and density on the midplane, respectively. The density and pressure scale height both are  the same as $h$.
In the present case, the disk is slightly deviated from an isothermal profile (see the $c_s$ profile in Figure \ref{fig:x0}) within the scale height, as an acceleration occurs at the expense of internal energy. As a consequence, we obtain a different isothermal pressure and density profiles, which are expressed as (see the middle panel of Figure \ref{fig:x0})
\begin{footnotesize}
  \beqn \label{eq:var-strut} p(r,z) = p_c (r) \exp{\left(\frac{-z^2}{2 (0.93h)^2}\right)};\ \ \  \rho(r,z) = \rho_c (r) \exp{\left(\frac{-z^2}{2 (1.2h)^2}\right)}.\eeqn
\end{footnotesize}
Here, we observe a different pressure ($h_p$) and density scale height, and these are $\sim 0.93h$ and 1.2$h$, respectively.
In addition, we notice that the  heights $h_p$ and  $z_f$ are both related  by $z_f = h_p^2/h$ as from the $\frac{1}{\rho}\frac{\partial p}{\partial r}$ curve in the right panel of Figure \ref{fig:x0}, $z_f \sim 0.87h$.
In general, we find that the $z_f$ decreases with increasing $f_v$.  Since this analysis is independent of the launching radius $r$, the above result is similar to the Paper I.

In the right panel of Figure \ref{fig:x0}, we show the variation of different quantities, which have a force dimension, $v_z\frac{\partial v_z}{\partial z}$, $v_r\frac{\partial v_r}{\partial r}$, $\frac{1}{\rho}\frac{\partial p}{\partial r}$, $F_r$ and $F_z$ as a function of height $z$. We observe $v_r\frac{\partial v_r}{\partial r}$, $v_z\frac{\partial v_r}{\partial z}$ $\ll$ $\frac{1}{\rho}\frac{\partial p}{\partial r}$ $\ll$ $F_r$; also,   $v_z\frac{\partial v_r}{\partial z}$ $\ll$ $F_z$ for $z < h$. 
Hence, within the $h_p$, the Keplerian limit is valid in the present formalism without external heating, and in general also for $f_v < 10$.


The velocity profiles $v_r, v_z$ and $c_s$ are shown in the left panel of
Figure \ref{fig:x0}. The sonic point ($v_r^2 \rightarrow \Gamma c_s^2$) occurs at a height of around $2.2 h$, or $z^{max} = 2.2 h$.
We notice that the sonic point condition (i) of discussion point (c) 
$\left( i.e., \left|\frac{1}{\rho}\frac{\partial p}{\partial z}\right| \approx \left|v_z \frac{\partial v_z}{\partial z}\right|;\  \left|\frac{1}{\rho}\frac{\partial p}{\partial r}\right| \approx \left|v_r \frac{\partial v_r}{\partial r}\right| \right)$ is satisfied.
For $z > 0.87 h(=z_f)$, $\frac{1}{\rho}\frac{\partial p}{\partial r}$ acts in the direction of the gravity. Here, $\frac{1}{\rho}\frac{\partial p}{\partial r} \ll F_r$ at $z \sim z^{max}$; hence, the disk material is rotationally bound.

{\tt Disk Photosphere ($r_{phot}$):} We find here that the  $z^{max}$ is $\sim 2.2h$, which slightly decreases with increasing $f_v$ (even for $f_v =0.001$, $z^{max} \sim 2.2h$) and slightly increases with increasing $r$. 
Hence, the Keplerian disk photosphere is $\sim 2.2h$.
The Keplerian value (or here initial value) of the flow variables at a  given $r$ is tuned with vertical hydrostatic equilibrium (see equation (\ref{eq:hydro}) with $x$ = 0) as $h = \frac{c_s}{v_\phi}r$ \cite[][]{Shakura-Sunyaev1973} with approximating an isothermal disk or 
$p \propto \exp\big(\frac{-z^2}{2 h^2}\big)$, and see \cite{Frank-etal2002} for other approximations, $\frac{\partial p}{\partial z} \approx \frac{p}{h}\ \&
\ z \sim h$.
This is the reason the disk photosphere is almost the same in units of $h$ over $r$, and the disk shape is concave as $h \propto r^{9/8}$. 
However, for this calculation, in principle, we should couple the model equation with the radiative transfer equations \cite[e.g.,][references therein]{Hubeny1990}, but to avoid the complexity we left this exercise for a future work.

As we can take any set of initial values of fluid variables, to understand the impact of equation  (\ref{eq:hydro}) on the model solutions, we take
arbitrary value of $p$ and $\rho$. 
We find that for a given $r$ if we change $p$, $\rho$ in such a way that $c_s$ is constant, then the photosphere remains the same, and if $c_s$ increases, the photosphere also increases, and vice versa.
For the Keplerian value of  $p$ and $\rho$ at a given $r$, if we increase $r$ only then after sufficient large $r$ the photosphere increases with increasing $r$, and it decreases with decreasing $r$ and shrinks to zero for appropriate small $r$ and further there is no solution.   


\subsection{Vertical Disk Structure for Fixed $x $ and $f_v$ at a Given $r$}
\begin{figure}
\centering
\begin{tabular}{lr}
 \includegraphics[width=0.44\textwidth]{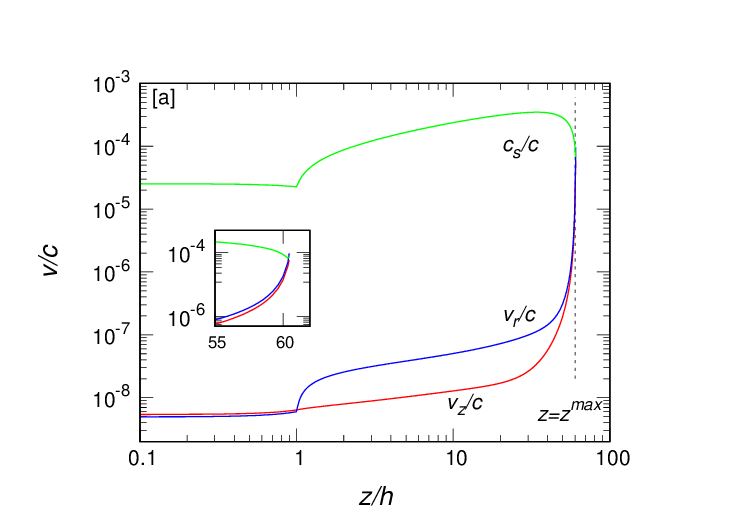}&\hspace{-1.490cm}
  \includegraphics[width=0.44\textwidth]{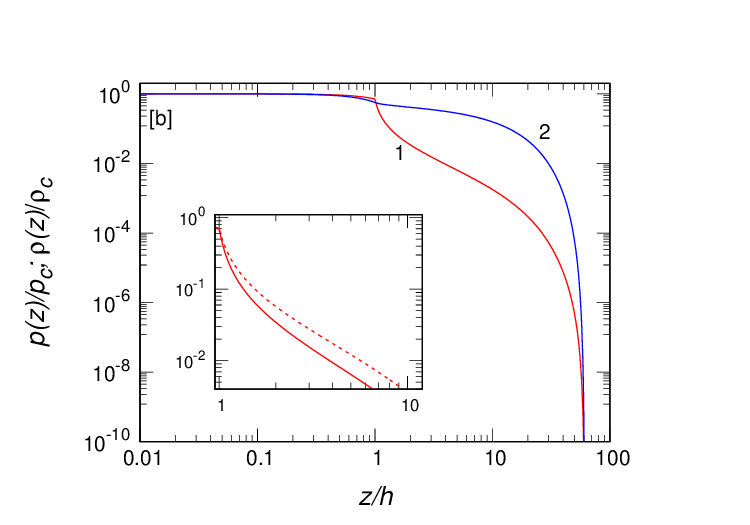}\\ 
  \includegraphics[width=0.44\textwidth]{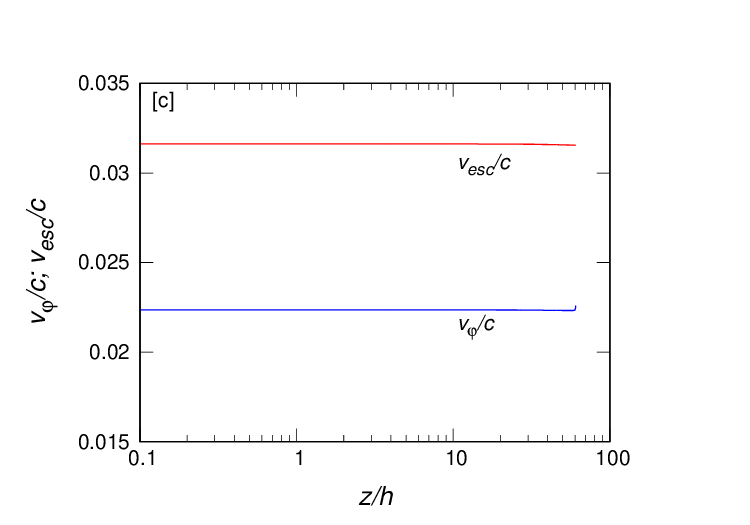}&\hspace{-1.490cm} 
  \includegraphics[width=0.44\textwidth]{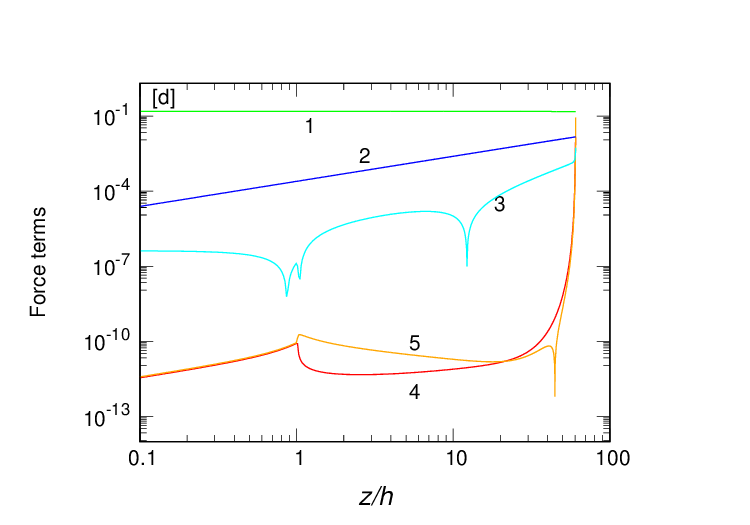}\\
\end{tabular}\vspace{-0.3cm}
\caption{The model solutions with  $z_s =h$, for $x$ = 7.8 $\times$ 10$^{-9}$ (or $z^{max} \sim$ 60$h$), and $r$ = 2000 $R_g$. Panels [a], [b], and [d] are same as the left, middle, and right panels of Figure \ref{fig:x0}. 
  Panel [c] shows the variations of $v_\phi$ and $v_{esc}$= $\sqrt{\frac{2GM}{\sqrt{r^2+z^2}}}$ with height. Here, for $z<z_s$ the solution is the same as the $x=0$ case as shown in Figure \ref{fig:x0}, and at $z \sim h$ the sharp change is due to the irradiation effect.
  In the inset figure of panel [b] the dashed curve is for $z_s$ = 0, $x$ = 7.0 $\times$ 10$^{-9}$ (or $z^{max} \sim$55$h$), shifted by $+h$ and lowered by a factor of 0.7, which shows that the sharp changes are consistent with the Paper I results.  
}
\label{fig:z60}
\end{figure}

\begin{figure}
\centering
\begin{tabular}{lcr}\hspace{-1.0cm}
  \includegraphics[width=0.44\textwidth]{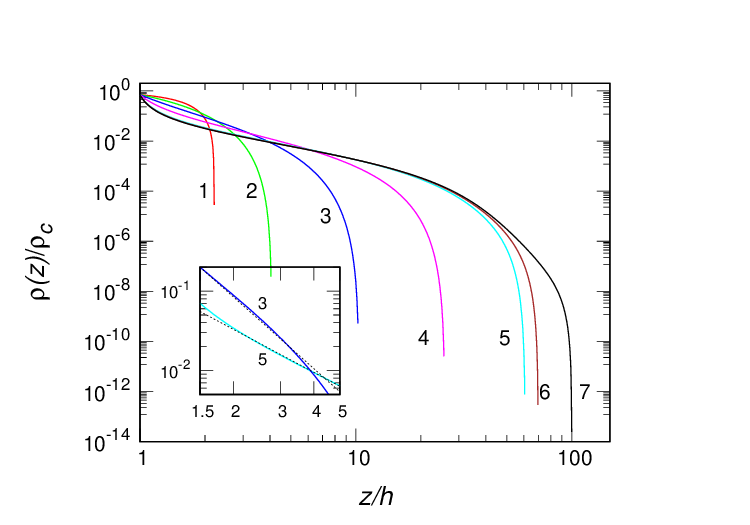}&\hspace{-1.60cm}
  \includegraphics[width=0.44\textwidth]{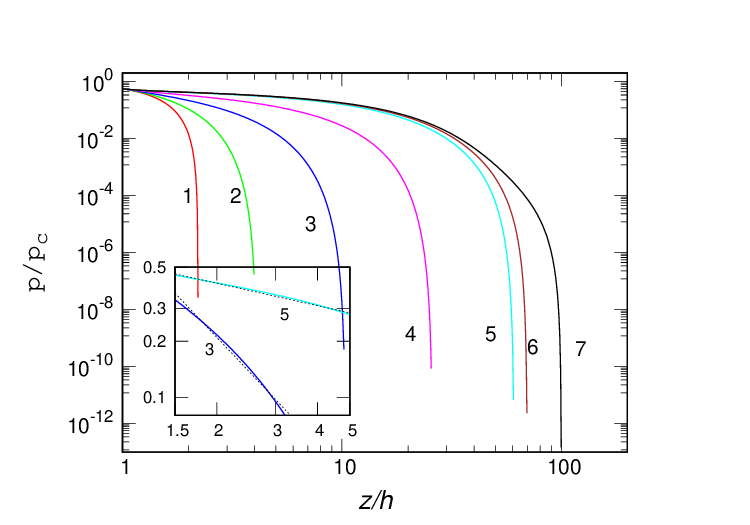}&\hspace{-1.6cm}
\includegraphics[width=0.44\textwidth]{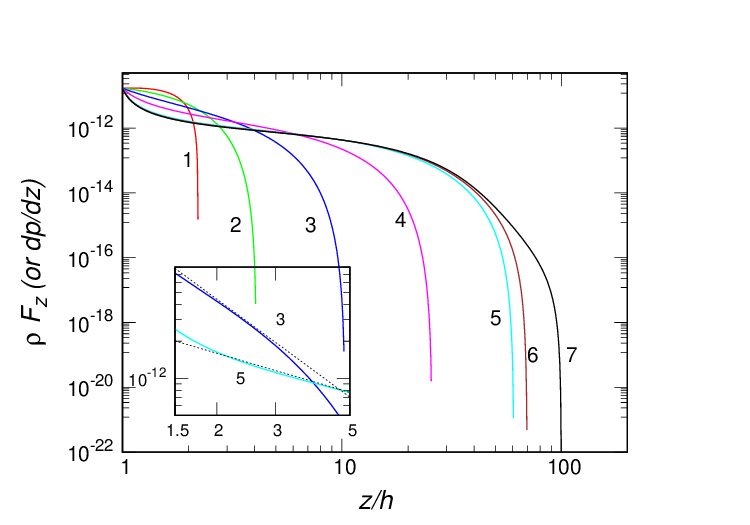} 
 \end{tabular}\vspace{-0.3cm}
\caption{The density, pressure, and $\frac{\partial p}{\partial z}$ profile in the vertical direction are shown for different $x$ with $z_s=h$ in the left, middle, and right panels, respectively. 
Here the curves 1, 2, 3, 4, 5, 8, and 7 are for $x$ ($z^{max}$) $\sim$ 0 (2.2$h$), 0.5 (4$h$), 1.4  (10$h$), 3.6 (25$h$), 7.8 (60$h$),  8.36 (70$h$), and 8.69 $\times$ 10$^{-9}$ (100$h$), respectively.
 The inset Figure shows the variation of the power-law index around $z=2h$. Here, the results for $z<h$ are not shown, as they are identical to the Figure  \ref{fig:x0}, and curve 5 of the left and middle panels is the same as curves 1 and 2 of panel [b] of Figure \ref{fig:z60}, respectively.
 } 
\label{fig:p_x-r2e3}
\end{figure}

\begin{figure}
\centering
\begin{tabular}{lcr}\hspace{-0.9cm}
  \includegraphics[width=0.44\textwidth]{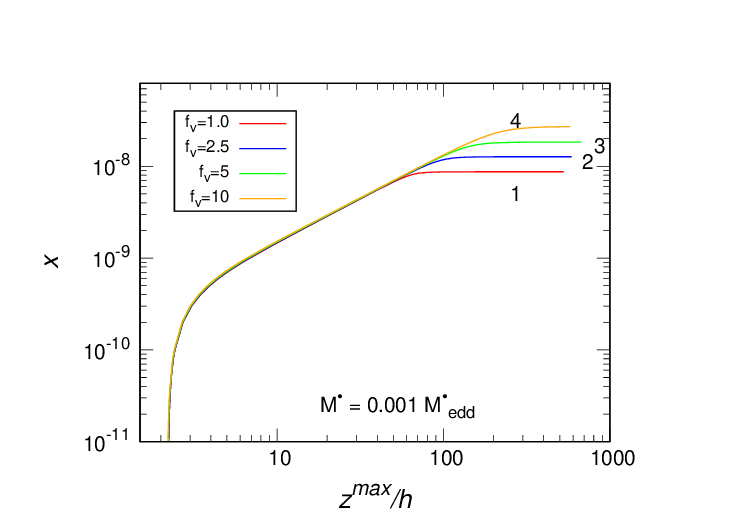} &\hspace{-1.59cm}
  \includegraphics[width=0.44\textwidth]{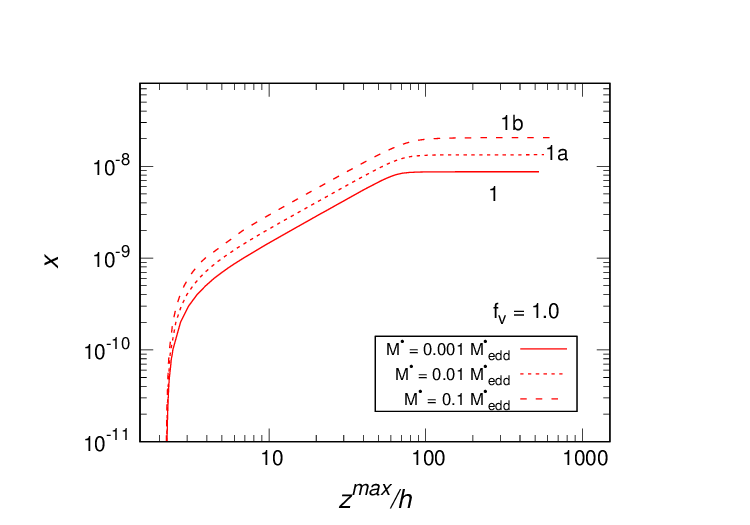}&\hspace{-1.59cm}
   \includegraphics[width=0.44\textwidth]{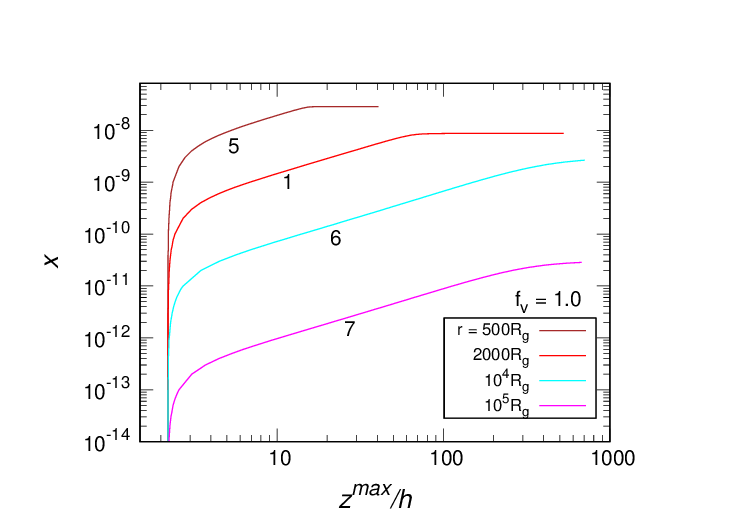} \\
\end{tabular}\vspace{-0.3cm}
\caption{The possible range of $x$ and the corresponding $z^{max}$ for the acceleration solutions of equation (\ref{eq:master}) with $z_s =h$. The left panel is for four different $f_v$ = 1, 2.5, 5, and 10 which are shown by curves 1, 2, 3, and 4, respectively, at fixed $r$ = 2000 $R_g$ and $\dot{M}$ = 0.001 $\dot{M}_{Edd}$. The middle panel is for $\dot{M}$ = 0.1, 0.01, and 0.001 $\dot{M}_{Edd}$ which are shown by curves 1b, 1a, and 1, respectively, at $r$ = 2000 $R_g$ and $f_v$ = 1. The right panel is for $r$ = 500, 2000, 10$^4$, and 10$^5$ $R_g$ which are shown by curves 5, 1, 6, and 7, respectively at $\dot{M}$ = 0.001 $\dot{M}_{Edd}$ and $f_v$ = 1} 
\label{fig:fv_x-r2e3}
\end{figure}

As stated, for the irradiation case, we first solve the model equations up to  height $z_s$ from the midplane with $x$ = 0; then for $z>z_s$ we solve it for a given $x$.
The results are shown in Figure \ref{fig:z60} for  $x$ = 7.8 $\times 10^{-9}$ (a limiting maximum value of $x$; see the curve 5 of Figure \ref{fig:p_x-r2e3} or the curve 1 of Figure \ref{fig:fv_x-r2e3}), $z_s$ = $h$ and $f_v = 1$. We obtain the sonic height $z^{max}$ $\sim 60h$ and $z_f$ $\sim 12.2h$. 
At $z = h$, we find a sharp change in fluid variables; this can be understood by horizontal shifting of the curve. If we shift these curves by $\frac{z}{h}-1$, then the sharp change vanishes and it appears similar to the respective curves of Paper I.
If one does a reverse exercise on the curves that have $z_s =0$, then a sharp change will appear at $z \sim h$, it is shown in the inset of Figure \ref{fig:z60}b  by the dashed curve, which is shifted by $\frac{z}{h}+1$ (for comparison, it is also lowered by a factor of 0.72) and computed with $z_s =0$, $x=7.0\times 10^{-9} (z^{max}=55h)$ and the rest of the parameters are the same as curve 1.


The numerical results for the density and pressure profile are shown in  Figure \ref{fig:z60}b.
To measure the change due to the irradiation from the height $z_s$, we define the pressure scale height $h_P^{irr}$ as
\beqn\label{eq:h-zs}
p(z=z_s+h_p^{irr})=e^{-1/2} p(z=z_s),
\eeqn
which is equivalent to the disk pressure scale height for $z_s$ = 0, i.e., $h_p=h_p^{irr}$, since for the $x=0$ case, by definition, $z_s$ = 0.
We find that for a given $x$ (or $z^{max}$) the $h_p^{irr}$ decreases with increasing $z_s$ which is mainly due to  $x$ starting to act in the large $F_z$ region with increasing $z_s$; also, the initial value of $c_s$ decreases with $z_s$.
Here, $h_p^{irr}  \sim 2.5h$ and similarly, the density scale height 
is 0.038$h$ (above the $z_s$).
At height $z = 2h$, the density decreases to the $\sim$0.034$\rho_c$,  the pressure decreases slightly  to $\sim 0.42 p_c$, and the sound speed is increased by $\sim$3.5 times from its midplane value, which reveals the external heating  interpretation for $x$ (but in the optically thin medium).
In  Figure \ref{fig:z60}d, for $z < \sqrt{h z_f}$ we find that $v_r\frac{\partial v_r}{\partial r}$, $v_z\frac{\partial v_r}{\partial z}$ $\ll$ $\frac{1}{\rho}\frac{\partial p}{\partial r}$ $\ll$ $F_r$; also,   $v_z\frac{\partial v_r}{\partial z}$ $\ll$ $F_z$. Hence, the equivalency of the present formalism to the Keplerian disk is still valid within the height $ \sqrt{h z_f}$, 
and one can initialize the flow variables to its  Keplerian values.  

The velocity profiles $v_r, v_z$, and $c_s$ are shown in Figure \ref{fig:z60}a.
In comparison to the  $x=0$ case, here $v_r$ and $v_z$ are accelerated more, and at $z = z^{max}$ their magnitudes are $v_z(r, z=z^{max})\ \sim 2.5c_s(r, z=0)$.
The variation of $v_\phi$ and escape velocity $v_{esc}$ $\left(=\sqrt{\frac{2 G M_c}{\footnotesize \sqrt{r^2+z^2}}}\right)$ are shown in Figure \ref{fig:z60}c, and at $z=z^{max}$, $v_\phi \gg v_r, \ v_z$.
Like $x=0$, the sonic point condition (i) (see the discussion point (c) of section \S\ref{sub:sol}) is fulfilled, which ensures a smooth solution around $z^{max}$ (see the inset figure of Figure \ref{fig:z60}a), and also the existence of an isobaric regime beyond the $z^{max}$.
Since $z_f \ll z^{max}$ and, at $z = z^{max}$, $\frac{1}{\rho}\frac{\partial p}{\partial r}$ $\sim$ 0.035$F_r$ (see Figure \ref{fig:z60}d), hence the matter will be ejected (as a wind) tangentially with speed $\sqrt{v_\phi^2+v_r^2+v_z^2}$. The wind is an equatorial wind with $v_{wind} \sim v_\phi$, and it will not escape the system as $v_\phi < v_{esc}$.

Next, we study the solution behaviors by varying $x$.
We obtain the solutions for seven different $x$ = 0, 0.5, 1.4, 3.6, 7.8, 8.36 and 8.69  $\times 10^{-9}$, and notice a different sonic point height for each $x$ which are  $z^{max}$ = 2.2$h$, 4$h$, 10$h$, 25$h$, 60$h$, 70$h$, and 100$h$, respectively (see also curve 1 of Figure  \ref{fig:fv_x-r2e3}).
We present the results in Figure \ref{fig:p_x-r2e3}, where we only study the
vertical structure of $\rho$, $p$ and $\frac{\partial p}{\partial z}$ which are shown in the left, middle, and right panels, respectively.
We note that for $z^{max}$ $<$ 60$h$, the profiles of $\rho$ and $p$ change significantly by varying $x$; however for $z^{max}$ $>$ 60$h$ the $\rho$ and $p$ profiles effectively do not change as their respective scale height is the same as the scale height at $z^{max}$ $\sim 60h$, e.g., $h_p^{irr} \sim 2.5h$ for $z^{max}$ $> 60h$.
Consequently, the $\frac{\partial p}{\partial z}$ profile and the internal energy density ($\frac{3}{2}\rho c_s^2$) of the fluid are effectively the same for $z^{max}$ $>$ 60$h$.
We reexamine the above by studying the variation of the power-law index around $z=2h$ which is shown in the inset of the respective Figure. 
The power-law index changes from -3.0 to -1.8 for $\rho$, from -1.9 to -0.4 for $p$, and from -2.0 to -0.8 for $\frac{\partial p}{\partial z}$, when the $z^{max}$ varies from $10h$ to $60h$. 
Since the $p$ and $\rho$ profiles do not change after $z^{max}$ = $60h$, we term this $z^{max}$ as a maximum physically possible $z^{max}$ and denote it as a $z^{max}_t$ and the corresponding $x$ as a $x^{max}$.

In general, we find a one-one mapping between $x$ and $z^{max}$, where $z^{max}$ increases with increasing $x$. In other words, with the interpretation of the external heating of $x$, the sonic point/height rises with the intensity of the external heating. 
The $x$ versus $z^{max}$ curve is shown in  Figure \ref{fig:fv_x-r2e3}, in which the left, middle, and right panels are obtained by varying $f_v$, $\dot{M}$, and $r$, respectively. In all panels, we note that first $x$ increases with $z^{max}$ (with a power law index $\sim$0.93 for $z^{max}$ $>$ 60$h$ for curve 1) but after some $z^{max}$ a very small increment in $x$ leads to a large increment in $z^{max}$ (see the horizontal region). 
We find that the starting $z^{max}$ of the horizontal region is the same as the $z^{max}_t$ (e.g., see the $z^{max}_t$ of Figure \ref{fig:p_x-r2e3} and here curve 1).
The $z^{max}_t$ increases with increasing either $f_v$, or $\dot{M}$, or $r$, while $x^{max}$ decreases with increasing $r$. Here, the curves 1, 2, 3, and 4 are for $f_v$ = 1, 2.5, 5, and 10; the curves 1b, 1a, and 1 are for $\dot{M}$ = 0.1, 0.01, and 0.001 $\dot{M}_{Edd}$; and the curves 5, 1, 6, and 7 are for $r$ = 500, 2000, 10$^4$, and 10$^5$ $R_g$, respectively, and the rest of the common parameters are the same as curve 1, or $f_v$ =1, $\dot{M} = 0.001\dot{M}_{Edd}$, $r=2000 R_g$ and $z_s =h$.

In the left panel of Figure \ref{fig:fv_x-r2e3},  $z^{max}_t$ is $\sim$60$h$, 85$h$, 130$h$, and 200$h$ for curves 1, 2, 3, and 4, respectively.
In addition, we find that the nonhorizontal regions of all curves overlap each other. It means that for any $f_v$ the same amount of external heat is needed to launch the wind from a particular height. 
To check it, we compute the internal energy density of the fluid for each $x$ \big($\int \rho c_s^2 dz /h$\big) and we find a similar overlapping region for all $f_v$ cases.
Since here the initial value of all fluid variables except $v_z$ is the same, the role of the higher $f_v$ is only to raise the wind launching height.

%


\begin{figure}
\centering
\begin{tabular}{lcr}\hspace{-0.9cm}
  \includegraphics[width=0.44\textwidth]{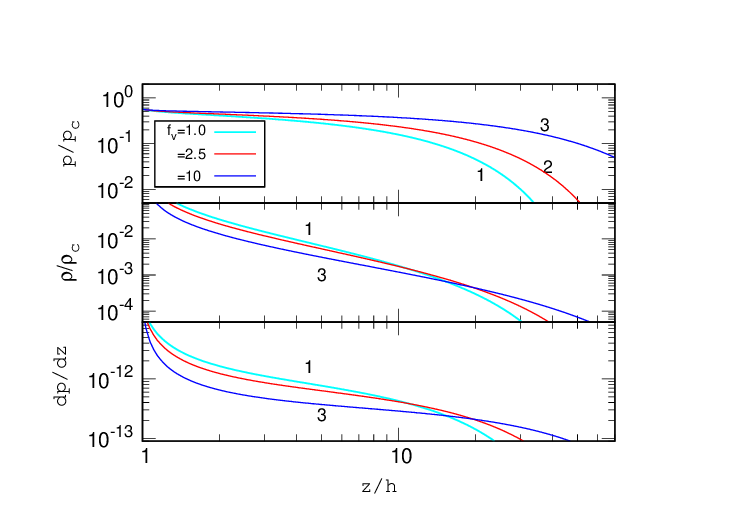} &\hspace{-1.59cm}
  \includegraphics[width=0.44\textwidth]{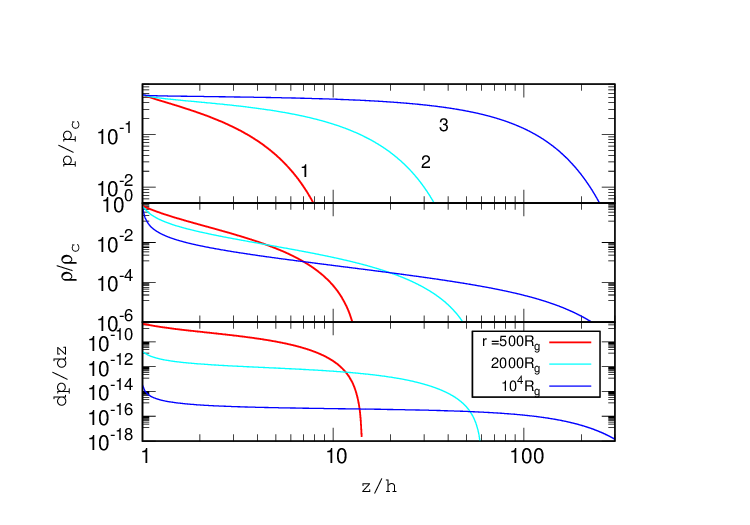}&\hspace{-1.59cm}
   \includegraphics[width=0.44\textwidth]{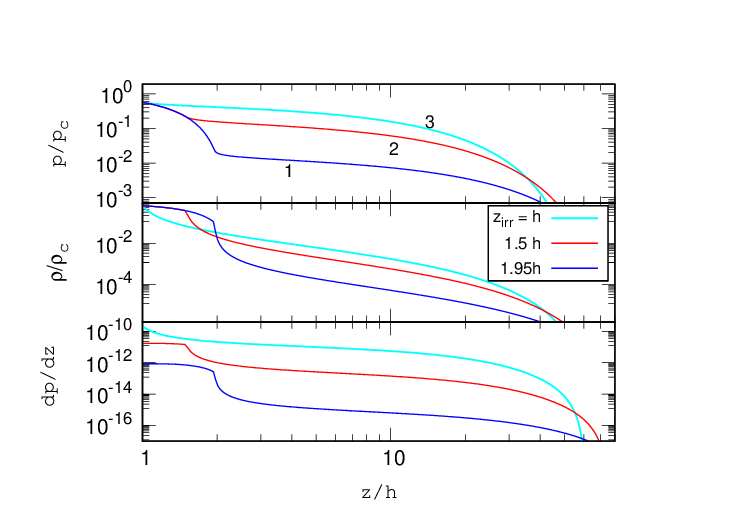} \\
\end{tabular}\vspace{-0.3cm}
\caption{The top, middle, and bottom panels are for the $p/p_c$, $\rho/\rho_c$ and $\frac{\partial p}{\partial z}$ profiles in the vertical direction for $x \rightarrow x^{max}$, respectively. The left panel is for different $f_v$ and the curves 1, 2, and 3 are for $f_v$ = 1, 2, and 10, respectively. The centre panel is for different $r$, and the curves 1, 2, and 3 are for $r$ = 500, 2000, and 10$^4$ $R_g$, respectively. The right panel is for different $z_s$, and the curves 1, 2, and 3 are for $z_s$ = 1.0, 1.5, and 1.95, respectively. The rest of the parameters are same as in Figure \ref{fig:z60}. Here curve 1 of the left, curve 2 of the middle, and curve 3 of the right panel are same as curve 5 of Figure \ref{fig:p_x-r2e3}, and for $2h<z<6h$ the power-law indexes of all curves are almost similar.    } 
\label{fig:x_max-r2e3}
\end{figure}

\subsection{Limiting Value  of $x$ and the Physically Accessible Regime of the Solutions}\label{sec:acces}
To understand the solution behavior around $x = x^{max}$ (or $z^{max}$ = $z^{max}_t$), we study the fluid variable profiles (mainly $p$, $\rho$ and $\frac{\partial p}{\partial z}$) at $x= x^{max}$.
In Figure \ref{fig:x_max-r2e3}, the top, middle, and bottom panels are for
$p$, $\rho$, and $\frac{\partial p}{\partial z}$, respectively.
These variables are studied by varying $f_v$ (left panel) and $r$ (middle panel). In addition, due to the uncertainty over $z_s$, we also study it by varying the $z_s$, the results are shown in the right panel, where the curves 1, 2, and 3 are for $z_s$  = 1.0, 1.5, and 1.95, respectively. As also mentioned earlier, here the initial value of $c_s$ decreases with increasing $z_s$.
We find that by increasing $f_v$ or $r$ the $h_p^{irr}$ and $z^{max}_t$ increase, while by increasing $z_s$ the  $h_p^{irr}$ decreases and $z^{max}_t$ increases.
Since for $z^{max}$=$z^{max}_t$ the pressure scale height is maximum, hence the increment in $z^{max}_t$ can be attributed by respective decrement in  $\frac{\partial p}{\partial z}$, as for the lowest  $\frac{\partial p}{\partial z}$ the pressure scale height is highest, and consequently a smaller density scale height (the same is noted here).

These can also be interpreted by using the equations (\ref{eq:hydro}) and (\ref{eq:cond2}). The equation (\ref{eq:cond2}) asserts that for larger initial value of $v_z$ (or $f_v$) or for a smaller $c_s$ the 
$\frac{\partial p}{\partial z}$ magnitude would be smaller; see the left and right panels of Figure, \ref{fig:x_max-r2e3}, respectively.
Equation (\ref{eq:hydro}) reveals that for a given parameter set  
$|\frac{\partial p}{\partial z}|$ would be smaller for a small $F_z$.
However, this situation cannot be achieved by simply varying $F_z$ for a given parameter set, therefore, we consider   
three different $r$ (see the middle panel), where the initial values of flow variables are different with $c_s^2 \propto r^{-3/4}$. Also, $F_z \propto \frac{z}{r^3}$ for $r \gg z$, and we obtain a similar trend, as predicted.
Curve 1 of the left, curve 2 of the middle, and curve 3 of the right panel are same as curve 5 of Figure \ref{fig:p_x-r2e3}. 
Finally, it appears that the $x^{max}$ or $z^{max}_t$ is associated wth the pressure and density profiles where $\frac{\partial p}{\partial z}$ $\propto z^{\sim -0.8}$; consequently, $\rho \propto z^{\sim -1.8}$ for $2h<z<6h$ and $z_s>h$, and these dependencies are almost similar to the  variation of either $f_v$ or $r$ or $z_s$.

\begin{figure}
\centering
\begin{tabular}{lr}
   \includegraphics[width=0.44\textwidth]{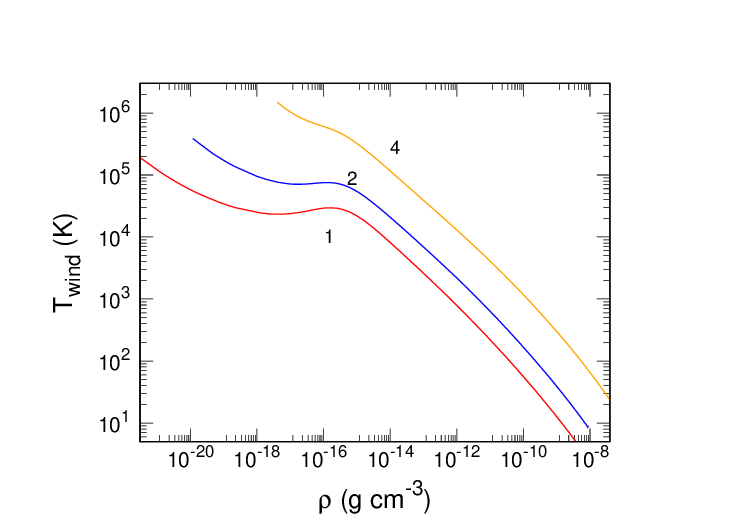}&\hspace{-1.590cm} 
 \includegraphics[width=0.44\textwidth]{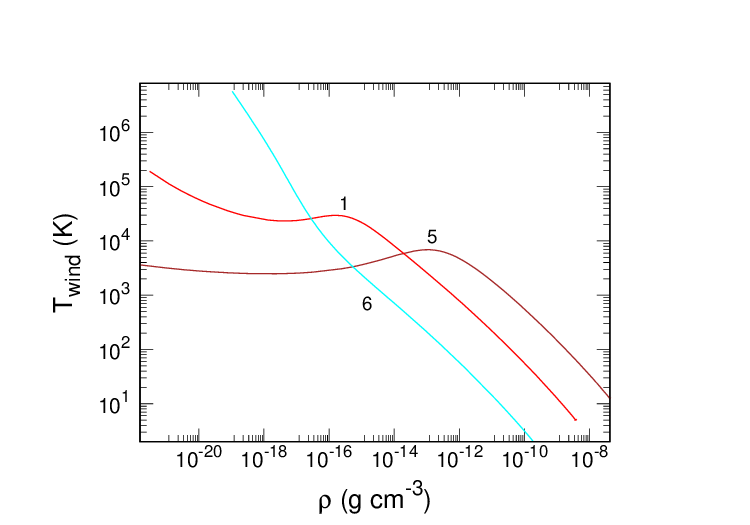}\\
 \end{tabular}\vspace{-0.3cm}
\caption{The density vs temperature curves at height $z = z^{max}$, in which the left panel is for a different $f_v$ at fixed $r$ and the right panel is for a different $r$ at $f_v=1$. 
  The curves are same as in Figure \ref{fig:fv_x-r2e3}. 
 } 
\label{fig:stab-r2e3}
\end{figure}

{\tt Physically accessible regime:}  For $z^{max}$ $>$ $z^{max}_t$, we noted earlier that the pressure and density profiles do not change effectively; also, the internal energy density is not enhanced (see Figure \ref{fig:p_x-r2e3}).  Therefore, these results indicate that the model solution after $z^{max}=z^{max}_t$ is not physically accessible.
In addition, the later results indicate that the $z^{max}_t$ seems to be an inflection point on the curve of flow variables as a function of $z^{max}$ (see Figures \ref{fig:w_fv-2e3}, or \ref{fig:wd_r-md}).
For clarity, we investigate the $\rho$-$T_{wind}$ curve at $z^{max}$ height, where $T_{wind}$ is the temperature of the fluid at $z=z^{max}$. 
The results are shown in Figure \ref{fig:stab-r2e3}, in which it is obtained by varying $f_v$ for fixed $r$ (in the left panel) and for a different $r$ at fixed $f_v=1$ (in the right panel).
The curves (and respective parameters) are the same as in Figure \ref{fig:fv_x-r2e3}.
Here, $x$ (or $z^{max}$) is increasing in the right-to-left direction, i.e., the bottom right corner corresponds to the $x = 0$ (or $z^{max} \sim 2.2h$).
In both panels, around $z^{max} = z^{max}_t$ the behavior of variations of the $\rho$-$T_{wind}$ curve with increasing $x$ deviates from the general trend (which is that by increasing $x$, $p$ and $\rho$ at $z^{max}$ decrease in such a way that $T_{wind}$ increases). And for $z^{max} \gg z^{max}_t$, where $x\sim x^{max}$, it again shows the general trend with smaller slope, which is an unphysical situation. Hence, these results again emphasize that the model solution after $z^{max}=z^{max}_t$ is not of a physical interest.

As noted earlier, for a given parameters set at $z^{max} = z^{max}_t$ the minimum wind density is  independent of $f_v$ also independent of the magnitude of $z^{max}_t$, here its magnitude is $1.6 \times 10^{-16}$ g $cm^{-3}$. 
Hence, the minimum wind density only depends on $F_z$ for a given $z_s$. 
Therefore, for a different $r$ it will change in the same ratio of $\rho_c$ (initial value of $\rho$) and $F_z$, or the minimum wind density $\propto$ $r^{-15/8}r^{-3}$, where the initial value of $\rho \propto r^{-15/8}$ for a given $\dot{M}, M_c, \alpha$, and $F_z \propto r^{-3}$.
We find the minimum wind density in the same ratio (shown in the  right panel), and it is $\sim$ $1.3 \times 10^{-13}$ and $0.6 \times 10^{-19}$ g $cm^{-3}$ for $r$ = 500 and 10$^4 R_g$, respectively.

\begin{figure}
\centering
\begin{tabular}{c}\hspace{-0.9cm}
  \includegraphics[width=0.6\textwidth]{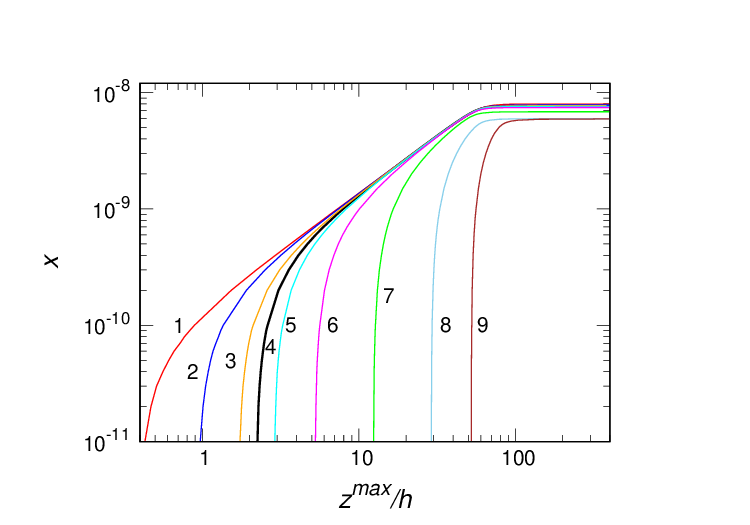}  \\
\end{tabular}\vspace{-0.3cm}
\caption{The $x$ vs $z^{max}$ curve with the arbitrary values of $p$ and $\rho$ at midplane 
  for $r=2000R_g$ and $\dot{M}=0.001\dot{M}_{Edd}$ in which curve 4 is for the  tuned value of $p (=p_c)$ and $\rho (=\rho_c)$ for the Keplerian disk. We consider 
 the $p$ and $\rho$ in the ratio of  $p_c$ and $\rho_c$, as $p_c \times a^{21/8}$ and $\rho_c \times a^{15/8}$, respectively, where $a$ is the number. Here, the curves 1, 2, 3, 4, 5, 6, 7, 8, and 9 are for $a$ = 0.01, 0.1, 0.5, 1, 2, 10, 100, 1000, and 5000, respectively. The $h$ corresponds to the Keplerian disk for curve 4. 
} 
\label{fig:r2e3surf}
\end{figure}

{\tt Disk photosphere $\&$ $z^{max}_t$ :} Since $z^{max}_t$ mainly depends on $f_v$, $c_s$ and $F_z$; and the disk photosphere is associated  with equation (\ref{eq:hydro}) or, in general, it depends on $c_s$ and  $F_z$ (see section \S\ref{sec:x0}). Next, we explore the connection existed  between $z^{max}_t$ and $r_{phot}$.
For this, we consider  a fixed $f_v$ (=1), and $r$ (=2000$R_g$, so $F_z$ is also). For simplicity we take $z_s$ = 0, however we note that the results are qualitatively the same for $z_s > h$ also.
For $c_s$, we take arbitrary values of $\rho$ and $p$ of this form, $\rho_c  \times a^{15/8}$ and  $p_c \times a^{21/8}$, respectively, where $a$ is the number.
The results are shown for nine different $a$ in Figure \ref{fig:r2e3surf}, where the curves 1, 2, 3, 4, 5, 6, 7, 8, and 9 are for $a$ = 0.01, 0.1, 0.5, 1, 2, 10, 100, 1000, and 5000, respectively.
Here, the curve 4 is for the Keplerian disk. 
Since the $r_{phot}$ is the height $z^{max}$ for $x$ =0, and in the $x$ versus $z^{max}$ curve it corresponds to the $z^{max}$ of the vertical region (see Figure \ref{fig:fv_x-r2e3}). For curves 1, 2, 3, 4, 5, 6, 7, 8, and 9 the $r_{phot}$ are $\sim$ 0.4$h$, 0.9$h$, 1.7$h$, 2.2$h$, 2.8$h$, 5.2$h$, 12.3$h$, 28.9$h$, and 52.1$h$, respectively.
For all curves, $z^{max}_t$ is $\sim 55h$, however, the corresponding $x^{max}$ slightly decreases with increasing $c_s$.
Particularly for curve 9 the disk photosphere is almost equal to the $z^{max}_t$, it means that without any irradiation/external heating
the disk fluid could rise up to the height $z^{max}_t$, only one has to increase the sound speed of the fluid. If one further increases the $c_s$ from the maximum $c_s$ (where $r_{phot}$ = $z^{max}_t$), the obtained $x$ versus  $z^{max}$ curve is similar to curve 9 with large  $r_{phot}$. 
The horizontal region of the $x$ versus  $z^{max}$ curve is not a physically accessible regime.
In the present model,  $z^{max}_t$ is the maximum height  where the disk can hold the fluid with maximum $c_s$ in hydrostatic equilibrium. 


\subsection{$\frac{1}{\rho}\frac{\partial p}{\partial r}$ $\&$ Shear Stress at Large Height} \label{sec:stres-dp}
\begin{figure}
\centering
\begin{tabular}{lr}
 \includegraphics[width=0.56\textwidth]{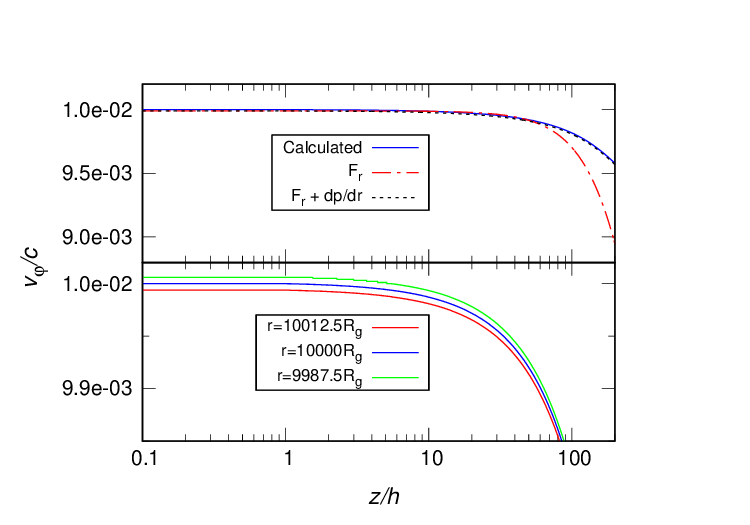}&\hspace{-1.80cm}
 \includegraphics[width=0.56\textwidth]{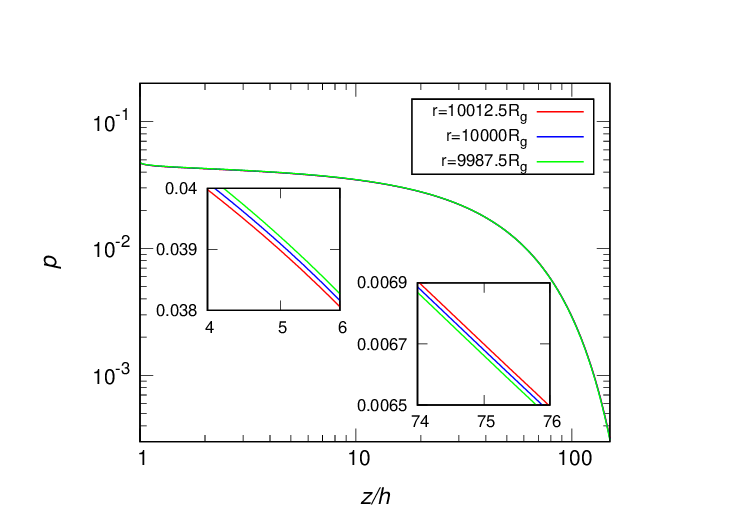}\\ 
 \end{tabular}\vspace{-0.3cm}
\caption{Left panel: $v_\phi/c$ vs $z/h$ curves, the top panel shows the contribution of the $\frac{1}{\rho}\frac{\partial p}{\partial r}$ term in support of the  rotation,  and the bottom panel shows the $\frac{\partial v_\phi}{\partial  r}$ or the gradient of $v_\phi$. In the top panel, the solid curve is for the model value of $v_\phi$ and the dotted-dashed and dashed curves are for the analytic ones using the relations $v_\phi^2 = rF_r$ and $v_\phi^2 = rF_r(1+(\frac{1}{\rho}\frac{\partial p}{\partial r})/F_r)$, respectively.
  In the bottom panel, the calculated $v_\phi$ is shown for three adjacent  $r$ = 10012.5, 10000, and 9987.5$R_g$. 
  Right panel: the pressure profile is shown for three adjacent $r$, to show the sign flip of $\frac{1}{\rho}\frac{\partial p}{\partial r}$
  which also asserts the calculation validity in the $r$-direction. Here, the three different $r$ are the same as the bottom left panel. The left and right insets are for a small range of $z$ with $z < z_f$ and $z >z_f$, respectively. Here $x$ = $1.5\times 10^{-9}$, and the rest of the parameters are the same as the curve 6 of Figure \ref{fig:fv_x-r2e3}.
} 
\label{fig:shear-r1e4}
\end{figure}

As the $\frac{1}{\rho}\frac{\partial p}{\partial r}$ flips the sign at height $z_f$ and above this height, it starts to support the fluid rotation along with $F_r$.
We first examine the variation in $v_\phi$ with height 
in view of the equation ( \ref{eq:lamb}). To examine at large height ($z\gg h$), we consider $r=10^4 R_g$ and $x = 1.5\times 10^{-9}$ and the rest of the parameters are same as the curve 6 of Figure \ref{fig:fv_x-r2e3}. 
The results are shown in the top left panel of Figure \ref{fig:shear-r1e4}, where $z_f$ $\sim 55h$ and $z^{max}$ $\sim 240h$, in which the solid curve is for a calculated one and dotted-dashed and dashed curves are an analytic ones that are computed using the relations $v_\phi^2 =rF_r$ and $v_\phi^2= rF_r(1+\frac{1}{\rho}\frac{\partial p}{\partial r}/F_r)$, respectively, however, for the later case, $\frac{1}{\rho}\frac{\partial p}{\partial r}$ is taken from the calculations. At large height, the $v_\phi$ differs significantly from the analytic one when the gravity alone supports the rotation.

Next, we examine the sign flip of $\frac{1}{\rho}\frac{\partial p}{\partial r}$ at a given $r$. For this, we compute $p$ at two adjacent annular radii  around $r=10^4R_g$, $r \pm \Delta r$. Therefore, this analysis also checks 
the consistency of the numerical setup of the model with radial (+ vertical) grid points.
The results are shown in the right panel of Figure  \ref{fig:shear-r1e4}, in which the red, blue, and green solid curves are for $\Delta r = +\frac{r}{800}$, 0, and $-\frac{r}{800}$, respectively, and the rest of the parameters are same as the top left panel. The left and right insets  are for $z<z_f$($4<\frac{z}{h}<6$) and $z>z_f$ ($76<\frac{z}{h}<78$), respectively.
The pressure gradient acts in the radially outward direction in the left inset (or $p(r+\Delta r,z)<p(r-\Delta r,z)$), while in the right inset, it is in radially inward (or $p(r+\Delta r,z)>p(r-\Delta r,z)$); i.e., it flips the sign at some height.
Hence, although we 
solve the equation along the $z$-axis for a given $r$, but the results are approximately consistent with those obtained with the use of 
radial grid point.

{\tt Shear stress:} 
Since the shear stress is generated due to the differential rotation, to check the consistency of the solution, we examine the presence of differential rotation at large height by computing the $v_\phi$ as a function of height for three adjacent annular radii. The results are shown in the bottom left panel of Figure  \ref{fig:shear-r1e4}, where the curves symbol and parameter set are the same as the right panel. We find that the differential rotations are present at large height, and $\frac{\partial v_\phi}{\partial r}(z)$ (or $\frac{\partial \Omega}{\partial r}(z)$) decreases  with increasing $z$.
In the present model, we assume that the  $\alpha$-prescription for viscosity within the scale height is also valid for any higher height.
For consistency, we examine a turbulent eddy of size $h'(z) \left(\approx \frac{c_s(z)}{v_\phi(z)}r \right)$ around a higher height, say $z=100h$, we find that the pressure decreases very slowly inside the eddy, so within this eddy, one can approximate  $p \approx c_s h' \rho r \frac{\partial \Omega}{\partial r}|_{eddy}$. Hence the assumption of $W_{\phi r}$ = $\alpha p$ a throughout the height is consistent in the solution. And, $W_{\phi r}$ will decrease with height as $p$, and it becomes almost negligible at the wind injection height $z^{max}$, hence, the wind is nonviscous. 

\subsection{General Wind Characteristics}
\begin{figure}
\centering
\begin{tabular}{lr}\vspace{-0.9cm}
 \includegraphics[width=0.44\textwidth]{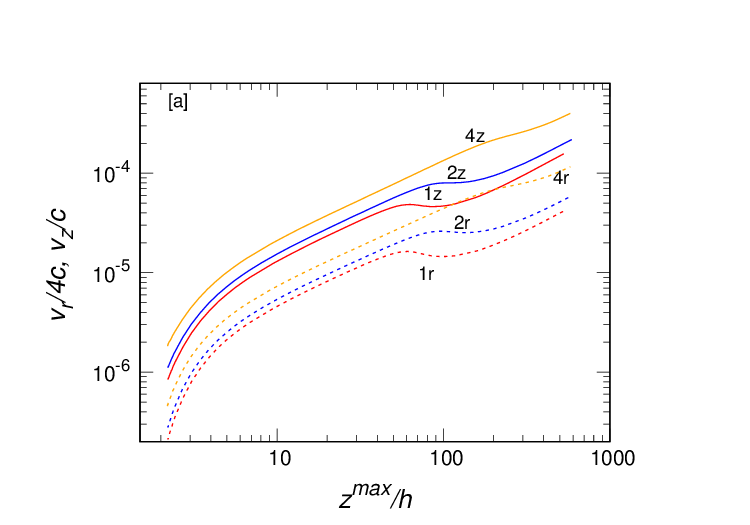}&\hspace{-1.590cm}
  \includegraphics[width=0.44\textwidth]{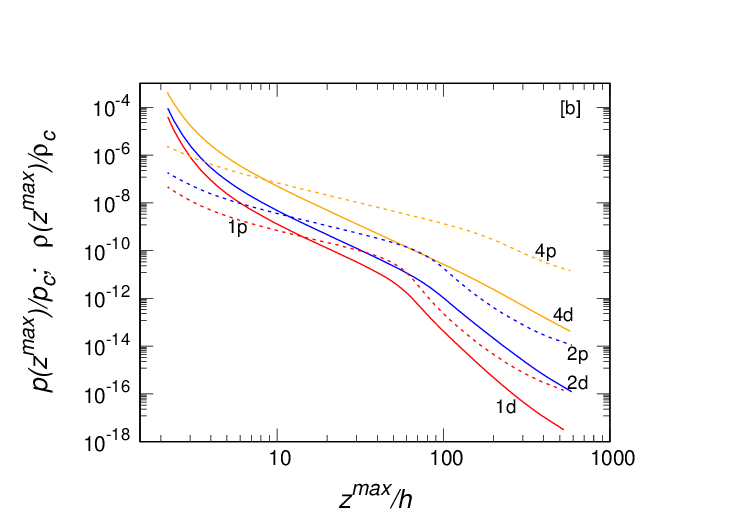}\\
  \includegraphics[width=0.44\textwidth]{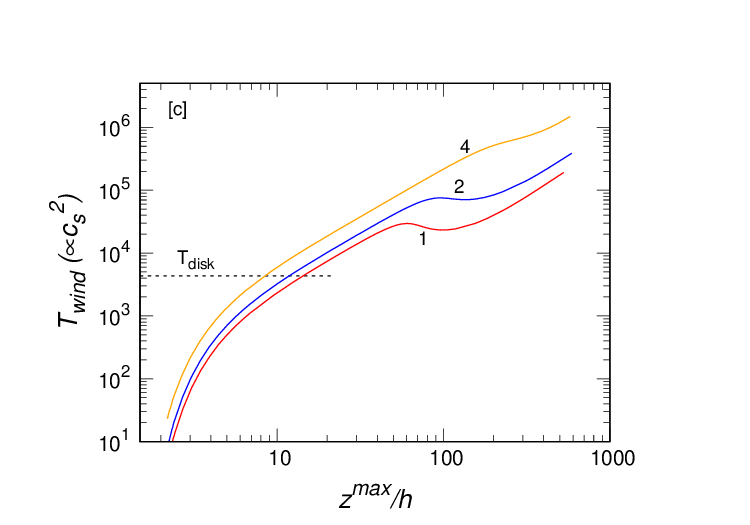}&\hspace{-1.590cm}
 \includegraphics[width=0.44\textwidth]{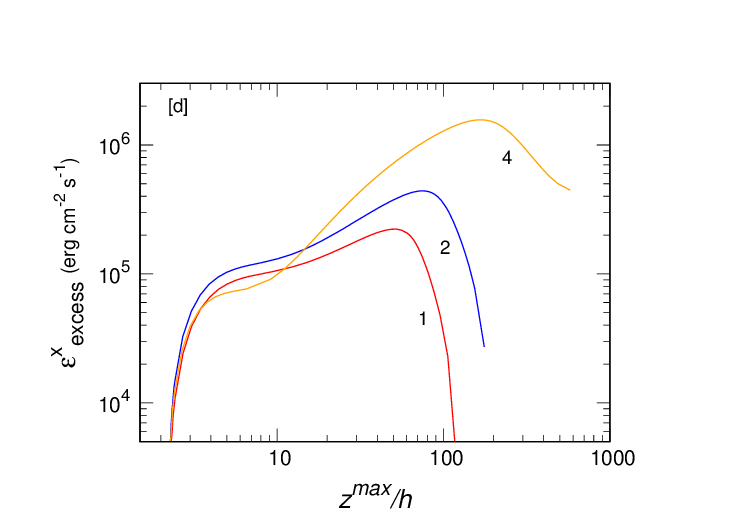}\\ 
\end{tabular}\vspace{-0.3cm}
\caption{The wind characteristics for three different values of $f_v$. Two velocities $v_r/4c$ (dotted curve or with suffix r) and $v_z/c$ (solid curve or with suffix z) are shown in panel [a]. Panel [b] is for pressure (dotted curve or with suffix p) and  density (solid curve or with suffix d), and panels [c] and [d] are for $T_{wind}$ and $\epsilon^x_{excess}$, respectively. In panel [c] the horizontal line is for $T_c$. Here, the curves symbol and parameters are the same as in Figure \ref{fig:fv_x-r2e3}.
}
\label{fig:w_fv-2e3}
\end{figure}

\begin{figure}
\centering
\begin{tabular}{lcr}\hspace{-1.2cm} 
  \includegraphics[width=0.44\textwidth]{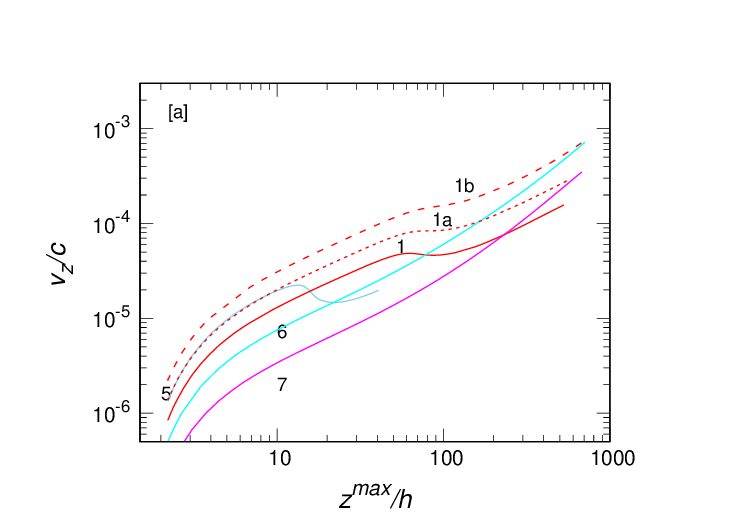}&\hspace{-1.590cm}
  \includegraphics[width=0.44\textwidth]{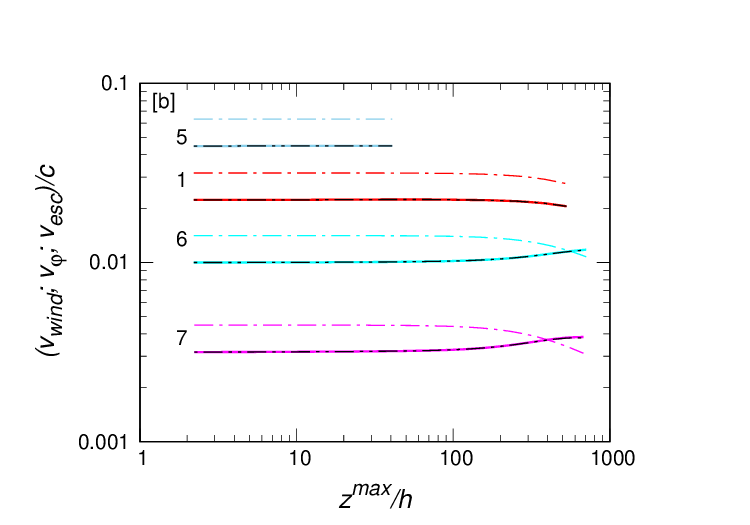}&\hspace{-1.590cm}
  \includegraphics[width=0.44\textwidth]{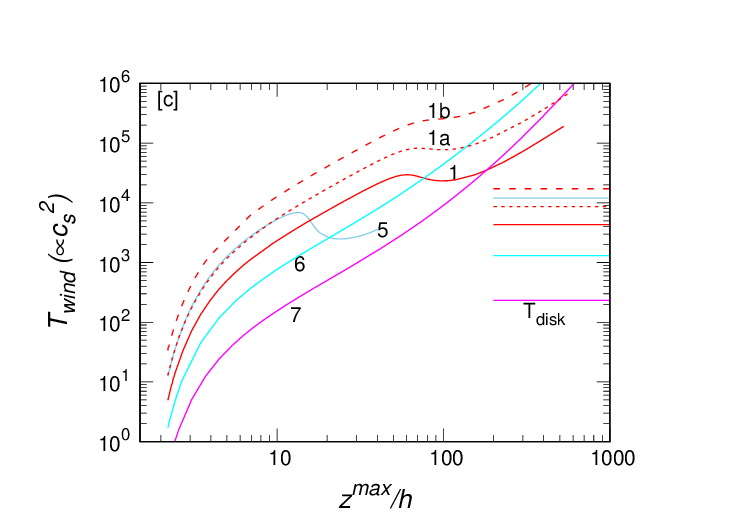}\\\hspace{-1.2cm}
  \includegraphics[width=0.44\textwidth]{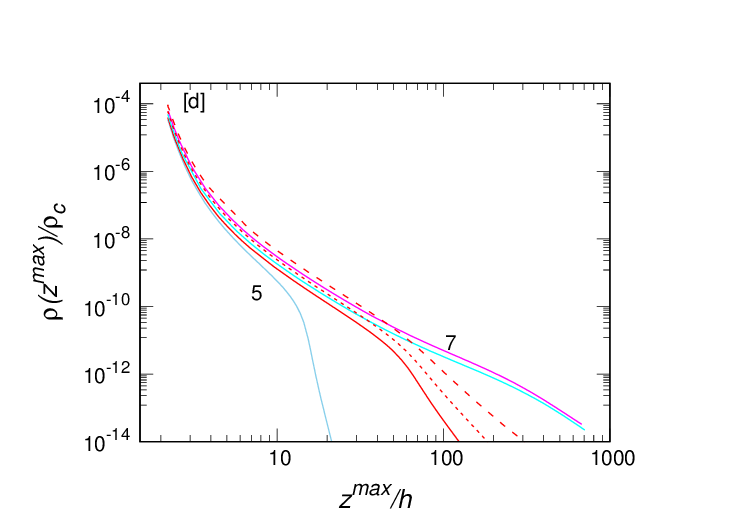}&\hspace{-1.590cm}
  \includegraphics[width=0.44\textwidth]{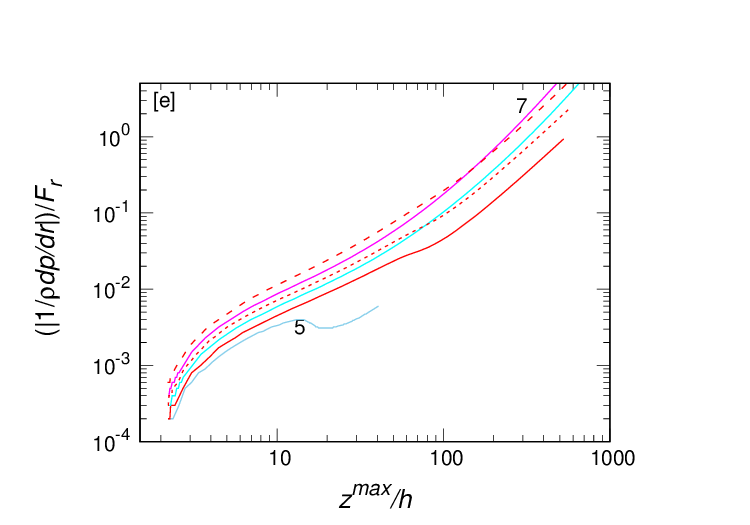}&\hspace{-1.590cm}
  \includegraphics[width=0.44\textwidth]{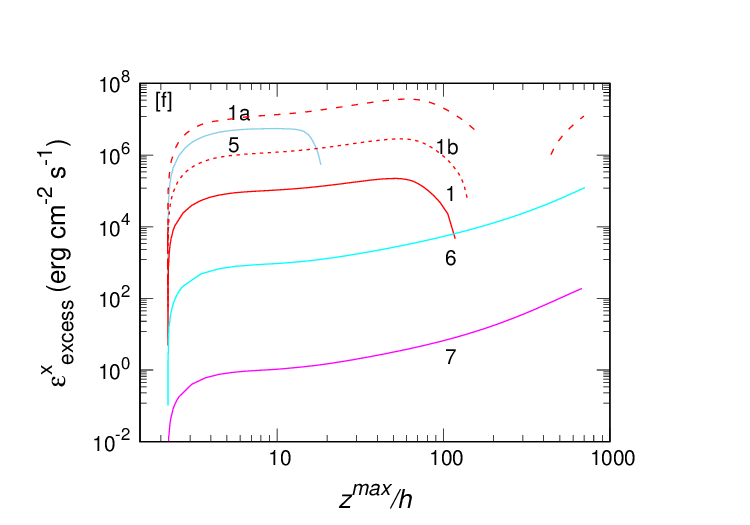}\\
\end{tabular}\vspace{-0.3cm}
\caption{The wind characteristics 
  by varying  $r$ and $\dot{M}$.
  Panels [a], [c], [d], [e], and [f] are for $v_z/c$, $T_{wind}$, $\rho/\rho_c$, $\frac{\frac{1}{\rho}\frac{dp}{dr}}{F_r}$, and $\epsilon^x_{excess}$, respectively.
The panel [b] is for three different velocities $v_{wind}/c$ (solid curve), $v_\phi /c$ (dotted curve), and $v_{esc}/c$ (dotted-dashed curve). In panel [c] the horizontal lines are for $T_c$. The curves symbol and parameters are the same as in Figure \ref{fig:fv_x-r2e3}.
}
\label{fig:wd_r-md}
\end{figure}

We now study the flow variables only at the sonic height (or explore the wind characteristics) by varying either $f_v$ or $\dot{M}$ or $r$. 
The results are shown in Figures \ref{fig:w_fv-2e3} and \ref{fig:wd_r-md}, where the curves and parameters are the same as in Figure \ref{fig:fv_x-r2e3}.
Figure \ref{fig:w_fv-2e3}a shows $v_z$ (solid curves) and $v_r$ (dotted curves).
The $\rho/\rho_c$ (solid curves) and $p/p_c$ (dotted curves) are shown in Figure \ref{fig:w_fv-2e3}b, and Figures \ref{fig:w_fv-2e3}c and \ref{fig:w_fv-2e3}d present
$T_{wind}$ and $\epsilon^x_{excess}$, respectively.
In Figure  \ref{fig:wd_r-md}, the panel [a] shows $v_z$ and the panel [b] is for $v_{wind}$ (solid curves), $v_\phi$ (double dotted-dashed curves), and $v_{esc}$ (single dotted-dashed curve). Panels [c], [d], [e] and [f] are for $T_{wind}$, $\rho/\rho_c$, $\big|\frac{1}{\rho}\frac{\partial p}{\partial r}\big|\big/F_r$ and $\epsilon^x_{excess}$, respectively. 

We note that around $z^{max}$ = $z^{max}_t$ almost all flow variables are starting to change the variation pattern from the previous one (i.e., for $z^{max} < z^{max}_t$), e.g., for curves 1, 2, 4, 1a, and 1b at height $z^{max}_t$, $v_z, v_r$ and $T_{wind}$ change from increasing behavior to almost constant. Hence, the $z^{max}_t$ seems to be an inflection point on these curves. 
We find that in the physically accessible regime (or $z^{max}<z^{max}_t$) for a given $z^{max}$, in general, the $v_z$, $v_r$ and $T_{wind}$ increase with either increasing $f_v$, or increasing $\dot{M}$, or decreasing $r$.
However, $\rho/\rho_c$ increases with increasing either $f_v$ or $\dot{M}$ or $r$. Consequently, the required external flux or corresponding $\epsilon^x_{excess}$ increases with either increasing $f_v$, or increasing $\dot{M}$, or decreasing $r$.
In addition, $\epsilon^x$ does not follow the $\frac{1}{r^2}$ (like $\epsilon^{irr}$) dependency, but it decreases faster than this, e.g., here $\frac{\epsilon^x(r=500R_g)}{\epsilon^x(r=10^4R_g)}$ $:\frac{\epsilon^x(r=2000R_g)}{\epsilon^x(r=10^4R_g)}$ $:\frac{\epsilon^x(r=10^5R_g)}{\epsilon^x(r=10^4R_g)}$ $\sim$ 6000:100:$\frac{1}{900}$.
In other words, to raise the fluid at a similar height in units of $h$ one needs a larger external flux (bolometric flux) for a smaller $r$ in comparison to the large $r$.



We find that $v_{wind} \sim v_\phi$, and $v_{wind}$ increases with decreasing $r$. The $v_{wind}$ exceeds from $v_{esc}$ at $z^{max}$ $\sim$ 550$h$ and 395$h$ for curves 6 and 7, respectively. Since for curve 6, $z^{max}_t$ is $\sim$ 650$h$, 
for the considered parameter sets
for $r<10^4 R_g$ the wind will not escape the system.
In addition, for larger $r$ ($r>10^4 R_g$), the wind will escape the system from the smaller height.
Within the physically accessible regime, at a given  $z^{max}$ the ratio $\big|\frac{1}{\rho}\frac{\partial p}{\partial r}\big|\big/F_r$ increases with increasing either $f_v$ or $\dot{M}$ or $r$. It means that the wind will  start to launch from smaller $z^{max}$ either for larger $f_v$ or $\dot{M}$ or $r$.
As noted earlier (see left panel of Figure \ref{fig:stab-r2e3}), the density for $z^{max} = z^{max}_t$ is constant for curves 1, 2, and 4, however, the pressure and temperature have larger values for the large $f_v$. The reason for this is that at $z = z^{max}$ first  one reaches an equipartition of energy state and second $v_z$ and $v_r$ increase with increasing $f_v$ which will increase the pressure so also temperature.

For a given parameter set, at some $z^{max}$ the $T_{wind}$ is the same as the $T_c$, and we term this  $z^{max}$ as a $z^{max}_{temp}$, and for $z^{max}$ $<$ $z^{max}_{temp}$, $T_{wind} < T_c$.
We find that the range of $z^{max}_{temp}$ for the considered parameter sets is $\sim(8h-15h$). In  general, $z^{max}_{temp}$  decreases with increasing either $f_v$ or $r$ or $\dot{M}$. Although in this range of $z^{max}$, $T_{wind} < T_c$, the value of $x$ is significant, and to reach the $z^{max}_{temp}$  a significant  amount of irradiated flux is required (see panel (f)).


\begin{figure}
\centering
\begin{tabular}{lr}\vspace{-0.9cm}
 \includegraphics[width=0.44\textwidth]{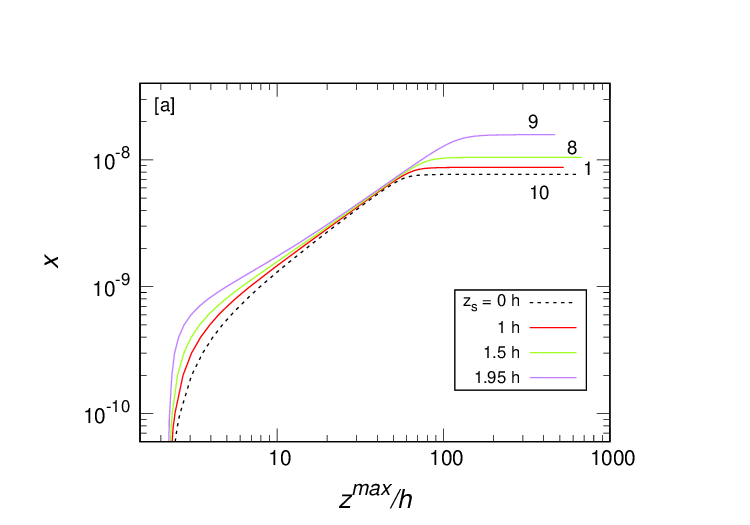}&\hspace{-1.490cm}
  \includegraphics[width=0.44\textwidth]{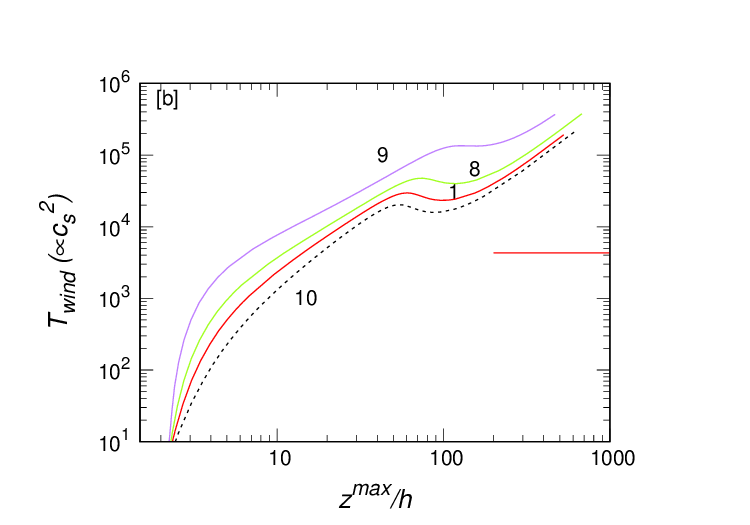}\\
  \includegraphics[width=0.44\textwidth]{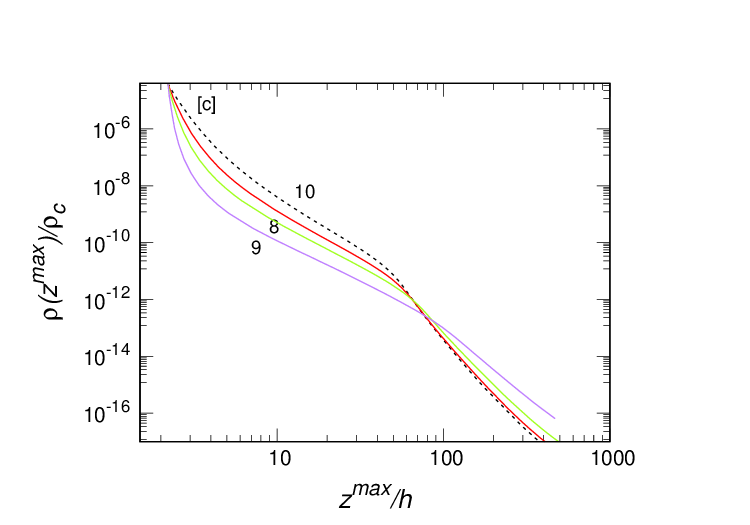}&\hspace{-1.490cm}
 \includegraphics[width=0.44\textwidth]{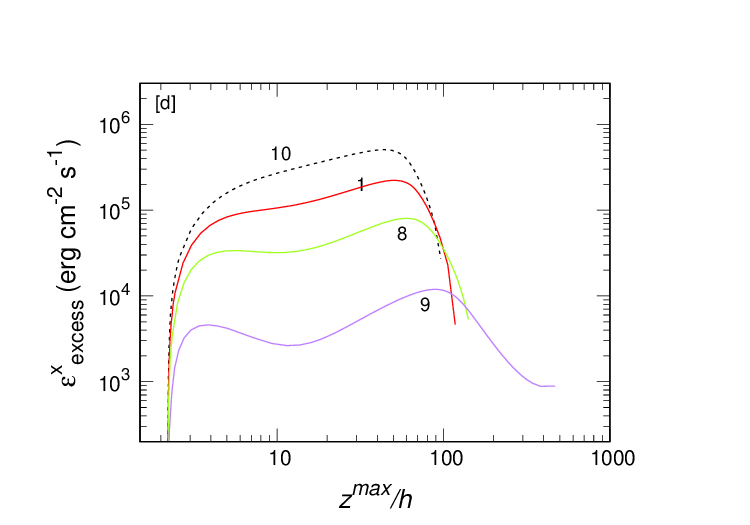}\\
\end{tabular}\vspace{-0.3cm}
\caption{Wind characteristics for four different $z_s$. Curves 10, 1, 8, and 9 are for $z_s$ = 0$h$, 1$h$, 1.5$h$, and 1.95$h$, respectively. Panel [a] is the $x$ vs $z^{max}$ curve. Panels [b], [c] and [d] are for $T_{wind}$, $\rho/\rho_c$ and $\epsilon^x_{excess}$, respectively. In panel [b], the horizontal line is for $T_c$. The rest are the same as in Figure \ref{fig:fv_x-r2e3}.
}
\label{fig:wd_zs-2e3}
\end{figure}

{\tt Irradiation equilibrium height $z_s$:}
Due to the uncertainty over $z_s$, in Figure \ref{fig:wd_zs-2e3} we study the wind characteristics for four different $z_s$. Here,  curves 10, 1, 8, and 9 are for $z_s$ = 0, 1, 1.5, and 1.95$h$, respectively. 
We find that $z^{max}_t$ increases with increasing $z_s$.
The $z^{max}_t$ is $\sim$ 55$h$, 60$h$, 70$h$, and 110$h$ for curves 10, 1, 8, and 9, respectively. 
In the physically accessible regime, for  a given $z^{max}$  the $T_{wind}$ increases with increasing $z_s$ while the density decreases.
Like density,  for  a given $z^{max}$ the $\epsilon^x_{excess}$ decreases with increasing $z_s$, or in other words, for larger $z_s$ one needs a comparatively small irradiated/external flux to launch the wind from almost the same height.
Particularly for curve 10, we still use the optically thin approximation, while in this case, one should also include the optically thick  approximation. Hence, the curve 10 does not show actual things, and here it is presented only for a comparison purposes. 

For the region $2.2h< z^{max} <3.5h$ we have  $\big(\frac{1}{\rho}\frac{\partial p}{\partial r}\big)/F_r <0.001$, where the fluid is either rotationally bound or just in the position to launch the wind. One can exclude this region for the wind analysis. Therefore,  the  $\epsilon^x_{excess}$ =$\epsilon^{irr}$ line will  intersect only
some range of curves among these series of curves (which are obtained by varying  $z_s$). One may approximately argue in the reverse way that for a given $\epsilon^{irr}$ this range of $z_s$ is viable, or in other words, for a given $\epsilon^{irr}$, an equilibrium height (from the midplane) 
would be established at this $z_s$.
For example, for $L_{bol} =8.5 \times 10^{41}$ erg s$^{-1}$ from equation (\ref{eq:irr_en}),  $\epsilon^{irr}$ $\sim$ $1.2\times 10^4$ erg cm$^{-2}$ s$^{-1}$ at $r=2000R_g$ and $\dot{M} =0.001\dot{M}_{Edd}$, and for this $\epsilon^{irr}$ the probable range of $z_s$ is $\approx$ (1.5$\pm$0.3)$h$.
However, as noted earlier,  $\epsilon^x_{excess}$ decreases faster than the  $\frac{1}{r^2}$ for a fixed $z^{max}$ in units of $h$, for a given $L_{bol}$ the $z_s$ would be different for  different $r$. In general,  the $z_s$ will decrease with increasing  $r$ for a fixed $z^{max}$ in units of $h$ and for a given $L_{bol}$.

%

\subsection{Accretion Disk Self-gravity}

The vertical structure of the thin accretion disk is formulated by neglecting the self-gravity  force (or the force exerted by the mass clump enclosed in volume $h^3$ on test mass $m_t$ at height $h$, $F_{clup}$) in comparison to the $F_z$.  The radius at which $F_{clup}$ = $F_z$, is termed as the disk self-gravity radius, $R_{sg}$. Since $F_{clup} = \frac{G (\rho h^3) m_t}{h^2}$ and $F_z = \frac{G M_{c} m_t}{r^2}\frac{h}{r}$; so at $r=R_{sg}$ we have $R_{sg} = \left(\frac{M_{c}}{\rho}\right)^{1/3}$. In the literature, $R_{sg}$ has been computed for different interested regions of the disk; e.g., \cite{Laor-Netzer1989} have obtained $R_{sg}$ for radiation-pressure-dominated region, and \cite{Collin-dumont1990} have computed $R_{sg}$ for five  different regions of the disk (in general, see also \citealt[][and references therein]{Lobban-King2022}).
  For the considered outer region, the $R_{sg}$ can be determined as  (using expression (5.59) of \cite{Frank-etal2002})
  \beqn R_{sg} \approx 4.0 \times 10^{19} \alpha^{56/90} \dot{M}_{16}^{-22/45} f^{-88/45} m_1^{1/3}  \qquad \quad (cm) \eeqn 
  where $\dot{M}_{16} = \dot{M}/(10^{16} \text{g $s^{-1}$}) $, $m_1 = M_{c}/M_\odot$.

In section \S\ref{sec:acces}, we have studied the variation of the disk photosphere at a given $r$ by varying the $\rho$, $p$ from its Keplerian value at that $r$. We found that  $r_{phot}$ decreases with decreasing $\rho$ and $p$ (or $T_c$; see Figure \ref{fig:r2e3surf}).
It means that by decreasing the temperature from $T_c$ at a given $r$ the $F_z$ starts to dominate over gas pressure from a lower height, as a result, $r_{phot}$ becomes smaller.
In the reverse way, if the vertical downward force only increases and $\rho$ and $p$ (or $T_c$) are remain fixed to its Keplerian value, then $r_{phot}$ will again decrease. Physically, the vertical downward force can be increased by disk self-gravity (or $F_{clup}$).
Therefore, in the present formalism to account for the disk self-gravity, we replace the $F_z$ term by $(f_z^{cl} + 1)F_z$ in the respective equation, where $f_z^{cl} = \frac{F_{clup}}{F_z} = \rho\left(\frac{r^3}{M_c}\right)$ for all $z$, however, in the following paragraph, we will constrain the magnitude of $f_z^{cl}$ as a function of $z$. 
Clearly, for $f_z^{cl} \ll 1$ the formalism will retain the form of the thin disk of \cite{Shakura-Sunyaev1973} where the disk self-gravity is unimportant.



\begin{figure}
\centering
\begin{tabular}{lcr}\hspace{-0.9cm}
    \includegraphics[width=0.44\textwidth]{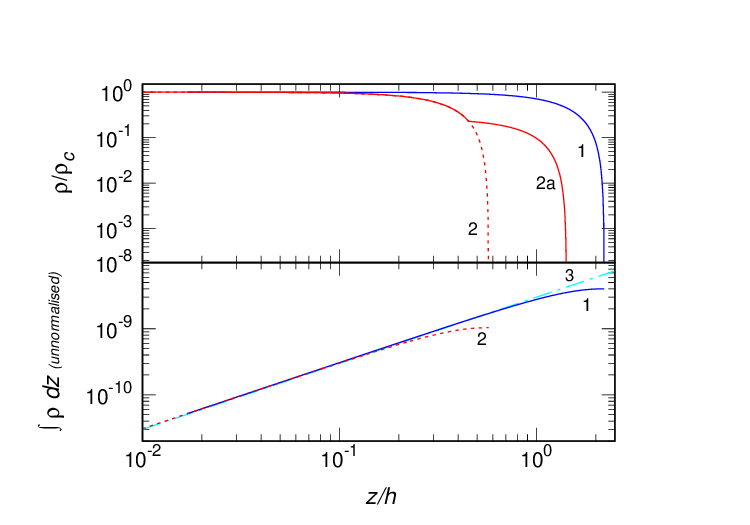}&\hspace{-1.590cm} 
    \includegraphics[width=0.44\textwidth]{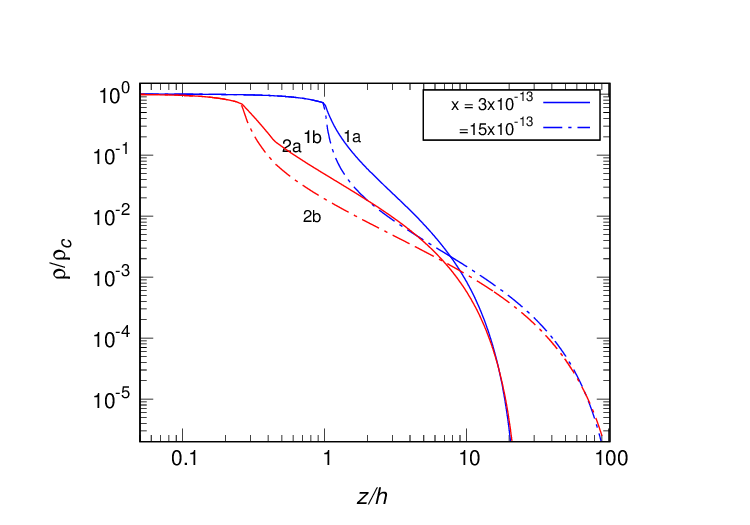}&\hspace{-1.590cm} 
      \includegraphics[width=0.44\textwidth]{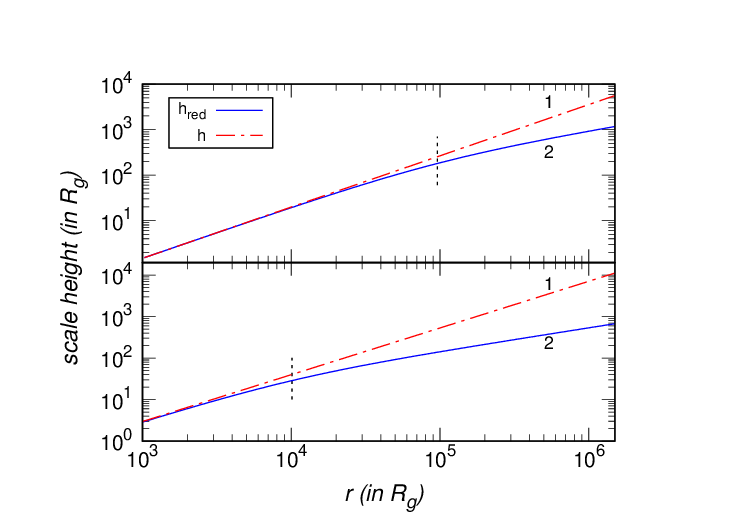}\hspace{-1.2cm}\\
\end{tabular}\vspace{-0.3cm}
\caption{ {\tt Left Panel:} for $x =$ 0. The top left panel is for the vertical structure of $\rho$ by varying the downward vertical force at $r = 10^6 R_g$, $\dot{M}=0.001\dot{M}_{Edd}$. Here, for curve 1 $\frac{F_{clup}}{F_z}$ is $\ll 1$, and for curve 2 $\frac{F_{clup}}{F_z}$ is =13.9. In the case of curve 2a, $\frac{F_{clup}}{F_z}$ is 13.9 for $z/h_{red} < 1.7$, and for $z/h_{red} >1.7$, $\frac{F_{clup}}{F_z} \ll 1$.
    The bottom left panel shows  $\int_{z=0}^z \rho dz$ (or the clump's mass per unit area $\Sigma_{clup}$) as a function of height for curves 1 and 2 of the top panel (which are shown here by the same notation). The curve 3  is $\propto z$.
    {\tt Middle Panel:} for $x \neq$ 0. The vertical profile of $\rho$ for two different $x$ with (curve 1) and without (curve 2) accounting for the disk self-gravity and $z_s$ =1. The curve symbols 1 and 2 are same as the left top panel. The curve suffixes a and b are for $x$ = 3  and 15 $\times 10^{-13}$, respectively. For curve 2 the disk self-gravity effect is accounted for same as the curve 2a of the top left panel.
    {\tt Right Panel:} for $x =$ 0. The disk scale height is a function of $r$, where curves 1 and 2 are  computed without and with accounting for the disk self-gravity or for $h$ and $h_{red}$, respectively. The top and bottom panels are for $\dot{M}$ = 0.001 and 0.1 $\dot{M}_{Edd}$ respectively, and the vertical line represents $R_{sg}$.
} 
\label{fig:sg_h}
\end{figure}

We first study the decrement of $r_{phot}$ (or the corresponding scale height as $r_{phot} \approx 2.2 h)$ by disk self-gravity with $f^{cl}_z =\frac{F_{clup}}{F_z} $ for all $z$. 
As expected, we note that due to disk self-gravity, the shape of vertical disk structure does not change, only the size of the vertical disk gets reduced, and we term the reduced disk photosphere and scale height as $r_{phot}^{red}$ and $h_{red}$, respectively. It still follows the equation (\ref{eq:var-strut}) and $r_{phot}^{red} \approx 2.2 h_{red}$.
  In the top left panel of Figure \ref{fig:sg_h} we show the vertical $\rho$ profile at $r=10^6 R_g$ for $\dot{M} = 0.001\dot{M}_{Edd}$, where $f^{cl}_z \sim 13.9$, in which the curves 1 and 2 are for $f_z^{cl}$ $\approx$ $\ll 1$ and 13.9, respectively for all $z$.
  For the reduced vertical disk structure (curve 2 or curve 2a), we note that $h_{red} \sim 0.26 h$, and  at $z = h_{red}$, $F_z > F_{clup}$ as at this height, in general, $F_z$ decreased by factor $\frac{h}{h_{red}}$ while the $F_{clup}$  decreased by factor $\left(\frac{h}{h_{red}}\right)^3$.
  To check the consistency of the assumption of $f^{cl}_z = \rho \left(\frac{r^3}{M_c}\right)$ for all $z$,
we compute the  $\int_{z=0}^z\rho dz$ (the clump's mass per unit area $\Sigma_{clup}$) as a function of height $z$ 
for curves 1 and 2 of top panel and show these in the bottom left panel by the same curve notations 1 and 2. In bottom left panel, curve 3 is a straight line; hence, within the scale height, the quantity $\Sigma_{clup}$ is $\propto z$ or varies linearly with respect to height, like the $F_z$. 
Above the disk scale height, the $\Sigma_{clup}$ increases slowly, and for $z \gtrsim 1.7 h$ (or $h_{red}$) it remains fixed; in other words, 
in this region, the $F_{clup}$ will  decrease with height.
Therefore, in a conservative manner without loss of generality, for the present calculations, we assume  $f_z^{cl}$ = $\frac{F_{clup}}{F_z}$ for $z < 1.7h_{red}$ and $f_z^{cl} \ll 1$ for $z \gtrsim 1.7 h_{red}$, and we label these conditions as (sg).
In the top left panel, the curve 2a is the vertical profile of $\rho$ using the condition (sg). 

As we note, when the self-gravity dominates over $F_z$, the vertical disk structure gets reduced, and for $z > h_{red}$ $F_z$ starts to dominate over the self-gravity (or $F_{clup}$). Above the reduced disk scale height, the dynamics of test mass will be determined predominantly by $F_z$.  
  We examine it with two different $x$ and $z_s = 1$ for curve 2a of the top left panel of Figure  \ref{fig:sg_h} (also for curve 1, as a reference). The results
  (the vertical structure of $\rho$) are shown in the middle panel of Figure  \ref{fig:sg_h}, where the curve suffixes a and b are for $x$ = 3 and 15 $\times 10^{-13}$; and the curves marked 1 and 2 are for curves 1 and 2a of the top left panel, respectively.
  We note that  for a given $x$ (i)  the $z^{max}$ is the same for both curves marked 1 and 2, and (ii) although the vertical structure of $\rho$ is different for curves marked 1 and 2, at $z = z^{max}$,  the density is same. This also holds for pressure.
  It means that in the case of dominance of self-gravity within the scale height, the wind launching height
and wind characteristics remain the same as when the self-gravity is neglected in the formalism.
Therefore, for a given $x$
the self-gravity effect does not alter the wind characteristics; it only reduce the disk scale height.

In the right panel of Figure  \ref{fig:sg_h} we compare $h_{red}$ and $h$ as a function of radius $r$ 
for two mass accretion rates $\dot{M} =$ 0.001 (top) and 0.1 (bottom) for $M_c = 10^8 M_\odot$. Here, $R_{sg}$ $\approx$ 9.6 $\times 10^4$ and $10^4$ $R_g$; and at $r = 2 \times 10^6 R_g$ the $\frac{h_{red}}{r}$ $\approx$ $\frac{1}{1450}$ and $\frac{1}{2550}$ for the top and bottom panels, respectively. We find that the $h_{red}$ increases with increasing $r$, or in other word the self-gravity does not destroy the concave shape of the disk, and hence the inner disk region can shine  on the outer accretion disk region.
In summary, the disk self-gravity only reduces the vertical size of the disk while maintaining its concave shape, and in the formalism, the disk self-gravity can be incorporated by replacing the $F_z$ with $(f_z^{cl} +1)$ terms, where $f_z^{cl}$ satisfies the condition (sg). The wind characteristics are not influenced by disk self-gravity as above the scale height, the  dynamics of the fluid is predominantly determined by $F_z$.
In the present work, our interest is to explore the wind characteristics, so we explore it without accounting for the self-gravity.

\section{Mass Inflow Rate and Wind  Outflow} \label{sec:wind-gen}

We compute the mass accretion rate as a function of $r$ in the presence of wind outflow for a wide range of the model free parameters. 
In the next section, we  compare the model results with observations for an LLAGN source, NGC 1097, which has an estimated SMBH mass of $1.2 \times 10^8 M_\odot$ (\citealp[][]{Lewis-Eracleous2006}, \citealp[see also][]{Onishi-etal2015}).
To explore the general results, here we also   consider  $M_c = 1.2 \times 10^8 M_{\odot}$. As for many LLAGNs (including this source), there is no estimation for $r_{acc}$ in the literature, for generality, we consider two different $r_{acc}$. Without loss of the general results, we assume that the outer radius of the thin disk  exists  at three-fourths of the $r_{acc}$,  $r_o^{thin}$ = $5 \times 10^5$ and $2 \times 10^6 R_g$ (which corresponds to the hot ISM temperature at $r_{acc}$ $\sim$1.2 and 0.3 keV, respectively, see equation (\ref{eq:rbondi})), as either a sufficient time or an appropriate circularization radius inside the circle of radius $r_{acc}$ or both is requisite to cool and condense the gas for attaining the required temperature and density for the thin disk structure for the given mass accretion rate.
We consider three different Bondi mass accretion rates at radius $r=r_o^{thin}$, $\dot{M}_{Bondi}$ = 0.01, 0.05, and 0.1 $\dot{M}_{Edd}$, in which (0.01, 0.05$\dot{M}_{Edd}$) and (0.01, 0.1 $\dot{M}_{Edd}$) are for $r_o^{thin}$ = $5 \times 10^5$ and $2 \times 10^6 R_g$, respectively. The hot ISM electron number densities are $n$ = $6.3 \times 10^2$ and $\sim 1.5 \times 10^2$  $cm^{-3}$ for $\dot{M}_{Bondi}$ = 0.05 and 0.1 $\dot{M}_{Edd}$, respectively (see equation \ref{eq:mdbondi}).
We obtain the model solutions from the outer radius to the inner radius $r_{in}^{thin}$ $\sim 10^3 R_g$ (\citealp[see][]{Storchi-etal2017}).  

\begin{figure}
\centering\vspace{-0.5cm}
\begin{tabular}{lcr}\hspace{-1.2cm}\vspace{-0.4cm}
  \includegraphics[width=0.44\textwidth]{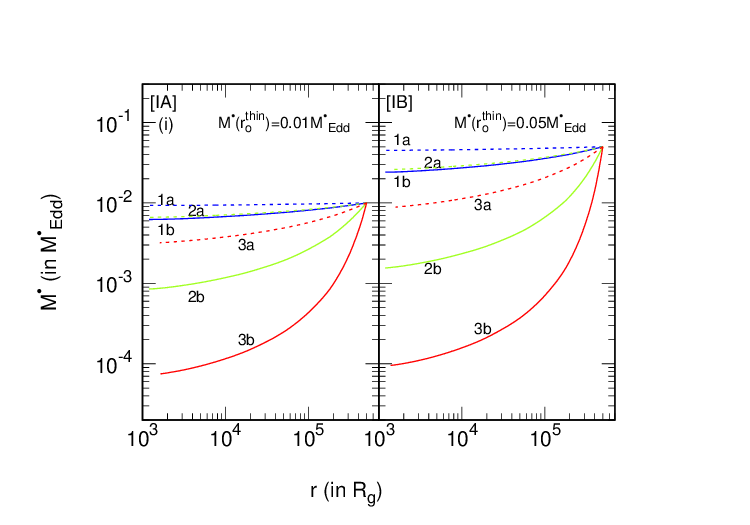}&\hspace{-1.90cm}
  \includegraphics[width=0.44\textwidth]{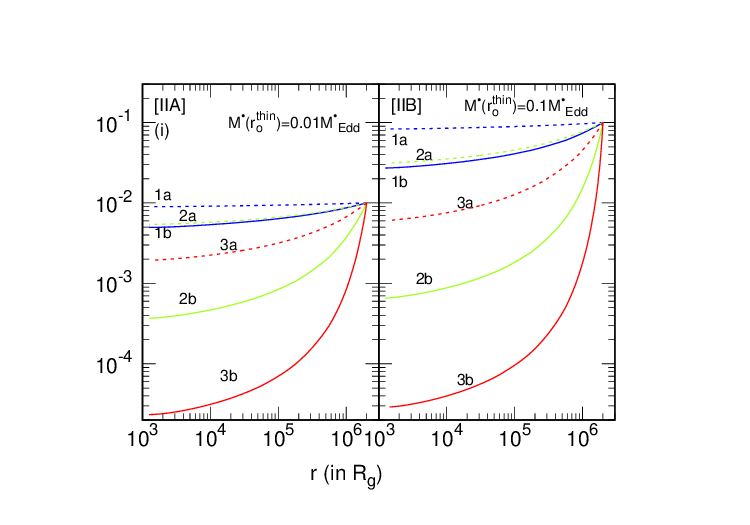}&\hspace{-1.90cm}\vspace{-0.2cm}
  \includegraphics[width=0.44\textwidth]{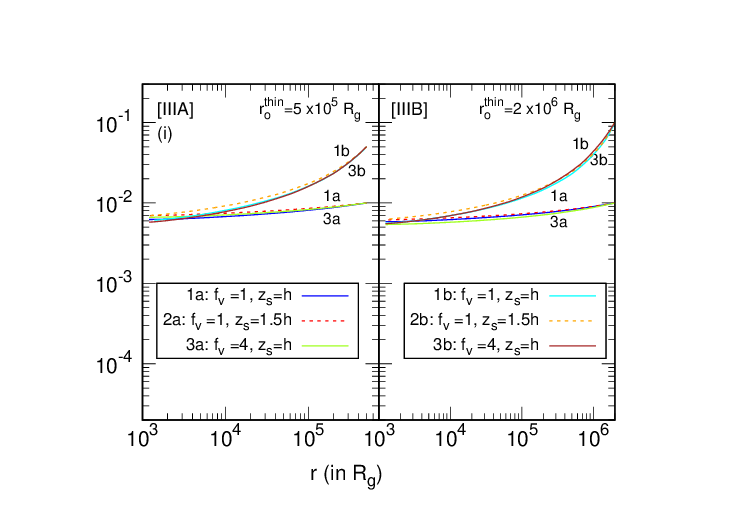} \\\hspace{-1.2cm}
  \includegraphics[width=0.44\textwidth]{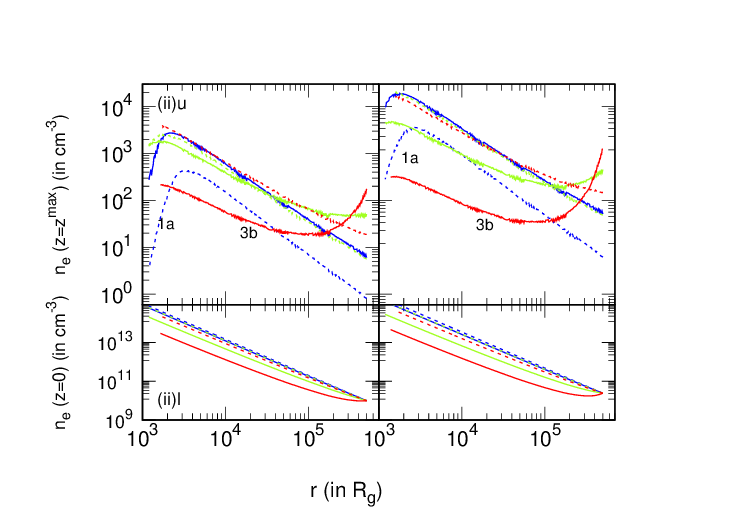}&\hspace{-1.90cm}
  \includegraphics[width=0.44\textwidth]{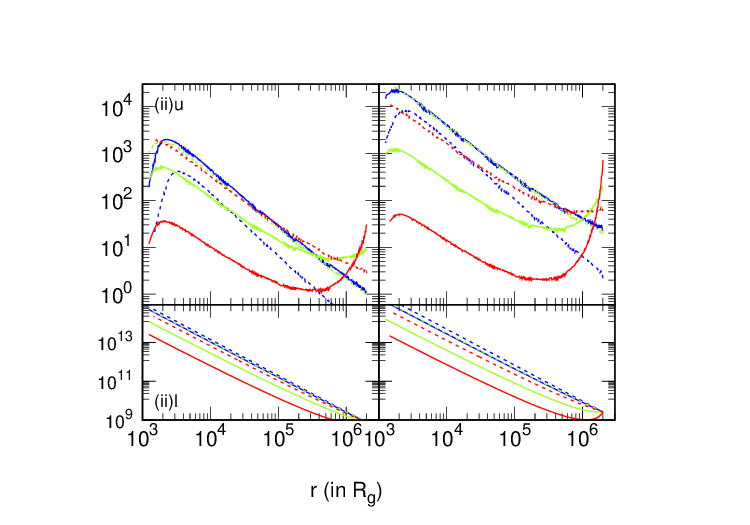}&\hspace{-1.90cm}\vspace{-0.5cm}
  \includegraphics[width=0.44\textwidth]{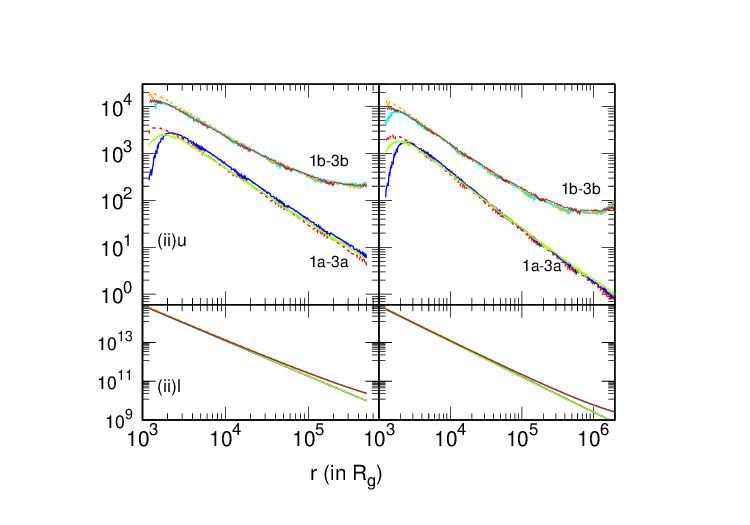} \\\hspace{-1.2cm}
  \includegraphics[width=0.44\textwidth]{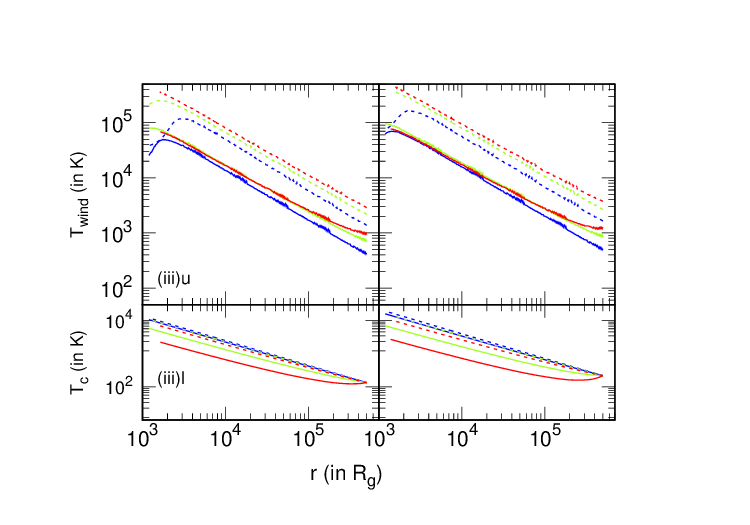}&\hspace{-1.90cm}
  \includegraphics[width=0.44\textwidth]{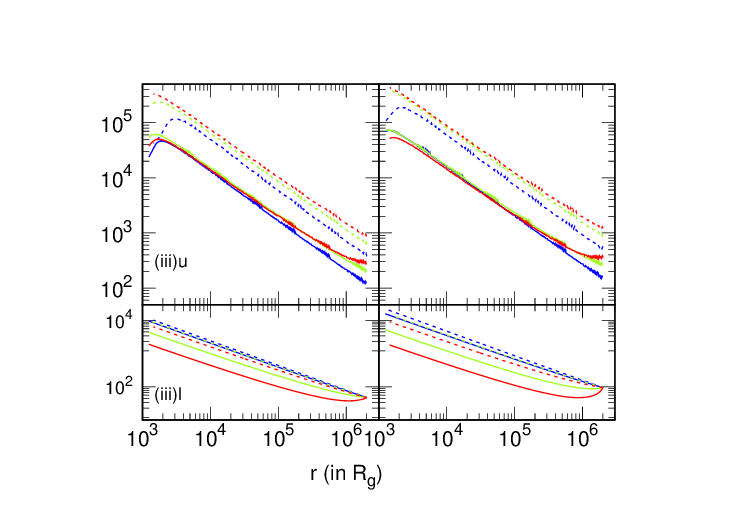}&\hspace{-1.90cm}\vspace{-0.5cm}
  \includegraphics[width=0.44\textwidth]{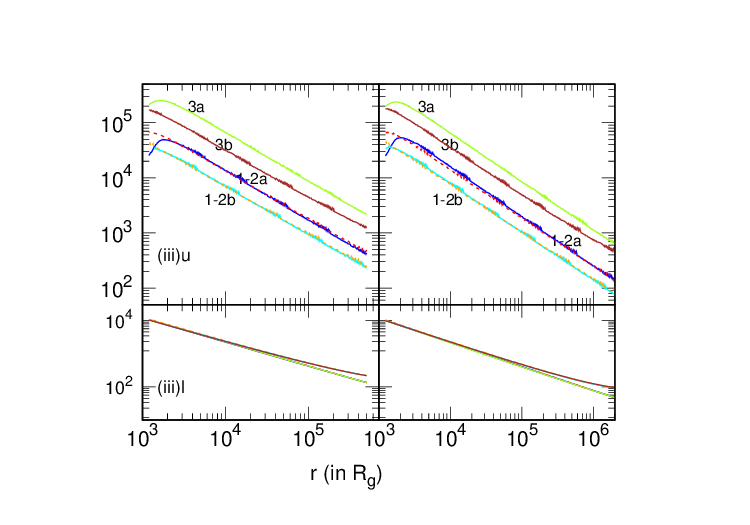} \\\hspace{-1.2cm}
  \includegraphics[width=0.44\textwidth]{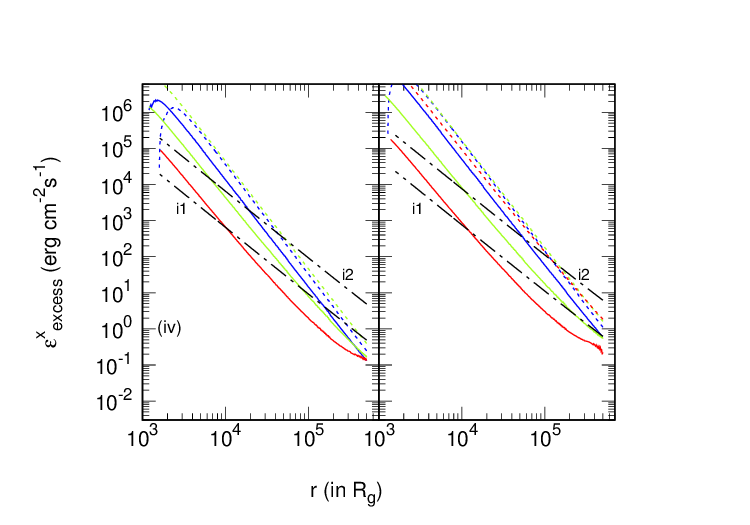}&\hspace{-1.90cm}
  \includegraphics[width=0.44\textwidth]{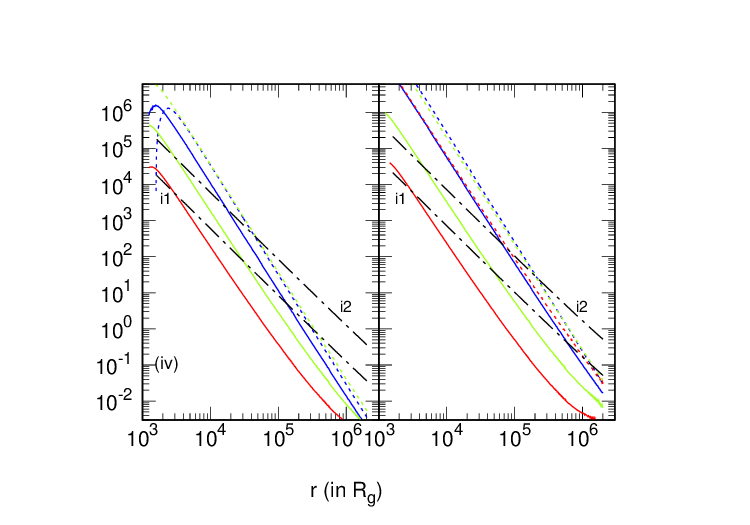}&\hspace{-1.90cm}
  \includegraphics[width=0.44\textwidth]{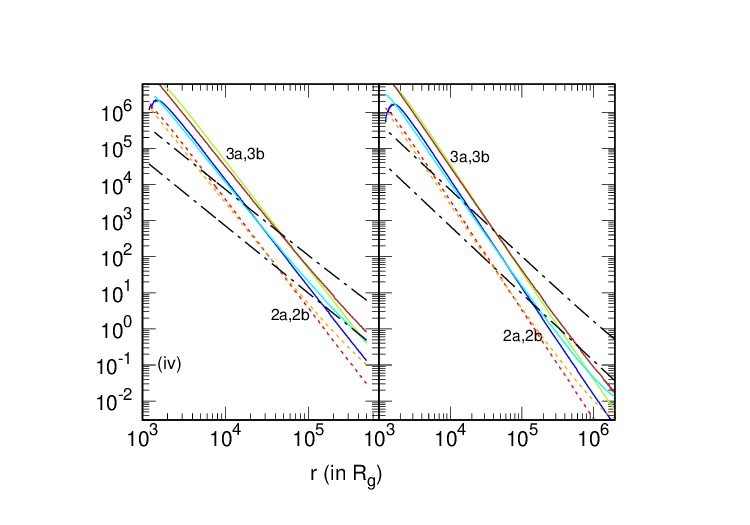}  \\
\end{tabular}\vspace{-0.5cm}
\caption{The mass accretion rate and the wind parameters as a function of $r$, in which the calculation is started from the outer radius of the thin disk with the Bondi accretion rate by varying the initial value of $v_z$ ($f_v$), wind ejection height ($f_z$), and $z_s$. Panels (columns) (I) and (II) are for $r_o^{thin}$ = $5\times10^5$ and $2 \times 10^6$ $R_g$, respectively and $z_s$ = $h$. Panels (IA) (left) and IB (right) are for $\dot{M}_{Bondi}$ = 0.01 and 0.05 $\dot{M}_{Edd}$, respectively, (IIA) and (IIB) are for $\dot{M}_{Bondi}$ = 0.01 and 0.1 $\dot{M}_{Edd}$, respectively.
  Curves 1, 2 and 3 are for $f_v$ = 1, 4, and 8, respectively, and the suffix a and b stand for $f_z$ = 0.2 and 0.1, respectively.
  The right column (III) is for a fixed mass accretion rate at $r_{in}^{thin}$ with having different $\dot{M}_{Bondi}$, $z_s$ and $f_v$, here $\dot{M}(r=r_{in}^{thin})$ $\sim$ 0.006$\dot{M}_{Edd}$.
  Panels (IIIA) and (IIIB) are for $r_o^{thin}$ = $5\times10^5$ and $2 \times 10^6$ $R_g$, respectively. The curves 1, 2, and 3 are for ($f_v$=1, $z_s$=h),  ($f_v$=1, $z_s$=1.5h), and   ($f_v$=4, $z_s$=h), respectively. Suffixes a and b are for $\dot{M}_{Bondi}$ = 0.01 and 0.05 $\dot{M}_{Edd}$ (in panel (IIIA)); and  0.01 and 0.1 $\dot{M}_{Edd}$ (in panel (IIIB)), respectively. 
  Rows (i), (ii)u, (ii)l, (iii)u, (iii)l, and (iv) are for $\dot{M}$, $n_e(wind)$, $n_e(disk)$, $T_{wind}$, $T_c$, and $\epsilon^x_{excess}$, respectively, here suffixes u and l stand for upper and lower panels.
  In row (iv), the dotted-dashed black curves i1 and i2 are for $\epsilon^{irr}$, in which for curve i1 $L_{bol}$ = 8.5 $\times 10^{41}$erg/s and curve i2 is curve i1 vertically shifted by a factor of 10. 
}
\label{fig:wind_gen}
\end{figure} 



The general results are shown in the left (panel (I)) and middle (panel (II)) columns of Figure
\ref{fig:wind_gen} which are for $r_o^{thin}$ = $5\times 10^5$ and $2\times 10^6$ $R_g$, respectively. Here, the curves 1, 2, and 3 are for $f_v$ = 1, 4, and 8; suffixes a and b are for $f_z$ = 0.1 and 0.2, respectively, and panels (IA), (IB), (IIA) and (IIB) are for $\dot{M}_{Bondi}$ = 0.01, 0.05, 0.01, and 0.1 $\dot{M}_{Edd}$, respectively.
The rows (i), (ii)u, (ii)l, (iii)u, (iii)l, and (iv) are for $\dot{M}$, $n_e(wind)$, $n_e(dis k)$, $T_{wind}$, $T_c$, and $\epsilon^x_{excess}$, respectively, where suffixes u and l stand for upper and lower panels and $n_e=\frac{\rho}{\mu m_p}$ is the hydrogen number density.
Respectively, $x$ and $v_{wind}$ are shown in the rows named (v) and (vi) of Figure \ref{fig:wind-gen1}.
In Figure  \ref{fig:wind-gen-h}, we show the respective $z^{max}$ in units of $h$.
In rows (iv) the dotted-dashed black curves i1 and i2 are for $\epsilon^{irr}$, in which the curve i1 is computed for $L_{bol}$ = 8.5$\times 10^{41}$ $erg$ $s^{-1}$ (the bolometric luminosity of NGC 1097 estimated by \cite{Nemmen-etal2014}) using equation (\ref{eq:irr_en}).
Curve i2 is obtained by raising the curve i1 with a factor of 10 to incorporate the uncertainties in the factor $C_{sph}$
\cite[see][]{King-Ritter1998} and  in the unspherical region of inner disk emitting region. 

In the  present model, the wind characteristics mainly depend on $z^{max}$ in units of $h$. However, for the observation point of view, we consider $f_z$ = 0.1 and 0.2, which reflect the viewing angle $i$ $\sim$ 84.2 and 78.6 degrees, respectively.
Since, in the Keplerian disk,  $\frac{r}{h(r)}$ $\propto r^{-1/8} \dot{M}^{-3/20}$, the $z^{max}$ (in units of $h$) will increase with decreasing $r$, particularly $\frac{z^{max}(r=r^{thin}_{in})}{z^{max}(r=r^{thin}_{o})}|_{\text{in unit of}\ h}$ $\sim$ 2.1 and 2.5 for panels (I) and (II), respectively for constant $\dot{M}$ (which is nearly true for curve 1a of all panels, see Figure  \ref{fig:wind-gen-h}).
This is the reason we note that,
in general, the $\dot{M}$ decreases faster around radius $r_o^{thin}$ and  comparatively much slower around radius $r_{in}^{thin}$.
For  example, in panel (IA) the $\dot{M}  \propto$ $r^{2.3}$ and  $r^{0.13}$ (at $\sim 2.5\times 10^5 R_g$) and $\propto$  $r^{0.3}$ and  $r^{0.054}$ (at $\sim 10^4 R_g$) for curves 3b and 1b, respectively.
As noted in Figures \ref{fig:w_fv-2e3} and \ref{fig:wd_r-md}, for a given $z^{max}$, the wind density is larger either for larger $f_v$  or for larger $\dot{M}$ or smaller $r$ when the rest of the parameters are fixed, and it also decreases with increasing $z^{max}$ (or $f_z$). These trends would be also true for $\dot{M}_{out}$ (see equation \ref{eq:m-out}). Also, for the considered parameter sets, we have $v_{wind}$ $\sim v_{\phi}$, or $v_{wind}$ $=\sqrt{c^2/r}$ ($r$ is in $R_g$). 
Consequently, 
$\dot{M}$ decreases more  either by increasing $f_v$ or by decreasing $f_z$ or for larger $\dot{M}$ at fixed $r$ (e.g., see at $r_{in}^{thin}$).
In addition, we find that for a given parameter,  $\dot{M}$ decreases more  by increasing $r_o^{thin}$ (see panels (I) and (II)).

We find that $n_e(disk)$ and $T_c$ both deviate from the $r$ dependency of the Keplerian disk, 
as here for all curves, $\dot{M}$ is not a constant. However, at a given $r$ both will have to take
the Keplerian values; therefore, the $n_e$(disk) and $T_c$ of curve 1a are larger (due to the higher $\dot{M}$) than the respective values of curve 1b, and the same is true for the curves 2 and 3 of all panels.
Interestingly, we note that for curve 3b (also curve 2b) the $n_e$ (wind) initially decreases with decreasing $r$ then after some $r$ it starts to increase like other curves. It is mainly due to a sharp decrement of $\dot{M}$ in this range of $r$, which leads a comparatively lower value of $n_e$ (disk) and which further makes a smaller $h$ (or in other words, the fixed $z^{max}$ corresponds to a large number of times  $h$).
However, in those ranges, the respective $T_{wind}$ deviates slightly from the
general trend while $\epsilon^x_{excess}$ deviates more, as seen by comparing the curves with suffixes a and b.
The $T_{wind}$ and  $\epsilon^x_{excess}$ of the curves with suffix a are comparatively higher than the magnitude of the respective curves with suffix b. 
For the considered parameter sets, on average, $n_e$(wind) is almost $10^{10}$  smaller than the respective $n_e$(disk), and $T_{wind} \sim10^4K$ (a completely ionized temperature for the hydrogen) occurs around the ranges of $r$ ($1-2\times 10^4 R_g$) and $(6-10 \times 10^4R_g$) for curves with suffixes a and b, respectively.
For a few curves (e.g., 1a, 1b, 2b) near $r_{in}^{thin}$, the wind parameters (like $n_e$(wind), $T_{wind}$) start to decrease with decreasing $r$, which corresponds to the situation $z^{max} > z^{max}_t$, so that the range of $r$ of the corresponding curves is not a physically interesting situation.




\section{Comparison with Observation}
NGC 1097 is an LLAGN in a low-ionization nuclear emission-line region, and
it exhibits broad double-peaked Balmer (H$\alpha$) lines that were monitored for more than two decades.  
These broad lines are believed to be generated in the thin accretion disk at $r \approx 550 R_g$; hence, the outer accretion disk 
is a thin disk \cite[e.g.,][and references therein]{Storchi-etal2003, Schimoia-etal2012, Schimoia-etal2015}.
\cite{Nemmen-etal2006} had explained the broadband SED in an inner RIAF plus an outer thin disk with transition radius $R_{tr}$ = 450 $R_g$, and having SMBH mass $M_c$ = 1.2 $ \times 10^8 M_\odot$. Their estimated mass accretion rate at $R_{tr}$ is $\sim 0.0064 \dot{M}_{Edd}$, and the computed bolometric luminosity is $L_{bol} = 8.5 \times 10^{41}$ erg $s^{-1}$ \cite[see also][]{Nemmen-etal2014}. 
However, in NGC 1097, indirect evidence for wind is growing, e.g., from Atacama Large Millimeter/submillimeter Array
(ALMA) observations, \citet[][and references therein]{Fathi-etal2013} have claimed that at  $r = 40$pc (or $\sim 7\times 10^6 R_g$) the molecular gas inflow rate is 0.033$\dot{M}_{Edd}$ and the molecular and ionized gas inflow rate is 0.073$\dot{M}_{Edd}$ in the nuclear spiral arm, which may also be a mass accretion rate at $r_o^{thin}$.
On other hand, one has from SED (optical to X-ray band) modeling $\dot{M} = 0.0064\dot{M}_{Edd}$ at $r \sim 450 R_g$, which indicates a wind outflow from the thin disk regime of NGC 1097.
We now study the decrement of the  mass accretion rate in a constraint manner.
We start the calculations from the $r_o^{thin}$ with  different $\dot{M}_{Bondi}$ (as in the previous section) and constrain $f_z$  for a given $\dot{M}_{Bondi}$ in such a way that $\dot{M}(r=r^{thin}_{in}) \sim 0.006\dot{M}_{Edd}$. We next constrain the wind characteristics by examining the energetics with the observed $L_{bol}$.

In the literature, we do not find the estimated $r_{acc}$ and $\dot{M}_{Bondi}$, we
perform the calculations with the same sets of free parameters, $r_o^{thin}$, $\dot{M}_{Bondi}$, $r_{in}^{thin}$ and $M_c$ of section \S\ref{sec:wind-gen} with a restriction on $\dot{M}$ at $r$ = $r_{in}^{thin}$ (or at $r=r_{in}^{thin}$, $\dot{M} \sim 0.006\dot{M}_{Edd}$).
In addition, to account for the smaller irradiation intensity, we also consider  $z_s$ in the more optically thin regime, $z_s$ = $1.5h$.
The results are shown in the right column (or panel (III)) of Figure \ref{fig:wind_gen}. Panels (IIIA) and (IIIB) are for $r^{thin}_o$ = 5 $\times 10^5$ and $2 \times 10^6 R_g$, respectively.
The curves 1, 2, and 3 are for ($f_v, z_s$) = (1, $h$), (1, $1.5h$), and (4, $h$), respectively. The suffixes a and b are for $\dot{M}_{Bondi}$ = 0.01 and 0.05 $\dot{M}_{Edd}$ in panel (IIIA), and for panel (IIIB) $\dot{M}_{Bondi}$ = 0.01 and 0.1 $\dot{M}_{Edd}$, respectively.
In panel (IIIA) the $f_z$ for curves 1a, 2a, 3a, 1b, 2b, and 3b are 0.1, 0.085, 0.2,  0.0675, 0.055, and 0.125 and for panel (IIIb)  are 0.11, 0.09, 0.2, 0.073, 0.059, and 0.136, respectively. In the present notations, the $\tan^{-1}(f_z)$ will reflect the viewing angle  $i$.
The different rows are the same as the corresponding row of panels (I) or (II).



We find that for a given $\dot{M}_{Bondi}$ the $\dot{M}$ versus $r$ curves are almost independent of $f_v$ and $z_s$ (see the respective curves a and b of both panels), mainly due to the fixed value of $\dot{M}$ at $r_{in}^{thin}$.
As in panels (I) and (II), here $\dot{M}$ also decreases faster around $r_o^{thin}$ and comparatively slower at the inner thin disk radius (see the right panel of Figure \ref{fig:wind-gen-h} for the corresponding $z^{max}$ in units of $h$).
In panel (IIIA), the $\dot{M} \propto $  $r^{0.13}$ and $r^{0.85}$ (at $r=2.5\times 10^5R_g$) and  $\dot{M} \propto $  $r^{0.06}$ and $r^{0.18}$ (at $r=10^4R_g$) for curves 1a and 1b, respectively.
In panel (IIIB), the $\dot{M} \propto $  $r^{0.13}$ and $r^{1.1}$ (at $r=10^6R_g$) and  $\dot{M} \propto $  $r^{0.06}$ and $r^{0.16}$ (at $r=10^4R_g$) for curves 1a and 1b, respectively.
There are two branches in the $n_e$(disk) and $T_c$ curves around $r_o^{thin}$, in which the lower branch corresponds to the lower value of $\dot{M}$ (or curves with suffix a) and the upper branch is for curves with suffix b.


We find that for $r \lesssim 0.1r_o^{thin}$ the $n_e$(wind) $\propto \frac{1}{r}$ in all curves with suffix b, and in the case of curves with suffix a,  $n_e$(wind) $\propto \frac{1}{r^{1.2}}$. 
For a given $\dot{M}$ profile, $T_{wind}$ does not depend on $z_s$ but it increases with increasing $f_v$.
For curve 1b of panel (IIIA), the $T_{wind}$ is equal to the $T_{c}$ around radius $r_o^{thin}$. Hence, in general, for $f_v =1$ one obtains $T_{wind} < T_c$ for $\dot{M} > 0.05\dot{M}_{Bondi}$. 
In the case of curve 1b (or curve 2b) of panel (IIIB), $T_{wind} < T_c$ for $r \gtrsim 1.5\times 10^6R_g$, and around $r_o^{thin}$ the $T_{wind}$ drops to $\sim$76 K (or $\sim$0.8$T_c$). Therefore,  for larger $\dot{M}_{Bondi}$, one will have a comparatively larger $n_e$(wind) and smaller $T_{wind}$ ($<T_c$) near $r_o^{thin}$, and these may  provide a favorable condition to form a molecular gas in this wind medium (here, we simply speculate about this, in future, we will study this in details).
In NGC 1097, a molecular gas has been observed on parsec-scale \cite[e.g.,][]{Fathi-etal2013, Izumi-etal2017}.
However, the estimated ranges of hydrogen number density and temperature  from the emission molecular line ratio are comparatively higher than the wind density and temperature of this work \cite[e.g.,][]{Izumi-etal2013}.
Finally, we compare the modeled values of $\epsilon^x_{excess}$ with $\epsilon^{irr}$ of the observed $L_{bol}$, and all considered parameter sets can launch the wind to some extent. Particularly, for curves 2, the wind can launch up to the radius $4\times 10^3$ and $\sim4\times 10^5 R_g$ for $\epsilon^{irr}$ curves i2 and i1, respectively.
The wind is an equatorial wind with $v_{wind} \sim v_\phi$.

Quantitatively, at two different radii $r =$ $10^5 R_g$ and $10^4 R_g$ the range of the wind hydrogen number density are ($\sim$20 $-$ 300 $cm^{-3}$) and ($\sim$300 $-$ 3000 $cm^{-3}$); the range of wind kinetic luminosity ($L_{wind} = \frac{1}{2}\dot{M}_{out} v_{wind}^2$) are  ($\sim 5 \times 10^{36} -  10^{38}$ erg s$^{-1})$ and ($\sim 2 \times 10^{37} - 3\times 10^{38}$ erg s$^{-1})$.  $L_{wind}$ is smaller than $L_{bol}$. In addition, the range of $n_e$(wind) is consistent with that obtained using MHD simulations for wind in the inner advection disk of LLAGNs M81* by \cite{Shi-etal2021}. In the
  wind medium 
  the emission/absorption lines can be generated by the photoionization process, and due to the equatorial wind these lines would be  blue- and redshifted. For the constraint range of $f_z$,  the viewing angle is  $i\gtrsim 85$ degrees, and the wind will appear almost along the disk plane. In the present analysis, we consider the arbitrary values for $r_o^{thin}$ and $\dot{M}_{Bondi}$, hence, in general the predictions can be feasible for any LLAGN with $\dot{M} < 0.005\dot{M}_{Edd}$ at $r=r_{in}^{thin}$.

In the present work we consider a constant $f_z$ (for $z^{max}$) over $r$, in general $f_z$ can vary with $r$ (e.g., a smaller $f_z$ near to the $r_o^{thin}$ and larger $f_z$ near to the $r_{in}^{thin}$, or the reverse) and it may provide a refined constraint on the model by comparison with observations.
In summary, the winds are present in the outer thin disk of NGC 1097 with speed Keplerian $v_\phi$ ($< v_{esc}$, so it does not escape the system), which can be visible almost along the disk ($i>85$ degrees). Therefore, the red-/blueshifted emission lines generated in the wind medium are visible mainly along the disk plane. In general, it would be true for a wide range of LLAGNs.  The wind medium around $r_o^{thin}$ may provide a favorable conditions for molecular gas formation at a  higher value of  $\dot{M}_{Bondi}$, which are consistent with observation of molecular gas at parsec scales in NGC 1097.

\begin{figure}
\centering
\begin{tabular}{l}
  \includegraphics[width=1.\textwidth]{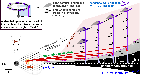}
 \end{tabular}\vspace{-0.3cm}
\caption{\footnotesize A cross-sectional schematic view of the thermal irradiated wind model from outer geometrically thin accretion disks in LLAGNs. Here the inner RIAF (of LLAGNs) shines on the outer disk, due to low-intensity irradiation, the irradiated energy is dissipated into the optically thin medium at height $z_s$, form the midplane (as shown here, and its effect is parameterized by a number $x$ through unbalancing hydrostatic equilibrium).
  Here, the photosphere of the Keplerian disk is either $2.2h$ (where the disk self-gravity is unimportant) or $2.2h_{red}$ (where the self-gravity dominates over $F_z$) . We solve the steady axisymmetric governing equations in cylindrical coordinates ($r,\phi,z$)  at a fixed $r$ along the $z$-axis (which is not a streamline, as at any $z$ the fluids are rotating under the influence of radial gravity $F_r$ and pressure gradient $\frac{1}{\rho}\frac{\partial p}{\partial r}$ components with velocity $\sqrt{v_r^2+v_\phi^2+v_z^2}$ \Big) with having tiny initial $v_z$ and with implementing $x$ for $z > z_s$ while for $z< z_s$ $x = 0$.
  We initialize the flow variable at the midplane by its Keplerian value. However, for $\frac{\Delta r}{r} \ll 1$ (see the right panel of Figure \ref{fig:shear-r1e4}) this approach reproduces approximately the similar solutions of the grid points (in $r,z$) approach.
  The smooth solution at sonic height ($z^{max}$) indicates an isobaric regime , also from the height $z_f (\ll z^{max}) $  $\frac{1}{\rho}\frac{\partial p}{\partial r}$ is started to support the fluid rotations, as a result of both conditions the wind will be launched at height $z^{max}$ with speed $v_{wind} =\sqrt{v_r^2+v_\phi^2+v_z^2}$ ($\sim v_\phi)$, the wind is an equatorial wind (see inset). 
  We start the computation from the outer radius of the thin disk $r_o^{thin}$ (or three-forths of the Bondi accretion radius $r_{acc}$) with Bondi mass accretion rate $\dot{M}_{Bondi}$ up to the inner radius $r_{in}^{thin}$ 
  for a fixed viewing angle $i$. 
  Here we show the wind ejected from $z^{max} = 0.2r$, also showing a  different ejection height (which will happen for a different irradiation intensity or different $x$). 
 } 
\label{fig:geo-cal}
\end{figure}

\section{Summary}

We have extended the Paper I work, mainly by studying the mass inflow rate as a function of radius in the presence of wind outflow,  and applied the model for the outer disk of LLAGNs (thin disk) while taking into account the inner disk (RIAF) irradiation,  where the calculations are started from the three-forths of the Bondi accretion radius $r_{acc}$ (or the outer radius of the thin disk $r_o^{thin}$) with Bondi mass accretion rate $\dot{M}_{Bondi}$.
We have also clarified some assumptions, especially due to the low-intense external heating at outer region from inner disk of LLAGNs the present formalism is only applicable for the optically thin medium, that is now the base of wind launching is not a midplane (like Paper I) but at some height from the midplane $z_s$ (also termed as an effective irradiation equilibrium height where the irradiated energy is almost deposited into the medium, see point 'a' of section \S\ref{sub:sol}, 
also see the Figure \ref{fig:geo-cal} for a schematic diagram of the model).
In particular, in LLAGNs, the high-energy emission and double-peaked broad $H\alpha$ emission lines reveal that it comprises of both types of accretion flow, an inner hot accretion flow (mainly RIAF) and an outer thin disk \cite[e.g.,][]{Ho2008, Storchi-etal2017}. The wind outflow is ubiquitous in LLAGNs, mainly inferred in an indirect way (like two independent methods estimate mass inflow rate at two radii, inner and outer, in which $\dot{M}$ at inner radius is smaller in comparison to the outer, e.g., in NGC 1097 $\dot{M} (r)$ $\sim$ 0.006$\dot{M}_{Edd}$ (at 450$R_g$ by SED modeling, \citealp[][]{Nemmen-etal2006}) and $\sim 0.07 \dot{M}_{Edd}$ (at 7 $\times 10^6 R_g$ by the estimation of mass accretion rate in nuclear spiral arm due to the presence of molecular clouds, \citealp[][]{Fathi-etal2013}), however for few cases it is interpreted indirectly by blueshifted absorption/emission lines \cite[e.g.,][]{Goold-etal2023}.


Similar to Paper I, we have considered a steady, axisymmetric disk in cylindrical coordinates and set up a formalism for a wind outflow 
along the $z$-axis at a given launching radius $r$ from the height $z_s$.
We have assumed a very small vertical speed $v_z$ in comparison to the sound speed $c_s$, and taken its magnitude as a ratio of the radial speed $v_r$; $v_z =f_v v_r \ll c_s$ at the midplane for a given $r$.
We have taken account of both tangential shearing stresses $W_{\phi r}$ and $W_{\phi z}$ and assumed that other shearing stress is negligible in comparison to the 
$W_{\phi r}$, $W_{\phi z}$; or $W_{rz} \sim$0.
Like the Keplerian disk, we adopted $\alpha$-prescriptions for $W_{\phi  r}$, $W_{\phi r} = \alpha p$ and assumed that at any height,  this is also valid, i.e., $W_{\phi r}(z)=\alpha p(z)$ (however, we find a consistency in the solution under this assumption, see section \S\ref{sec:stres-dp}). 
Our interest for the solution is the outer region of the disk (where the gas pressure dominates over radiation pressure, $p \gg p_{rad}$), and  we incorporate the irradiation effects of the inner region into the outer, which can only unbalance the hydrostatic equilibrium as the pressure due to the irradiation $p_{rad}^{irr}$ $<$ $p_{rad}$ $\ll p$. It is parameterized by $x$, where $x$ = 0 reflects that the flows are in vertical mechanical equilibrium. 
As the Keplerian disk, we have assumed that within the scale height the medium is optically thick (where opacity is mainly due to the free-free absorption) and the viscous generated heat radiates out immediately in the vertical direction by blackbody emission.
As in the present case, the irradiated energy gets almost absorbed in the optically thin medium at height $z_s (\gtrsim h)$, so it does not contribute to the increase in the blackbody temperature. 
We have $v_z \ll c_s$, and in this limit, the present formulation behaves like a Keplerian disk, at least around the midplane. Therefore, for the initial value of the flow variable, we take the corresponding  Keplerian value at a given $r$. 
We have solved the model equations for $z< z_s$ with $x$ = 0, and for $z\gtrsim z_s$ with a given $x$ (see the Figure \ref{fig:geo-cal} for a schematic diagram for the model).

The present framework is equivalent to the Keplerian disk for  $x=0$ and $f_v \ll 1$. For the Keplerian disk, if one computes either the quantity $\big[-F_r(z)+\frac{\lambda^2}{r^3}|_{z=0} +\frac{1}{\rho}\frac{\partial p}{\partial r}|_{z=0}\big]$\big($\equiv \frac{1}{\rho}\frac{\partial p}{\partial r}(z)$ for the constant $\lambda$ within the scale height\big) or $\frac{1}{\rho}\frac{\partial p}{\partial r}(z)$ in the present model at a given $r$ then they flip the sign at height $z_f \sim 0.92h$.
In other words, for $z<z_f$ the radial component of the pressure gradient acts in a radially outward direction like the Keplerian disk, while for $z>z_f$ it supports the rotation of fluid.
In general, the sign flip height $z_f$ increases with increasing $x$ for a given parameter set. 
The sonic point ($v_z^2 \sim v_r^2 \rightarrow \Gamma c_s^2$) provides two conditions for the smooth solution, in which the condition (i), $|\frac{1}{\rho}\frac{\partial p}{\partial z}| \approx |v_z \frac{\partial v_z}{\partial z}|$ and  $|\frac{1}{\rho}\frac{\partial p}{\partial r}| \approx |v_r \frac{\partial v_r}{\partial r}|$, states that fluids reach to the equipartition of the energy state, and there is no pressure gradient above the sonic point. The sonic height is the maximum attainable height by fluids and termed as $z^{max}$, and $z^{max}$ increases with increasing $x$.
We find $z_f \ll z^{max}$. For $z>z_f$ the $\frac{1}{\rho}\frac{\partial p}{\partial r}$ and $F_r$ both support the fluid rotation, and at $z^{max}$ the fluid reaches the isobaric regime (or there is no pressure gradient); therefore, if  
$\frac{1}{\rho}\frac{\partial p}{\partial r} \ll F_r$ at $z^{max}$ then $F_r$ can support the rotations  alone, and the fluid would be rotationaly bound.
In other cases (or sufficiently large  $\frac{1}{\rho}\frac{\partial p}{\partial r}$ in comparison to the $F_r$), the $F_r$ cannot support the rotations alone, and the fluid would be ejected from the $z^{max}$ as a wind outflow with velocity $(v_r,v_\phi,v_z)$ and speed $v_{wind}$= $\sqrt{(v_r^2+v_\phi^2+v_z^2)}$.
For the considered parameter sets, here we note that 
$v_{wind}\sim v_\phi$, that is, the wind is an equatorial wind. 

For the $x=0$ case, we obtain the disk photosphere $r_{phot}\sim 2.2h$, which increases very slightly with  decreasing $f_v$, and is also independent of $r$. 
The Keplerian disk photosphere is $2.2h$, and the disk shape is concave as $h \propto r^{9/8}$.
Here, the vertical profiles of pressure $p$ and density  $\rho$ are isothermal (like Keplerian), but both have different scale heightd (unlike the Keplerian disk).
For nonzero $x$, and for the given parameters set, the $p$ scale height increases with increasing $x$ (or $z^{max}$), while the $\rho$ scale height decreases (see left and middle panels of Figure \ref{fig:p_x-r2e3}), where the scale height is defined by equation (\ref{eq:h-zs}) for a given $z_s$. 
Interestingly.  
the scale height of $p$ and $\rho$ do not change after some large $x$ (or effectively profile also) we term this $x$ as a $x^{max}$ and corresponding $z^{max}$ as  $z^{max}_t$.
The $x^{max}$ can also be depicted by a horizontal region in $x$ versus $z^{max}$ curves (see Figure \ref{fig:fv_x-r2e3}), where a very small increment in $x$ leads to a large increment in $z^{max}$. Therefore,  we argue that the model solution with $z^{max}>z^{max}_t$ is not a physical (see also Figure \ref{fig:stab-r2e3}).
In addition, we note that $z^{max}_t$ is related to the disk photosphere, in which $z^{max}_t$ at a given $r$ is a maximum possible $r_{phot}$ where the gravity can hold  the gas with the highest sound speed in hydrostatic equilibrium (see Figure \ref{fig:r2e3surf}). 

We have explored the wind characteristics (or in other words, the fluid variables as a function of $z^{max}$) by varying the free
parameter, either initial $v_z$ (or $f_v$), or mass accretion rate $\dot{M}$, or $r$ while keeping the rest parameters are constant.
The $z^{max}_t$ increases with increasing $f_v$, however, the $x$ vs $z^{max}$ curves for different $f_v$ overlap each other (see left panel of Figure \ref{fig:fv_x-r2e3}), and the wind density at $z^{max} = z^{max}_t$ is constant for all $f_v$. It signifies that the role of $f_v$ is only to raise the $z^{max}_t$.
The $z^{max}_t$ increases with increasing $r$. Interestingly, for the considered parameter set,  at $z^{max} \sim z^{max}_t$, $v_{wind}$ becomes equal to the escape velocity $v_{wind}\sim v_{esc}$ for $r=10^4R_g$. For $r<10^4 R_g$ the wind cannot escape the system. And for $r>10^4R_g$ the $z^{max}$, where the wind starts to escape the system, decreases with increasing $r$. Since for higher $z^{max}$, where $v_{wind}>v_{esc}$ there is still $v_{wind} \sim v_\phi$ (see Figure \ref{fig:wd_r-md}), the escaped wind (which is an equatorial wind, and ejected in all directions, see inset of Figure \ref{fig:geo-cal}) material may contaminate the rotation curve of its galaxy on kiloparsec scales.

In general, $v_z$, $v_r$, $T_{wind}$, and $|\frac{1}{\rho}\frac{\partial p}{\partial r}|/F_r$ increase with increasing $z^{max}$, while $p$ and $\rho$ decrease with increasing $z^{max}$.
Within the physically accessible regime, for a given $z^{max}$ the $v_z$, $v_r$, $T_{wind}$, and $\epsilon^x_{excess}$ increase with either increasing $f_v$, or increasing $\dot{M}$, or decreasing $r$. And $\rho/\rho_c$, and $|\frac{1}{\rho}\frac{\partial p}{\partial r}|/F_r$ increase with increasing either $f_v$ or $\dot{M}$, or $r$.
Interestingly, for the considered parameter sets, $T_{wind}$ becomes smaller than the midplane disk temperature $T_c$ for (on average) $z^{max} < 11h$.
We have also studied the wind characteristics by varying $z_s$. Particularly at a given $r$ within the physically accessible regime for a given $z^{max}$ the $\epsilon^x_{excess}$ decreases with increasing $z_s$. For a given $L_{bol}$, one can find irradiation flux $\epsilon^{irr}$ for a given $r$ using equation (\ref{eq:irr_en}). By comparing $\epsilon^{irr}$ with the computed $\epsilon^x_{excess}$ for different $z_s$ one can estimate the desired range of $z_s$  where the irradiated energy is completely deposited into the medium.
In addition, we found that  $\epsilon^x_{excess}$ decreases faster than $\frac{1}{r^2}$, so for a given $L_{bol}$ the $z_s$ would be different for different $r$.

In AGNs, in general, the disk self-gravity force starts to dominate over $F_z$ for $r > R_{sg}$ ($R_{sg}$ : disk self-gravity radius); evidently (mainly by estimation of the size of the broad line region, e.g., \citealp{Collin-etal2006}), the thin disk is present beyond the $R_{sg}$. However, the thin disk is formulated by neglecting the disk self-gravity. As we noted (in section \S \ref{sec:acces}) that by reducing the initial value of pressure and density (or temperature) from  their Keplerian value, $F_z$ starts to dominate over the pressure gradient from the lower height, so to incorporate the self-gravity effect in the present formalism, we have replaced the $F_z$ by the term \big($\frac{F_{clup}}{F_z}+1\big)F_z$, where $F_{clup}$ is the self-gravity force due to the mass enclosed in volume $h^3$ on the disk. We found that due to disk self-gravity, the concave shape of the thin disk is unchanged, only its vertical size is reduced. And in the new vertical structure above the scale height ($h_{red}$) the $F_z$ again starts to dominate over disk self-gravity. Further, the wind characteristics remain unaffected from its effect (see Figure \ref{fig:sg_h}).

We have studied the general trend of mass inflow rate in the presence of wind outflow and the corresponding wind characteristics as a function of $r$ at $z_s=h$ for two different $z^{max}$ with $f_z$ = 0.1 and 0.2 (where $z^{max} = f_z r$). We have started the computation with two different outer radii of the thin disk $r_o^{thin}$ ($\sim r_{acc}$) = $5\times 10^5$ and 2$\times 10^6$ $R_g$ and for each $r_o^{thin}$ we have two different $\dot{M}_{Bondi}$, namely, (0.01, 0.05 $\dot{M}_{Edd}$) and (0.01, 0.1 $\dot{M}_{Edd}$), respectively with an inner radius of thin disk $r_{in}^{thin}$ =$10^3R_g$.  
In the present model, the wind parameters mainly depend on $z^{max}$ in units of $h$. And in the Keplerian disk $\frac{r}{h(r)} \propto r^{-1/8} \dot{M}^{-3/20}$, 
for a given $f_z$, the $z^{max}$ (in units of $h$) for higher $r$ is smaller in comparison to the lower $r$ (see Figure \ref{fig:wind-gen-h}). As a result, $\dot{M}$ decreases faster around $r_o^{thin}$ and comparatively much slower around $r_{in}^{thin}$ (see first row of Figure \ref{fig:wind_gen}).  
In general,  $\dot{M}$ decreases with a faster rate by either increasing $f_v$ or decreasing $f_z$.
The wind is an equatorial wind with speed $v_{wind}$ $\sim v_\phi$, blowing out in all directions, hence blue/redshifted emission/absorption lines are expected  in the wind medium.

We have constrained the model parameters with observations of LLAGN  NGC 1097, where one has $\dot{M} (r=450R_g$) = 0.0064 $\dot{M}_{Edd}$ (from SED modeling) and $L_{bol} = 8.5 \times 10^{41}$ erg s$^{-1}$ \cite[e.g.][]{Nemmen-etal2006}.
As in the literature there is no estimation of $r_{acc}$ and $\dot{M}_{Bondi}$, and we considered the same sets of $r_o^{thin}$, $\dot{M}_{Bondi}$ and $r_{in}^{thin}$ as the general case and took three sets for ($f_v,z_s)$ = (1,$h$), (1,1.5$h$) and (4,$h$).
We constrained $f_z$ in such a way that $\dot{M}(r=r_{in}^{thin})$ $\sim 0.006 \dot{M}_{Edd}$.  
For higher $\dot{M}_{Bondi}$, the approximate hydrogen number density for wind $n_e$(wind) $\propto r^{-1}$ for $r< 0.1 r_o^{thin}$ and for lower $\dot{M}_{Bondi}$ the $n_e$(wind) $\propto r^{-1.2}$ for all $r$ (see the second row of the right column of Figure \ref{fig:wind_gen}).
Importantly, for $\dot{M}_{Bondi} \sim 0.05\dot{M}_{Edd}$, the temperature of the wind medium $T_{wind} \sim T_c$  at $r=r_o^{thin}$. Therefore, for $\dot{M}_{Bondi} > 0.05\dot{M}_{Edd}$, $T_{wind}$ would be smaller than $T_c$ near $r_o^{thin}$, and here $T_{wind} = 0.8T_c$ at $r_o^{thin}$ = 2$\times 10^6R_g$ (or $\sim 11$pc) for $\dot{M}_{Bondi} = 0.1\dot{M}_{Edd}$, which may provide a favorable conditions to form molecular gas.
In NGC 1097, the ALMA observation traced the $HCN$ molecule around a 40 pc radius \cite[][]{Fathi-etal2013}, and the model results are consistent with molecular observations and predict a higher $\dot{M}_{Bondi}$ ($ > 0.05\dot{M}_{Edd}$) for NGC 1097.
We compared the computed internal energy flux enhancement of fluid ($\epsilon^x_{excess}$) for a given $x$ 
with irradiated flux $\epsilon^{irr}$ (of observed $L_{bol}$), and all parameter sets can launch the wind up to some extent of radius.
Quantitatively, at two different radii, $r =$ $10^5 R_g$ and $10^4 R_g$ the range of the wind hydrogen number density is ($\sim$20 $-$ 300 $cm^{-3}$) and ($\sim$300 $-$ 3000 $cm^{-3}$), respectively, which is consistent with that obtained using MHD simulations for wind from the inner advection disk for LLAGN M81* by \cite{Shi-etal2021}.
The wind is an  equatorial wind as  $v_{wind}$ $\sim v_\phi$, and the constraint $f_z$ depicts its viewing angle $i>85$ degrees. That is, the wind will be visible almost along the disk plane, and it would be a general characteristic for the LLAGNs provided  $\dot{M} (r=R_{tr}) < 0.005 \dot{M}_{Edd}$.

\section*{Acknowledgements} The author thanks an anonymous referee for their comments 
and suggestions. The author is partly supported by the Dr. D.S. Kothari Post-Doctoral Fellowship (201718-PH/17-18/0013) of the University Grant Commission (UGC), New Delhi. The author thanks Banibrata Mukhopadhyay and his group members (2019-2020) for discussions and critical comments.

\section*{Data availability statement}No data sets are analyzed.

%
%

\appendix

\section{Few plots of Figure \ref{fig:wind_gen} 
}

\begin{figure}
\centering 
\begin{tabular}{lr}\vspace{-0.4cm}
  \includegraphics[width=0.44\textwidth]{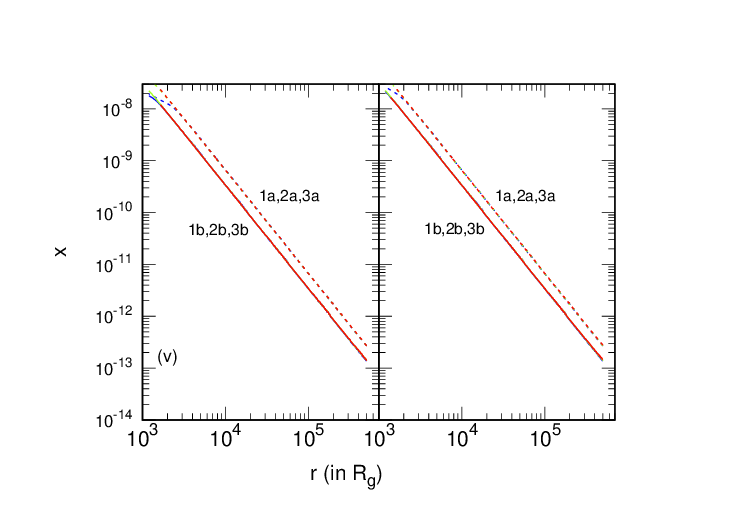}&\hspace{-1.490cm}
  \includegraphics[width=0.44\textwidth]{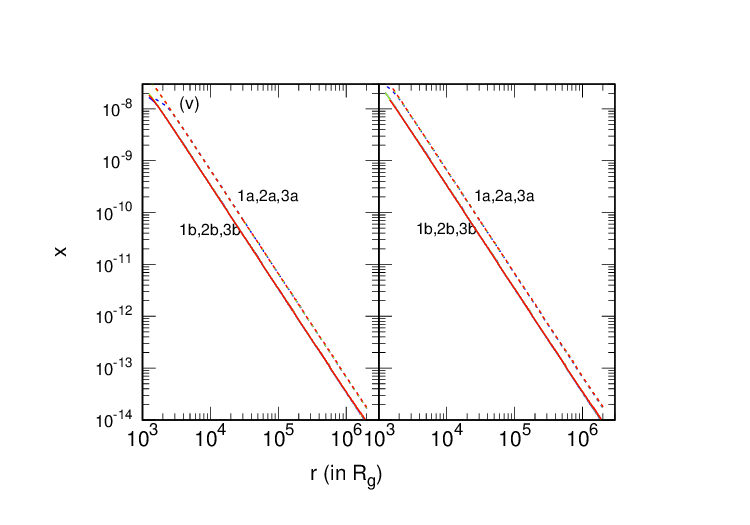}\\\vspace{-0.4cm}
  \includegraphics[width=0.44\textwidth]{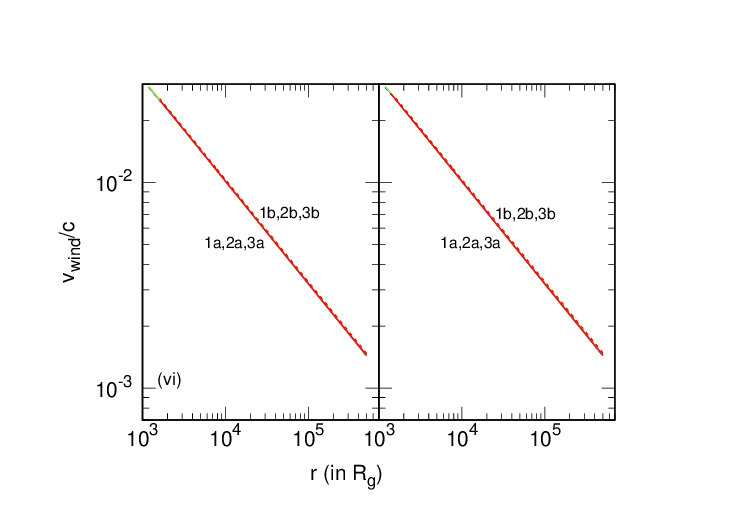}&\hspace{-1.490cm}
  \includegraphics[width=0.44\textwidth]{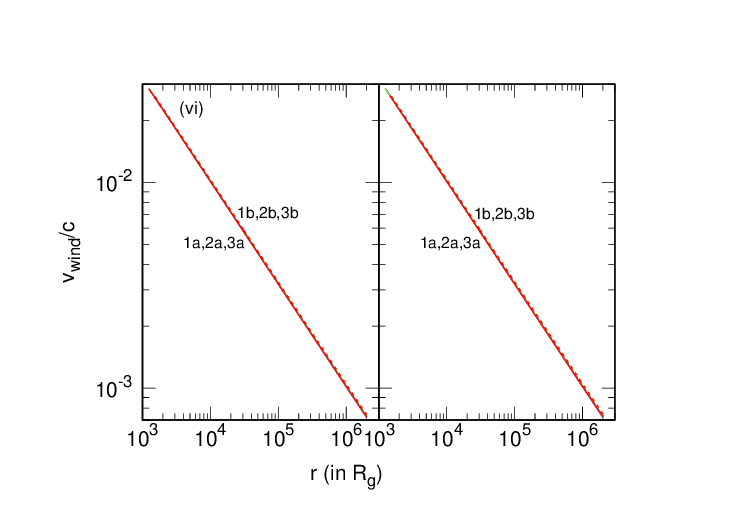}\\\vspace{-0.4cm}
\end{tabular}\vspace{-0.3cm}
\caption{The upper and lower rows are for $x$ and $v_{wind}$ of panels (I) and (II) of Figure \ref{fig:wind_gen}. The rests are the same  as panels (I) and (II) of Figure  \ref{fig:wind_gen}. 
}
\label{fig:wind-gen1}
\end{figure}

\begin{figure}
\centering 
\begin{tabular}{lcr}\vspace{-0.4cm}\hspace{-1.20cm}
  \includegraphics[width=0.44\textwidth]{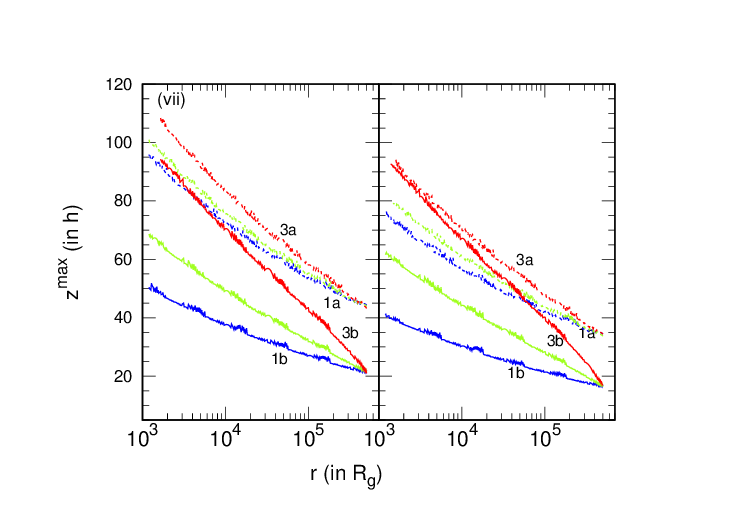}&\hspace{-1.60cm}
  \includegraphics[width=0.44\textwidth]{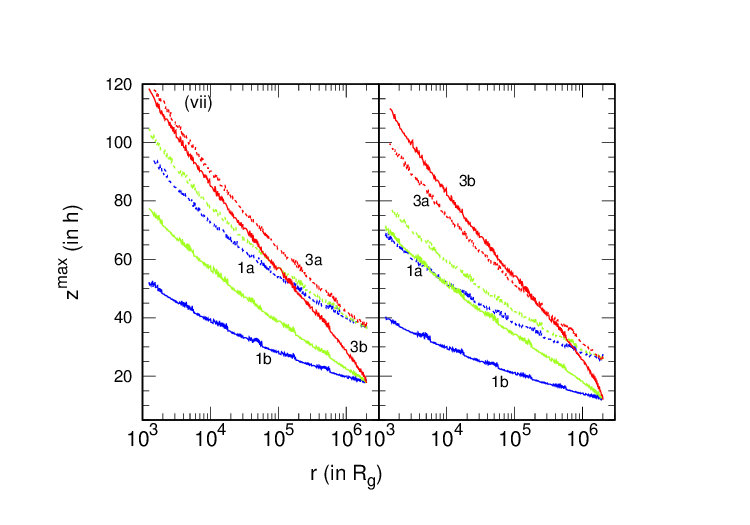}&\hspace{-1.60cm}
  \includegraphics[width=0.44\textwidth]{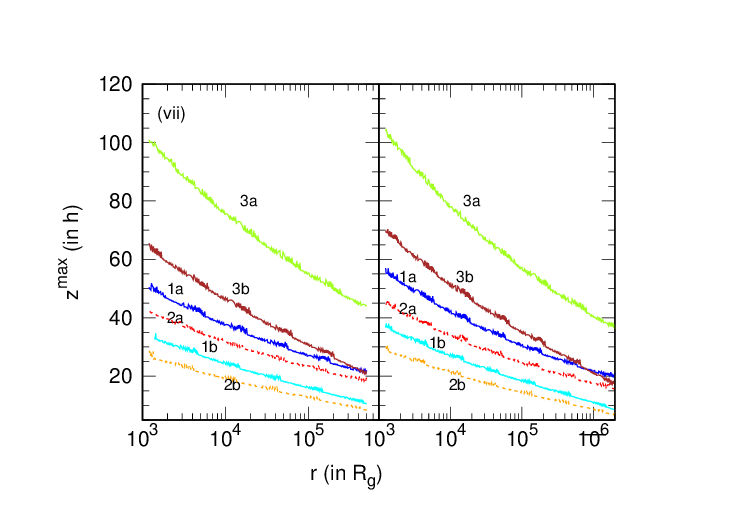}\\
\end{tabular}\vspace{-0.3cm}
\caption{The $z^{max}$ (in units of $h$, however here $z^{max}$ is estimated by fixing the viewing angle $i$, see equation \ref{eq:i}) of Figure \ref{fig:wind_gen}. The rests are the same  as Figure  \ref{fig:wind_gen}. 
}
\label{fig:wind-gen-h}
\end{figure}

\def\aap{A\&A}%
\def\aapr{A\&A~Rev.}%
\def\aaps{A\&AS}%
\def\aj{AJ}%
\def\actaa{Acta Astron.}%
\def\araa{ARA\&A}%
\def\apj{ApJ}%
\def\apjl{ApJ}%
\def\apjs{ApJS}%
\def\apspr{Astrophys.~Space~Phys.~Res.}%
\def\ao{Appl.~Opt.}%
\def\aplett{Astrophys.~Lett.}%
\def\apss{Ap\&SS}%
\def\azh{AZh}%
\def\bain{Bull.~Astron.~Inst.~Netherlands}%
\def\baas{BAAS}%
\def\bac{Bull. astr. Inst. Czechosl.}%
\def\caa{Chinese Astron. Astrophys.}%
\def\cjaa{Chinese J. Astron. Astrophys.}%
\def\fcp{Fund.~Cosmic~Phys.}%
\def\gafd{Geophys.\ Astrophys.\ Fluid Dyn.}
\def\gca{Geochim.~Cosmochim.~Acta}%
\def\grl{Geophys.~Res.~Lett.}%
\def\iaucirc{IAU~Circ.}%
\def\icarus{Icarus}%
\def\jcap{J. Cosmology Astropart. Phys.}%
\def\jcp{J.~Chem.~Phys.}%
\def\jfm{JFM}
\def\jgr{J.~Geophys.~Res.}%
\def\jqsrt{J.~Quant.~Spec.~Radiat.~Transf.}%
\def\jrasc{JRASC}%
\def\mnras{MNRAS}%
\def\memras{MmRAS}%
\def\memsai{Mem.~Soc.~Astron.~Italiana}%
\def\na{New A}%
\def\nar{New A Rev.}%
\def\nat{Nature}%
\def\natas{Nature Astronomy}%
\def\nphysa{Nucl.~Phys.~A}%
\def\pasa{PASA}%
\def\pasj{PASJ}%
\def\pasp{PASP}%
\def\physrep{Phys.~Rep.}%
\def\physscr{Phys.~Scr}%
\def\planss{Planet.~Space~Sci.}%
\def\pra{Phys.~Rev.~A}%
\def\prb{Phys.~Rev.~B}%
\def\prc{Phys.~Rev.~C}%
\def\prd{Phys.~Rev.~D}%
\def\pre{Phys.~Rev.~E}%
\def\prl{Phys.~Rev.~Lett.}%
\def\procspie{Proc.~SPIE}%
\def\qjras{QJRAS}%
\def\rmxaa{Rev. Mexicana Astron. Astrofis.}%
\def\sgg{Stud.\ Geoph.\ et\ Geod.}
\def\skytel{S\&T}%
\def\solphys{Sol.~Phys.}%
\def\sovast{Soviet~Ast.}%
\def\ssr{Space~Sci.~Rev.}%
\def\zap{ZAp}%
\def\memsai{Memorie della Societa Astronomica Italiana}

\bibliographystyle{aasjournal}
\bibliography{wi-llagn}

\end{document}